\documentclass[11pt,a4paper]{article}
\usepackage[english]{babel}
\usepackage[utf8]{inputenc}
\usepackage{epsf}
\usepackage{cite}
\usepackage{graphicx}
\usepackage{subcaption} 

\usepackage{verbatim} 

\usepackage{amsmath}
\usepackage{amssymb}
\usepackage{mathrsfs} 

\usepackage{mathbbol}

\usepackage{amsfonts}
\usepackage{bbding}
\usepackage{slashed}

\usepackage[
colorlinks=true
,urlcolor=black
,anchorcolor=black
,citecolor=black
,filecolor=black
,linkcolor=black
,menucolor=black
,linktocpage=true
,pdfproducer=medialab
,pdfa=true
]{hyperref}
\usepackage{color} 
\usepackage[left=2cm,right=2cm,top=2cm,bottom=2cm]{geometry}
\usepackage{tikz}
\usetikzlibrary{shapes,arrows,positioning,automata,backgrounds,calc,er,patterns}
\usepackage[compat=1.1.0]{tikz-feynman}
\usepackage{multirow,multicol,gensymb}

\vspace*{1cm}
\allowdisplaybreaks
\begin{document}

\begin{center}
\vspace*{1mm}
\vspace{1.3cm}
{\Large\bf
Flavour and precision probes of 
a class of scotogenic models 
}

\vspace*{1.2cm}

{\bf A.~Darricau, H.~Lee, J.~Orloff and A.~M.~Teixeira }

\vspace*{.5cm}
Laboratoire de Physique de Clermont Auvergne (UMR 6533), CNRS/IN2P3,\\
Univ. Clermont Auvergne, 4 Av. Blaise Pascal, 63178 Aubi\`ere Cedex,
France

\end{center}

\vspace*{5mm}
\begin{abstract}
\noindent
We address the phenomenological impact of a well-motivated class of 
scotogenic models regarding flavour and electroweak precision observables. In particular, and for the case of a ``T1-2-A'' variant, we carry out a full computation of the next-to-leading order corrections to leptonic Higgs and $Z$-boson decays. In view of the evolution in the tension between theory and observation regarding the anomalous magnetic moment of the muon, we revisit previously drawn conclusions on operator dominance and constraints on the parameter space. 
Finally, we consider the role of $H \to \mu\mu$ decays (and other flavour conserving Higgs decays), as well as precision observables in probing this class of models at future colliders.
\end{abstract}

\newpage
\section{Introduction}
Despite its many successes, the Standard Model (SM) of particle physics is plagued by several theoretical issues, and more importantly, it cannot explain several observations. 
The latter include neutrino oscillation data, the observed baryon asymmetry of the Universe (BAU) and its cold dark matter (DM) abundance. In recent years, several tensions between the SM theoretical predictions and experiments have emerged, further strengthening the case for considering models of New Physics (NP). 
Many well-motivated extensions have been proposed to address the SM observational problems and ease its theoretical caveats; among them the neutrino portal offers a particularly promising path.
The so-called scotogenic models thus emerge as very promising candidates of NP, by proposing a simultaneous explanation to the smallness of light (active) neutrino masses and viable dark matter candidates~\cite{Ma:2006km,Fraser:2014yha}: in such a framework, discrete symmetries are imposed to stabilise the lightest neutral NP particle - which is the dark matter candidate. As a consequence, neutrino mass generation occurs at higher order (loop-level), leading to a natural suppression of light neutrino masses.

In view of their theoretical appeal, and of the associated phenomenological (and cosmological) impact, scotogenic models have been intensively studied in recent years. Departing from the original minimal model~\cite{Ma:2006km}, numerous realisations and variants have been proposed (see e.g.~\cite{Restrepo:2013aga}), many aiming at augmenting the testability potential, be it concerning dark matter, or then regarding implications for the lepton sector, including observables such as charged lepton flavour violation (cLFV)~\cite{Toma:2013zsa,Vicente:2014wga,Rocha-Moran:2016enp,Avila:2019hhv,Baumholzer:2019twf,Ahriche:2020pwq,DeRomeri:2021yjo,Boruah:2021ayj,Liu:2022byu} and lepton number violation (LNV)~\cite{Mandal:2019oth,Bonilla:2019ipe,Ma:2021eko,DeRomeri:2022cem}. 
An extended field content, with potentially new CP violating couplings has also motivated studies of leptogenesis (and hence an explanation of the BAU).

Recently, a very complete and ambitious scotogenic extension of the SM has been proposed: building from an already promising construction~\cite{Sarazin:2021nwo} (whose parameter space was nevertheless severely compromised due to severe conflicts with bounds from cLFV searches), a variant of the so-called ``T1-2-A'' setup was explored. 
As extensively detailed in~\cite{Alvarez:2023dzz}, such a NP construction, in which the SM is extended via scalar (one doublet and a singlet) and fermionic fields (Dirac doublet and two fermion singlets), successfully allows to address all SM observational problems and furthermore offers a solution to the then existing tension in the anomalous magnetic moment of the muon, $\Delta a_\mu$. 
Moreover, and as many other scotogenic realisations, this variant of the minimal ``T1-2-A'' setup leads to potentially falsifiable cLFV predictions, which add to its overall appeal.

Although in many extensions of the SM lepton sector neutrino oscillation data and cLFV observables often play a leading role in probing important regimes, electroweak precision observables (EWPO) can also be at the source of important constraints (see, for instance~\cite{Abada:2023raf}). The current experimental precision is already very good, and the future prospects - especially in view of the projected precision of the next electron-positron collider (as is the case of FCC-ee, in its $Z$-pole run) - imply that these observables should also be considered upon assessing the viability of many NP models.

By taking as a starting point the above mentioned variant of the minimal ``T1-2-A'' setup, in this work we propose to consider the impact of the new states to several EW precision observables, including oblique parameters, ratios sensitive to the violation of lepton flavour universality, invisible $Z$ decays, leptonic Higgs decays, among others. In view of the existence of potentially large couplings in the model (as the scalar trilinear coupling), one could in principle expect sizeable contributions to the above referred quantities. We have thus carried out a full computation of all relevant one-loop contributions to the decay widths, without any simplifying approximations. As we will subsequently discuss, the comparison of the predicted decay widths (and ratios of different flavoured final states) with the available data might open interesting paths to test scotogenic models, especially in view of the expected sensitivity of the FCC-ee.  

For completeness, we also investigate the prospects for cLFV decays of both $Z$ and Higgs bosons, and revisit the predictions for the cLFV leptonic modes (including radiative and 3-body decays as well as neutrinoless conversion in nuclei). In particular, we extensively discuss how certain regimes (and operator dominance) might be indirectly affected by the requirement of saturating the discrepancy in the muon anomalous magnetic moment. It is important to recall that the previous comparison of the experimental world average on the muon anomalous magnetic moment with the SM prediction proposed in the ``White paper''~\cite{Aoyama:2020ynm} (relying on a data-driven computation of the hadronic vacuum polarisation contribution) suggested a very strong discrepancy, close to $5\sigma$. However, and in view of the evolution of first-principle computations of the hadronic contributions, the tension has been dramatically reduced, now standing close to $1\sigma$\cite{Budapest-Marseille-Wuppertal:2017okr,RBC:2018dos,Giusti:2019xct,Shintani:2019wai,FermilabLattice:2019ugu,Gerardin:2019rua,Borsanyi:2020mff,Lehner:2020crt,Aubin:2022hgm,Boccaletti:2024guq}.

We highlight that our phenomenological approach is distinct from that of~\cite{Alvarez:2023dzz}. Instead of requiring that the present ``T1-2-A'' variant accounts for an explanation to all of the SM observational issues (i.e. oscillation data, the observed BAU, and the DM relic density) and further explain the (existing) tensions in $(g-2)_\mu$, we relax some of the former requirements (as for instance explaining the BAU), and rather pursue a thorough exploration of a vast parameter space, discussing the prospects for a number of observables. In particular, and in view of the evolution in the SM prediction of $(g-2)_\mu$, we emphasise how previous assessments on the parameter space (preferred regimes) might now reflect the change in paradigm.
For instance, certain strong correlations between cLFV observables might become less pronounced upon having a SM-like $(g-2)_\mu$. Our study of both lepton flavour conserving and lepton flavour violating Higgs and $Z$ decays (as well as oblique parameters) suggests some interesting hints to probe this scotogenic model at future colliders\footnote{Close to the completion of this work, the analysis of~\cite{deSouza:2025uxb} was presented; while there is indeed a significant overlap in the topics considered, it is important to emphasise that~\cite{deSouza:2025uxb} primarily focuses on novel techniques to efficiently explore a high dimensional parameter space, and relies on public software for the phenomenological analysis. We will further highlight the common (distinct) points of both studies throughout the manuscript.}.

The manuscript is organised as follows: in Section~\ref{sec:model} we introduce the model considered in this work. Section~\ref{sec:observable} discusses the flavour observables which will be studied, along with their current experimental bounds and projected future sensitivities. Section~\ref{sec:numeric} describes the numerical scan methodology and presents the results. Section~\ref{sec:conclusion} summarises the main findings. Several appendices offer complementary information and detailed technical aspects relevant for the study.

\section{Brief description of the model}\label{sec:model}
As mentioned in the Introduction, we adopt the model which was explored in~\cite{Alvarez:2023dzz}, relying on the extension of the SM field content via an SU(2)$_L$ scalar doublet and a real scalar singlet (respectively, $\eta$ and $S$), two Majorana fermion singlets ($F_{1,2}$) and finally Dirac fermions, which are doublets under SU(2)$_L$, $\Psi_{1,2}$. For completeness, the new fields and the associated SU(2)$_L \times$U(1)$_Y$ charges are summarised in Table~\ref{table:NPcontent}. 
In order to ensure the stability of the lightest neutral state of the NP spectrum - and hence to have a potentially viable DM candidate - a discrete $Z_2$ symmetry is enforced; under the latter, all the SM fields are even, while the new states are odd, effectively preventing the decay of the lightest NP state. 
\begin{table}[h!]
\centering\renewcommand{\arraystretch}{1.4} 
\begin{tabular}{c|cc|cccc}
\hline
\hline
Field & $\eta$ & $S$ & $F_{1}$ & $F_{2}$ & $\Psi_{1}$ & $\Psi_{2}$\\ 
\hline
SU$(2)_L$ & $\mathbf{2}$ & $\mathbf{1}$ & $\mathbf{1}$ & $\mathbf{1}$ & $\mathbf{2}$ & $\mathbf{2}$ \\
U$(1)_Y$ & $1$ & $0$ & $0$ & $0$ & $-1$ & $1$  \\
\hline
\hline
\end{tabular}
\caption{Additional field content of the ``T1-2-A'' scotogenic model variant (cf.~\cite{Alvarez:2023dzz}). All fields are odd under the new $Z_2$ symmetry.} 
\label{table:NPcontent}
\renewcommand{\arraystretch}{1} 
\end{table}
Below we discuss the new interactions, and the extended fermion and scalar spectra. We also notice that concerning lepton number, only even fields under the $Z_{2}$ symmetry (i.e. SM lepton fields) do carry it, in agreement with usual SM attribution.

\subsection{An extended spectrum}\label{sec:model:spectrum}
Extending the SM interaction Lagrangian,  additional terms now encode the interactions of the NP fields with the SM leptons, gauge bosons and the Higgs. Denoting the fermionic doublets as $\Psi_1 = (\Psi_1^0, \Psi_1^-)^T$ and $\Psi_2 = ({\Psi_2^-}^c, -{\Psi_2^0}^c)^T$, one has 
\begin{align}\label{eq:lagrangian:fermion}
\mathcal{L}_\text{fermion}\, &= i\, (\overline{\psi_i} \,\gamma^\mu\, D_\mu\, \psi_i + \overline{F_i} \,\gamma^\mu \,D_\mu \,F_i) \nonumber \\ 
&- M^\nu_{\alpha \beta}\, \overline{\nu_\alpha^c} \,\nu_\beta + M_\psi \,\bar{\psi_1} \,\tilde{\psi_2} - \frac{1}{2} \,{M_F}_{ii} \,\overline{F_i^c}\, F_i + y_{1 i}^*\, \overline{F_i} \,\Phi^\dagger \,\tilde{\psi_1} + y_{2 i}^* \,\overline{F_i} \,\Phi \,\psi_2^c \nonumber\\
&- g_\psi^\alpha \,\tilde{\bar{\psi_2}} \,L^\alpha_L S - g_{F_i}^\alpha \,\widetilde{\overline{L_L^\alpha}}\, \eta F_i - g_R^\alpha \overline{e_R^\alpha} \,\eta^\dagger \,\Psi_1 + \text{H.c.}\, ,
\end{align}
in which we follow the usual notation for the SM fields, with
$L$ and $e^c$ corresponding to the left- and right-handed lepton multiplets, and $\tilde H = i \sigma_2 H^*$ (similarly for $\eta$) as well as $\tilde{\psi} = i \sigma_2 \psi^c$.
Greek indices correspond to lepton flavours ($\alpha, \beta=e, \mu, \tau$), while in Eq.~(\ref{eq:lagrangian:fermion}) $i,j=1,2$ denote generations of fields\footnote{After electroweak symmetry breaking, latin indices will generically denote mass eigenstates.}.  From the above interactions, it is also clear that the couplings $g_\Psi$, $g_F$ and $y_i$ do violate lepton number conservation.
After electroweak symmetry breaking (EWSB), the SM lepton spectrum is enlarged by five new fermions: a charged heavy (Dirac) fermion, $\Psi^\pm$, and four Majorana neutral fermions, $\chi^0_i$. As can be inferred from the mass matrices, in addition to $M_\Psi$ and $M_{F}$, Dirac mass terms arise from Yukawa interactions between the SM Higgs and the neutral components of $\Psi$ and $F$.
The interaction and physical (mass) eigenstates are related via the unitary mixing matrix $U_{\chi}$:
\begin{equation}\label{eq:chi:M:Uchi}
    \{\chi^0\}^T \, =\, U_\chi^T\, \{ F_1, F_2, \Psi^0_1, (\Psi^0_2)^c \}^T\, 
    \quad \text{in which}\quad 
    M_{\chi^0}\, =\, U_\chi\, M_{\chi^0}^\text{diag}\, U^T_\chi\,.
\end{equation}
In the above, we have introduced the neutral fermion mass matrix, $M_{\chi^0}$, which in the interaction basis is given by 
\begin{equation}\label{eq:chi0:matrix}
M_{\chi^0} \, =\, \begin{pmatrix}
        M_{1} & 0 & \frac{v}{\sqrt{2}}\, y_{11} & \frac{v}{\sqrt{2}} \,y_{21} \\
        0 & M_{2} & \frac{v}{\sqrt{2}}\, y_{12} & \frac{v}{\sqrt{2}} \,y_{22} \\
        \frac{v}{\sqrt{2}} \,y_{11} & \frac{v}{\sqrt{2}} \,y_{12} & 0 & M_{\Psi} \\
        \frac{v}{\sqrt{2}}\, y_{21} & \frac{v}{\sqrt{2}} \,y_{22} & M_{\Psi} & 0 
    \end{pmatrix}\,.
\end{equation}
 The mass of the charged heavy Dirac fermion is trivially given by $M_{\psi}$. Other than the SM scalar interactions ($M_H^2|H|^2 + \lambda_H |H|^4$), new terms are present in the scalar potential; these concern (self-) interactions of the new scalar and singlet, as well as further terms reflecting the possible interactions with the SM Higgs doublet, 
\begin{align}\label{eq:Vscalar}
\mathcal{V}_\text{scalar}\, &= \frac{1}{2} M_S^2\, S^2 + \frac{1}{2} \lambda_{4 S} \,S^4 + M_\eta^2 \,|\eta|^2 + \lambda_{4 \eta} \,|\eta|^4 + \frac{1}{2} \lambda_S \,S^2 |H|^2 + \frac{1}{2} \lambda_{S \eta} \,S^2 |\eta|^2 \nonumber \\ &
+ \lambda_{\eta} \,|\eta|^2\,|H|^2 + \lambda_{\eta}^\prime \,|\eta H^\dagger|^2 + \frac{1}{2} \lambda_{\eta}^{\prime \prime} \left[ \left(H \eta^\dagger \right)^2 +\text{H.c.} \right] + \alpha \,S \left[ H \eta^\dagger + \text{H.c.} \right].
\end{align}
After EWSB, (only) the SM Higgs develops its vacuum expectation value, and the scalar sector comprises the following states
\begin{equation}
    H = \begin{pmatrix}
        G^{+} \\
        \frac{1}{\sqrt{2}} \left( v + h^{0} + i G^{0} \right)
    \end{pmatrix}, \quad
    \eta = \begin{pmatrix}
        \eta^{+} \\ \frac{1}{\sqrt{2}} \left( \eta^{0} + i A^{0} \right)
    \end{pmatrix}
    \,,\quad
    S\,,
\end{equation}
recalling that $S$ is a real singlet scalar. The physical scalar sector is thus composed of charged and three neutral states. Under the assumption of CP-conservation in the scalar sector, i.e. taking $\alpha$ and $\lambda_\eta^{\prime \prime}$ to be real, there is no mixing between scalar and pseudoscalar neutral bosons. 
In the interaction basis, the scalar mass matrix is thus given by
\begin{equation}\label{MPhi:def}
    M_{\phi}^{2} = \begin{pmatrix}
        M_{S}^{2} + \frac{1}{2} v^{2} \lambda_{S} & v \alpha & 0 \\
        v \alpha & M_{\eta}^{2} + \frac{1}{2} v^{2} \lambda_{L} & 0 \\
        0 & 0 & M_{\eta}^{2} + \frac{1}{2} v^{2} \lambda_{A}
    \end{pmatrix}\,,
\end{equation}
in which $\lambda_{L,A} = \lambda_{\eta} + \lambda_{\eta}^{\prime} \pm \lambda_{\eta}^{\prime\prime}$. The above mass matrix can be diagonalised as
\begin{equation}
    \{\phi_1, \phi_2, A_0\}^T \, =\, U_\phi^T\, \{ S_0, \eta_0, A_0 \}^T\, 
    \quad \text{in which}\quad 
    M_{\phi}^2\, =\,U_\phi\, ({M_{\phi}^{\text{diag}}})^2\, U^T_\phi\,,
\end{equation}
with the unitary mixing matrix $U_{\phi}$ parametrised as 
\begin{equation}\label{eq:UPhi:def}
    U_\phi = 
    \begin{pmatrix}
    \cos{\theta_S} & \sin{\theta_S} & 0\\
    - \sin{\theta_S} & \cos{\theta_S} & 0\\
    0 & 0 & 1\\
    \end{pmatrix}\,,
\end{equation}
and the angle $\theta_S$ given by
\begin{equation}
    \cos{\theta_S} = \frac{\text{sign}(\alpha)}{\sqrt{2}} \sqrt{1 + \frac{ M_\eta^2 - M_S^2 + \frac{1}{2} \left( \lambda_L - \lambda_S \right) v^2 }{\sqrt{4 \alpha^2 v^2 + \left( M_\eta^2 - M_S^2 + \frac{1}{2} \left( \lambda_L - \lambda_S\right) v^2\right)^2}}}\,.
\end{equation}

\subsection{Higher-order neutrino mass generation}
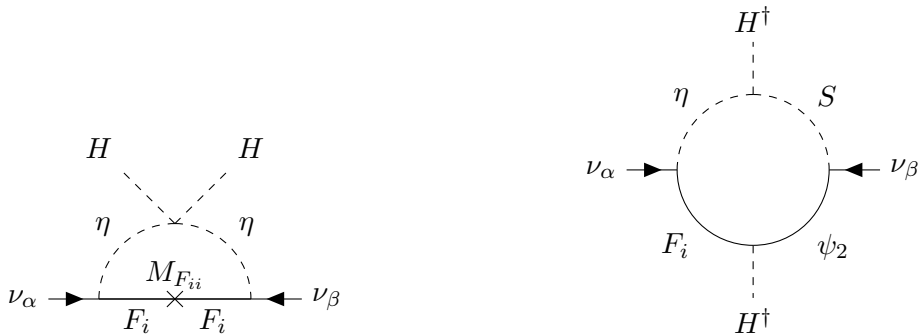
\begin{figure}[h!]
    \centering
    \begin{subfigure}[b]{0.48\textwidth}
        \centering
        \begin{tikzpicture}
        \begin{feynman}
        \vertex (a) at (0,0) {\(\nu_\alpha\)};
        \vertex (b) at (1,0);
        \vertex (c) at (2,0);
        \vertex (d) at (2,1);
        \vertex (e) at (3,0);
        \vertex (f) at (4,0) {\(\nu_\beta\)};
        \vertex (g) at (3,2) {\( H\)};
        \vertex (h) at (1,2) {\( H\)};
        \diagram* {
        (a) -- [fermion] (b),
        (b) -- [plain, edge label'=\( F_i\)] (c),
        (c) -- [plain, edge label'=\( F_i\)] (e),
        (b) -- [scalar, quarter left, edge label=\( \eta\)] (d),
        (d) -- [scalar, quarter left, edge label=\( \eta\)] (e),
        (b) -- [insertion = 0.5, edge label=\( M_{F_{ii}}\)] (e),
        (f) -- [fermion] (e),
        (d) -- [scalar] (g),
        (d) -- [scalar] (h)
        };
        \end{feynman}
        \end{tikzpicture}
    \end{subfigure}
    \begin{subfigure}[b]{0.4\textwidth}
        \centering
        \begin{tikzpicture}
        \begin{feynman}
        \vertex (a) at (0,0) {\(\nu_\alpha\)};
        \vertex (b) at (1,0);
        \vertex (c) at (2,-1);
        \vertex (d) at (2,1);
        \vertex (e) at (3,0);
        \vertex (f) at (4,0) {\(\nu_\beta\)};
        \vertex (g) at (2,2) {\( H^\dagger\)};
        \vertex (h) at (2,-2) {\( H^\dagger\)};
        \diagram* {
        (a) -- [fermion] (b),
        (b) -- [plain, quarter right, edge label'=\( F_i\)] (c),
        (c) -- [plain, quarter right, edge label'=\( \psi_2\)] (e),
        (b) -- [scalar, quarter left, edge label=\( \eta\)] (d),
        (d) -- [scalar, quarter left, edge label=\( S\)] (e),
        (f) -- [fermion] (e),
        (d) -- [scalar] (g),
        (c) -- [scalar] (h)
        };
        \end{feynman}
        \end{tikzpicture}
    \end{subfigure}
    \caption{One-loop diagrams contributing to neutrino masses (in the interaction basis).}
    \label{fig:NMassInt}
\end{figure}

As common to all scotogenic realisations, the discrete symmetry ensuring the stability of the lightest neutral NP particle precludes neutrino mass generation at the tree-level; loop-level generation, together with the smallness of the new couplings thus allows for a more natural explanation of neutrino masses and oscillation data in general. 
For the particular case of scotogenic models usually categorised under the so-called ``T1-2'' topology (i.e. two additional fermions and two new scalars, cf.~\cite{Restrepo:2013aga}), and for the case of two extra singlets (scalar and fermionic),
the leading order contributions to Majorana neutrino masses can be realised at one-loop, as shown in Fig.~\ref{fig:NMassInt} (in which the new fields in the loop are taken in the interaction basis).

In turn, this gives rise to contributions to the neutrino masses of the form  
$\overline{\nu_{\beta}^{c}} \left( \mathcal{M}_{\nu} \right)_{\beta \alpha} \nu_{\alpha}$; 
the neutrino mass matrix can be cast in a compact manner relying on a generalised matrix of ``couplings''
$\mathcal{G}$, and on the contributions arising from the exchange of the new massive fields in the loop,
$\mathcal{M}_{L}$~\cite{Alvarez:2023dzz}:
\begin{equation}\label{eq:Gmatrix}
    \mathcal{M}_{\nu} = \mathcal{G}^T \mathcal{M}_{L} \mathcal{G}, \quad \text{where} \quad \mathcal{G} = \begin{pmatrix}
        g_{\psi}^{e} & g_{\psi}^{\mu} & g_{\psi}^{\tau} \\[5pt]
        g_{F_{1}}^{e} & g_{F_{1}}^{\mu} & g_{F_{1}}^{\tau} \\[5pt]
        g_{F_{2}}^{e} & g_{F_{2}}^{\mu} & g_{F_{2}}^{\tau}
    \end{pmatrix}\,.
\end{equation}
The detailed decomposition and computation of $\mathcal{M}_{L}$ is presented in Appendix~\ref{app:neutrinomass};
the full computation of the neutrino mass matrix $\mathcal{M}_{\nu}$ relies on a modified Casas-Ibarra parameterisation~\cite{Casas:2001sr,Basso:2012voo}, and the approach followed is also detailed in Appendix~\ref{app:neutrinomass}. As mentioned in the latter, we use the most recent data from the NuFit collaboration~\cite{Esteban:2024eli}, and for illustration purposes mostly consider a normal ordering of the neutrino spectrum.

\subsection{Dark matter candidates}
If odd under the new $Z_{2}$ symmetry, the lightest state of the extended neutral fermion and scalar sectors are potential DM
candidates: these include the lightest CP-even scalar $\phi_{1}$, the CP-odd scalar $A^{0}$ and the lightest fermion $\chi_{1}^{0}$. In addition to having a viable relic density~\cite{Planck:2018vyg}, the lightest stable neutral particle must comply with constraints arising from numerous direct and indirect DM searches.
In order to do so, we rely on {micrOMEGAs}~\cite{Alguero:2023zol}, which provides a full evaluation of the dark matter prospects of a given NP model, offering the computation of the relic density, as well as of direct and indirect detection cross-sections (which can be then confronted with available experimental bounds and future prospects). In this analysis, we consider a point to have a valid dark matter candidate if its relic density lies within the $3 \sigma$ bound from theoretical uncertainties\footnote{In~\cite{Alvarez:2023dzz}, it is argued that theoretical uncertainties are higher than the experimental one by considering electroweak radiative corrections~\cite{Boudjema:2014gza,Harz:2016dql}.},
\begin{equation}\label{eq:omega:planck}
    \Omega_{\operatorname{CDM}} \,h^2 \,= \,0.120 \pm 0.012\,,
\end{equation}
while also having a spin-independent direct detection cross section below the LUX-ZEPLIN~\cite{LZ:2022lsv} experimental limit at 95\% CL.

\section{Flavour and electroweak observables in the scotogenic ``T1-2-A'' variant}\label{sec:observable}

The extended spectrum and new interactions present in this model open the door to many interesting new phenomena, among them contributions to cLFV leptonic observables and to the muon anomalous magnetic moment, as discussed in~\cite{Alvarez:2023dzz}.
Here, we revisit many of the latter observables, especially in light of the evolution concerning $\Delta a_\mu$. Moreover, we also investigate cLFV $Z$ and Higgs boson decays. 
As mentioned in the Introduction, we also study the prospects for a number of EW precision observables, including oblique parameters, invisible $Z$ and Higgs decays, and ratios of decays sensitive to the violation of lepton flavour universality. We have accordingly derived the full expressions for the observables (carrying out renormalisation when necessary). In the following subsections, we address the distinct classes of observables, focusing on the new contributions; we also provide (approximate) analytic expressions for the considered processes. 

\subsection{Anomalous magnetic moment of the muon}\label{sec:AMM}
We begin our discussion by considering this model's contributions to the anomalous magnetic moment of the muon. While in previous studies emphasis was given to the consequences of saturating a discrepancy between SM prediction and observation nearing $5\sigma$, here we propose to re-assess a new context, in which the SM is in good agreement with experiment. 

\begin{figure}[h!]
    \centering
        \begin{tikzpicture}
        \begin{feynman}
        \vertex (a) at (0,0) {\(\ell_\alpha\)};
        \vertex (b) at (1,0);
        \vertex (c) at (2,0);
        \vertex (d) at (3,0) {\(\ell_\beta\)};
        \vertex (e) at (2.3,0.2);
        \vertex (f) at (2.3,1.2) {\(\gamma\)};
        \diagram* {
        (a) -- [fermion] (b),
        (b) -- [fermion, half left, edge label=\( \psi^\pm\)] (c),
        (b) -- [scalar, half right, edge label'=\( \phi^k\)] (c),
        (c) -- [fermion] (d),
        (e) -- [boson] (f)
        };
        \end{feynman}
        \end{tikzpicture}
   \hspace*{20mm}
        \begin{tikzpicture}
        \begin{feynman}
        \vertex (a) at (0,0) {\(\ell_\alpha\)};
        \vertex (b) at (1,0);
        \vertex (c) at (2,0);
        \vertex (d) at (3,0) {\(\ell_\beta\)};
        \vertex (e) at (2.3,0.2);
        \vertex (f) at (2.3,1.2) {\(\gamma\)};
        \diagram* {
        (a) -- [fermion] (b),
        (b) -- [plain, half right, edge label'=\( \chi_{_i}\)] (c),
        (b) -- [scalar, half left, edge label=\( \eta^\pm\)] (c),
        (c) -- [fermion] (d),
        (e) -- [boson] (f)
        };
        \end{feynman}
        \end{tikzpicture}
        \mbox{\hspace*{31mm}
        (a)
        \hspace*{51 mm}
    (b)
    \hspace*{30mm}
        }
    \caption{New contributions to radiative lepton processes in the physical basis: for $\alpha = \beta =\mu$, NP contributions to the muon anomalous magnetic moment; for $\alpha \neq \beta$, cLFV radiative decays. On the left (right), neutral scalar-charged fermion (neutral fermion-charged scalar) exchange.}
    \label{fig:radiative:AMM:cLFV}
\end{figure}
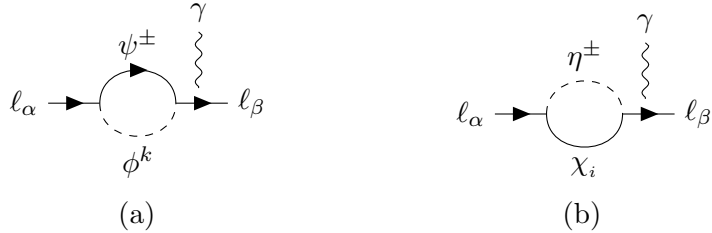

The new fields present are at the source of contributions to radiative lepton transitions, be it flavour conserving (as the case of magnetic moments) or flavour violating (which we will address in a subsequent section). In both cases (that is $\ell_\alpha \to \ell_\beta \gamma$, with $\alpha = \beta$ or $\alpha \neq \beta$), the leading diagrams are schematically depicted in Fig.~\ref{fig:radiative:AMM:cLFV}.

For both types of transitions, and relying on the effective field theory (EFT) approach, the coefficient of the  relevant dimension-six term, i.e. $\propto  c_{R}^{\alpha \beta} \, \overline{\ell}_{\beta} \,\sigma_{\mu\nu} \, P_{R}\, \ell_{\alpha} \,F^{\mu\nu}$
(the dipole contribution) can be cast  
as~\cite{Alvarez:2023dzz}
\begin{eqnarray}\label{eqn:wilson_cRij}
    c_{R}^{\alpha \beta} & = & \sum_i \frac{e}{64 \pi^2 M_{\chi_{_i}}^2} \left[ (\Gamma^{\beta i}_L )^*\, \Gamma^{\alpha i}_R\, M_{\chi_{_i}}\, f_{\chi,2}(x_i) + \left( m_{\ell_\beta} ( \Gamma^{\beta i}_L )^*\, \Gamma^{\alpha i}_L + m_{\ell_\alpha} ( \Gamma^{\beta i}_R )^* \,\Gamma^{\alpha i}_R \right) g_2(x_i) \right] \nonumber \\
    & - & \sum_k \frac{e}{64 \pi^2 M_{\phi_{_k}}^2} \left[ ( \Gamma^{\beta k}_L )^*\, \Gamma^{\alpha k}_R \,M_{\phi_{_k}} \,f_{\phi,2}(x_k) + \left( m_{\ell_\beta} ( \Gamma^{\beta k}_L )^* \,\Gamma^{\alpha k}_L + m_{\ell_\alpha}( \Gamma^{\beta k}_R )^* \, \Gamma^{\alpha k}_R \right) g_2(x_k) \right].
\end{eqnarray}
In the above, $e$ is the electric charge, $m_{\ell_\alpha}$ denotes the charged lepton mass, and $\Gamma_{L,R}$ the left- and right-handed interactions between physical states. For the transitions on Fig.~\ref{fig:radiative:AMM:cLFV}, these are respectively given by
\begin{align}\label{eq:GammaLR}
    \Gamma_L^{\alpha k}  &=  - g_\psi^\alpha\, U_\phi^{1 k}\,, \nonumber \\
    \Gamma_R^{\alpha k} & = \frac{\left(g_R^\alpha \right)^*}{\sqrt{2}} 
 \left( U_\phi^{2 k} + i U_\phi^{3 k} \right)\,, \nonumber \\ 
 \Gamma_L^{\alpha i} & =  g_{F_1}^\alpha \left(U_\chi^{1 i} \right)^* + g_{F_2}^\alpha \left(U_\chi^{2 i} \right)^* \,, \nonumber \\
    \Gamma_R^{\alpha i} & = \left(g_R^\alpha \right)^* 
 U_\chi^{3 i}\,,
\end{align}
and stem from the following interactions 
\begin{equation}\label{eq:phys:Gamma:lag}
    \mathcal{L} \supset \overline{\ell_\alpha}\,\left(\Gamma_L^{\alpha k} \,P_L + \Gamma_R^{\alpha k} \,P_R \right)\,\phi_k \,\psi^- + \dots + \overline{\ell_\alpha}\, \left(\Gamma_L^{\alpha i}\, P_L + \Gamma_R^{\alpha i} \,P_R\right) \,\eta^- \,\chi_i\,,
\end{equation}
with $\alpha, \beta$ denoting the charged lepton flavour, and $i=1-4$ and $k=1-3$ physical neutral fermion and scalar eigenstates; the couplings $g_{\psi}$, $g_{R}$ and $g_{F_i}$ were introduced in Eq.~(\ref{eq:lagrangian:fermion}). Moreover, $g_{F_i}$ and $g_{\psi}$ enter the generalised coupling matrix, see Eq.~(\ref{eq:Gmatrix}).
Other than an overall transposition (a consequence of the convention used for the diagonalisation of the new scalar and fermion mass matrices), one finds a small discrepancy when compared to~\cite{Alvarez:2023dzz} ($\Gamma_R^{\alpha i}$ depends on $U_\chi^{3 i}$ and not on its complex conjugate).
The associated loop functions are given by
\begin{align}
       f_{\chi,2}(x) &= \frac{x^{2} - 1 - 2x\ln (x)}{
       (x-1)^{3}}\, , \quad \forall x \neq 1\, ;
       \quad 
        f_{\chi,2}(1) = \frac{1}{3} ,\nonumber \\
       f_{\phi,2}(x) &= \frac{x^{2} - 4x + 3 + 2\ln (x)}{
       (x-1)^{3}}\, , \quad 
       \forall x \neq 1\, ;
       \quad 
        f_{\phi,2}(1) = \frac{2}{3} ,\nonumber \\
        g_2(x) &= \frac{x^{3} - 6x^{2} + 3x + 2 + 6x\ln (x)}{(x-1)^{4}}
        \, , \quad 
        \forall x \neq 1\, ;
       \quad 
        g_2(1) = \frac{1}{12} \,, 
\end{align}
in which $x_i = M_{\eta^\pm}^2/M_{\chi_{_i}}^2$, $x_k = M_\psi^2/M_{\phi_k}^2$ denote the mass ratios of the new particles in the loop.

As usual, the diagonal elements of the dipole coefficient $c_{R}^{\alpha \alpha}$ are at the source of the anomalous magnetic and electric dipole moments of charged leptons (respectively real and imaginary parts).
The NP contributions to the anomalous magnetic moment (corresponding to the potential discrepancies between SM prediction and observation) are thus given by 
\begin{equation}
    \Delta a_{\alpha} \,= \,-4 \frac{m_{\ell_{\alpha}}}{e} \, \text{Re} \left( c_{R}^{\alpha\alpha} \right)\,.
\end{equation}
In what follows we will consider two illustrative benchmark values for the tension between the SM prediction and observation in $(g-2)_\mu$, $\Delta a_{\mu}$:
\begin{align}
        \Delta a_{\mu} &= \left( 25.1 \pm 5.9 \right) \times 10^{-10} \quad \left( 4.2\sigma \right)\,, \\
        \Delta a_{\mu} &= \left( 10.7 \pm 7.0 \right) \times 10^{-10} \quad \left( 1.5\sigma \right)\,,
\end{align}
and compare the constraints that each of the above imposes on the model's parameters, as well as on the regimes for the cLFV observables, which we proceed to discuss.

\subsection{Leptonic cLFV transitions and decays: radiative, 3-body and conversion in muonic atoms}\label{sec:cLFV:lepton}

We briefly introduce the present NP model's contributions to a set of cLFV observables. The corresponding current bounds and projected future sensitivities are summarised in~Table~\ref{tab:cLFV_current_future_sensitivity}. 

\renewcommand{\arraystretch}{1.3}
\begin{table}[h!]
    \centering
    \hspace*{-2mm}{\small\begin{tabular}{|c|c|c|}
    \hline
    Observable & Current bound & Future sensitivity  \\
    \hline\hline
    $\text{BR}(\mu\to e \gamma)$    &
    \quad $<1.5\times 10^{-13}$ \quad (MEG II~\cite{MEGII:2025gzr})   &
    \quad $6\times 10^{-14}$ \quad (MEG II~\cite{Baldini:2018nnn}) \\
    $\text{BR}(\tau \to e \gamma)$  &
    \quad $<3.3\times 10^{-8}$ \quad (BaBar~\cite{Aubert:2009ag})    &
    \quad $3\times10^{-9}$ \quad (Belle II~\cite{Kou:2018nap})      \\
    $\text{BR}(\tau \to \mu \gamma)$    &
     \quad $ <4.2\times 10^{-8}$ \quad (Belle~\cite{Belle:2021ysv})  &
    \quad $10^{-9}$ \quad (Belle II~\cite{Kou:2018nap})     \\
    \hline
    $\text{BR}(\mu \to 3 e)$    &
     \quad $<1.0\times 10^{-12}$ \quad (SINDRUM~\cite{Bellgardt:1987du})    &
     \quad $10^{-15(-16)}$ \quad (Mu3e~\cite{Blondel:2013ia})   \\
    $\text{BR}(\tau \to 3 e)$   &
    \quad $<2.7\times 10^{-8}$ \quad (Belle~\cite{Hayasaka:2010np})&
    \quad $5\times10^{-10}$ \quad (Belle II~\cite{Kou:2018nap})     \\
    $\text{BR}(\tau \to 3 \mu )$    &
    \quad $<1.9\times 10^{-8}$ \quad (Belle II~\cite{Belle-II:2024sce})  &
    \quad $5\times10^{-10}$ \quad (Belle II~\cite{Kou:2018nap})     \\
    & & \quad$5\times 10^{-11}$\quad (FCC-ee~\cite{Abada:2019lih})\\
    \hline
    $\text{CR}(\mu- e, \text{N})$ &
     \quad $<7 \times 10^{-13}$ \quad  (Au, SINDRUM~\cite{Bertl:2006up}) &
    \quad $10^{-14}$  \quad (SiC, DeeMe~\cite{Nguyen:2015vkk})    \\
    & &  \quad $2.6\times 10^{-17}$  \quad (Al, COMET~\cite{Krikler:2015msn,COMET:2018auw,Moritsu:2022lem})  \\
    & &  \quad $8 \times 10^{-17}$  \quad (Al, Mu2e~\cite{Bartoszek:2014mya})\\    
    \hline
    \end{tabular}}
    \caption{Current experimental bounds and future sensitivities on relevant leptonic cLFV observables. Notice that limits are given at $90\%\:\mathrm{C.L.}$, and that Belle II projected sensitivities rely on an integrated luminosity of $50\:\mathrm{ab}^{-1}$.}
    \label{tab:cLFV_current_future_sensitivity}
\end{table}
\renewcommand{\arraystretch}{1.}

\paragraph{Radiative decays: $\pmb{\ell_\alpha \to \ell_\beta \gamma}$}
Following the above discussion and in agreement with the findings of~\cite{Hisano:1995cp}, the rates of the cLFV radiative decays can be expressed as
\begin{equation}
        \text{BR}( \ell_{\alpha} \rightarrow \ell_{\beta} \gamma ) \,= \,\frac{m_{\ell_{\alpha}}^{3}}{4\pi\Gamma_{\ell_{\alpha}}} \left(|c_{R}^{\alpha\beta} |^{2} + | c_{R}^{\beta\alpha} |^{2} \right)\,,
    \label{eqn:cLFV:radiative}
\end{equation}
in which $\Gamma_{\ell_{\alpha}}$ is the charged lepton decay width. 
In view of the very similar structure of the NP processes at the source of (new) contributions to $(g-2)_\mu$ and radiative decays, it is only natural to expect that conflicts may arise from a simultaneous attempt to saturate a sizeable $\Delta a_\mu$ while complying with available bounds on cLFV transitions. This will be further discussed upon presentation of the numerical results (in Section~\ref{sec:results:cLFV}). 

\paragraph{Three-body decays: $\pmb{\ell_\alpha \to 3 \ell_\beta}$}

A number of distinct operators can contribute to 
$\ell_\alpha \to 3 \ell_\beta$ decays\footnote{Notice that here we will only consider the $\ell_\alpha \to 3 \ell_\beta$ decays for simplicity (and not the generic and more involved $\ell_\alpha \to  \ell_\beta \ell_\gamma \ell_\gamma^\prime$ transitions.}: in addition to photon-penguins, one can have $Z$ and Higgs mediated penguins, as well as box diagrams. 
We have computed the full amplitudes for the transitions, and below we present the analytic expression (following the notation introduced in~\cite{Abada:2014kba}):
\begin{align}
    \label{eqn:analytic_lto3lp}
        \text{BR}\left( \ell_{\alpha} \rightarrow 3\ell_{\beta} \right) &= \frac{m_{\ell_{\alpha}}^{5}}{512\pi^{3} \Gamma_{{\ell_{\alpha}}}} \left[ e^{4} \left( \lvert K_{2}^{L} \rvert^{2} + \lvert K_{2}^{R} \rvert^{2} \right) \left( \frac{16}{3} \ln \frac{m_{\ell_{\alpha}}}{m_{\ell_{\beta}}} - \frac{22}{3} \right) \right. \nonumber \\
        &+ \frac{1}{24} \left( \lvert A_{LL}^{S} \rvert^{2} + \lvert A_{RR}^{S} \rvert^{2} \right) + \frac{1}{12} \left( \lvert A_{LR}^{S} \rvert^{2} + \lvert A_{RL}^{S} \rvert^{2} \right)\nonumber  \\
        &+ \frac{2}{3} \left( \lvert \hat{A}_{LL}^{V} \rvert^{2} + \lvert \hat{A}_{RR}^{V} \rvert^{2} \right) + \frac{1}{3} \left( \lvert \hat{A}_{LR}^{V} \rvert^{2} + \lvert \hat{A}_{RL}^{V} \rvert^{2} \right) + 6 \left( \lvert \hat{A}_{LL}^{T} \rvert^{2} + \lvert \hat{A}_{RR}^{T} \rvert^{2} \right) \nonumber \\
        &+ \frac{e^{2}}{3} \left( K_{2}^{L} \,A_{RL}^{S*} + K_{2}^{R} A_{LR}^{S*} + \text{c.c} \right) - \frac{2e^{2}}{3} \left( K_{2}^{L}\, \hat{A}_{RL}^{V*} + K_{2}^{R} \,\hat{A}_{LR}^{V*} + \text{c.c} \right) \nonumber \\
        &- \frac{4e^{2}}{3} \left( K_{2}^{L} \,\hat{A}_{RR}^{V*} + K_{2}^{R}\, A_{LL}^{V*} + \text{c.c} \right) \nonumber \\
        &\left. -\frac{1}{2} \left( A_{LL}^{S} \,A_{LL}^{T*} + A_{RR}^{S}\, A_{RR}^{T*} + \text{c.c} \right) -\frac{1}{6} \left( A_{LR}^{S} \,\hat{A}_{LR}^{V*} + A_{RL}^{S} \,\hat{A}_{RL}^{V*} + \text{c.c} \right) \right]\,,
\end{align}
in which
\begin{equation}
  \hat{A}_{XY}^{V} = A_{XY}^{V} + e^{2} K_{1}^{X}\,, \quad \text{with }\quad X, Y = L, R\,. 
\end{equation}
In the above, $A$ (and $\hat{A}$) correspond to the 4-fermion ($4 \ell$) form factors, which can be decomposed into photon, $Z$ and Higgs penguin as well as box contributions.
Let us mention that the anapole (dipole) $\gamma$-penguin 
contributions are referred to as $K_{1}^{L/R}$ ($K_{2}^{L/R}$) where $K_{2}^{R} = 2 c_{R}/m_{\ell_{_\alpha}}$ and $L \leftrightarrow R $. 
The detailed (analytical) expressions of the remaining operators can be found in Appendix~\ref{sec:FF}.

\paragraph{Neutrinoless muon-electron conversion in nuclei}
Muon-electron conversion in the presence of nuclei offers the most promising future prospects for a cLFV discovery, cf.~Table~\ref{tab:cLFV_current_future_sensitivity}, and is thus one of the most sensitive probes of NP in the lepton sector.
The analytic expression for the conversion rate can be cast as:
\begin{align}
        \text{CR}\left( \mu \rightarrow e, \text{Nucleus}(A,Z) \right) &=  \frac{1}{\Gamma_{\text{capt}}}\times
        \frac{p_{e} \,E_{e} \,m_{\mu}^{3} \,G_{F}^{2} \alpha^{3}\, Z_{\text{eff}}^{4} \,F_{P}^{2}}{8\pi^{2} \,Z} \nonumber\\
        &\times \left[ \left| \left( Z+N \right) \left( g_{LV}^{\left( 0 \right)} + g_{LS}^{\left( 0 \right)} \right) + \left( Z-N \right) \left( g_{LV}^{\left( 1 \right)} + g_{LS}^{\left( 1 \right)} \right) \right |^{2} \right. \nonumber\\
        &+ \left. \left | \left( Z+N \right) \left( g_{RV}^{\left( 0 \right)} + g_{RS}^{\left( 0 \right)} \right) + \left( Z-N \right) \left( g_{RV}^{\left( 1 \right)} + g_{RS}^{\left( 1 \right)} \right) \right |^{2} \right] \,.
\end{align}
In addition to quantities already introduced, in the above $p_{e}$ and $E_{e}$ are the momentum and energy of the electron, with $\Gamma_{\text{capt}}$ the total muon capture rate. Also, $G_F = e^2\sqrt{2}/(8 s_w^2 M_W^2)$ denotes the Fermi constant, with $s_w = \sin \theta_w$ (the weak angle). The nucleus is characterised by the number of protons and neutrons ($Z$ and $N$ respectively), and $Z_{\text{eff}}$ is the effective atomic charge~\cite{Chiang:1993xz}.
The effective couplings, $g_{XK}^{\left(0,1\right)}$, can be decomposed as
\begin{align}
    g_{XK}^{\left(0\right)} &= \frac{1}{2} \sum_{q=u,d,s} \left( g_{XK(q)} \,G_K^{q,p} + g_{XK(q)} \,G_K^{q,n}\right), 
    \quad 
    g_{XK}^{\left(1\right)} = \frac{1}{2} \sum_{q=u,d,s} \left( g_{XK(q)} \,G_K^{q,p} - g_{XK(q)} \,G_K^{q,n}\right),
\end{align}
where $X=L,R$ and $K=S,V$. The nucleon form factors $G_K$~\cite{Kosmas:2001mv} are
\begin{eqnarray}
    G_V^{(u,p)} = G_V^{(d,n)} = 2\,, \quad
   & G_V^{(d,p)} = G_V^{(u,n)} = 1\,, \quad
   & G_V^{(s,p)} = G_V^{(s,n)} = 0\,,
    \nonumber\\
   G_S^{(u,p)} = G_S^{(d,n)} = 5.1\,, \quad
 &   G_S^{(d,p)} = G_S^{(u,n)} = 4.3\,, \quad
  &  G_S^{(s,p)} = G_S^{(s,n)} = 2.5\,.
\end{eqnarray}
Finally, the $g_{XK(q)}$ coefficients can be written
as
\begin{align}
    g_{LV(q)} &= \frac{\sqrt{2}}{G_F} \left[ e^2 Q_q \left( K_1^L - K_2^R \right) - \frac{C^{Zqq}_L + C^{Zqq}_R}{2 M_Z^2} F_L\right], 
    \quad g_{RV(q)} = g_{LV(q)} \,\,\,(L \leftrightarrow R),
    \\
    g_{LS(q)} &= - \frac{\sqrt{2}}{G_F} \frac{1}{2} \left( C_{\ell\ell qq}^{SLL} + C_{\ell\ell qq}^{SLR} \right), \quad
    g_{RS(q)} = g_{LS(q)} \,\,\,(L \leftrightarrow R),
\end{align}
in which $Q_q$ is the quark electric charge. Here, $C^{Zqq}_{L,R}$ denotes the tree level $Z$ coupling to the quarks and $C_{\ell\ell qq}^{SXY}$ corresponds to the scalar ($H$) penguin contributions (due to the subdominant role of the latter, we do not provide the corresponding expressions, but they are taken into account upon the numerical studies). It is also important to notice that the new fields present in this scotogenic ``T1-2-A'' variant do not couple to quarks, so that no box contributions are present for neutrinoless muon-electron conversion.
As before, the expressions for the $Z$ penguin operators $F_{L/R}$ as well as the anapole operators $K_1^{L/R}$ are given in Appendix~\ref{sec:FF}.

\subsection{cLFV $\pmb{Z}$- and Higgs boson decays}\label{sec:cLFV:Z-H}
Although in general the most promising channels for the discovery of NP in the lepton sector are the rare cLFV muon transitions and decays, in view of the promising potential of a future electron-positron collider as FCC-ee (both as a Higgs factory and as a Tera-$Z$ facility), we also address cLFV $Z$ and Higgs decays. In what follows we compute the amplitudes for the decays $Z\to \ell_\alpha^\pm \ell_\beta^\mp$ and $H\to \ell_\alpha^\pm \ell_\beta^\mp$ ($\alpha \neq \beta$); subsequently, we will address the NP contributions to flavour-conserving decays. In Table~\ref{tab:cLFV-Z-H-decays}, we summarise the bounds from current searches and the projected experimental sensitivities.

\renewcommand{\arraystretch}{1.3}
\begin{table}[h!]
    \centering
    \hspace*{-2mm}{\small\begin{tabular}{|c|c|c|}
    \hline
    Observable & Current bound & Future sensitivity  \\
    \hline
    \hline  
    $\mathrm{BR}(Z\to e^\pm\mu^\mp)$ & \quad$< 4.2\times 10^{-7}$\quad (ATLAS~\cite{Aad:2014bca}) & \quad$\mathcal O (10^{-10})$\quad (FCC-ee~\cite{Abada:2019lih})\\
    $\mathrm{BR}(Z\to e^\pm\tau^\mp)$ & \quad$< 4.1\times 10^{-6}$\quad (ATLAS~\cite{ATLAS:2021bdj}) & \quad$\mathcal O (10^{-10})$\quad (FCC-ee~\cite{Abada:2019lih})\\
    $\mathrm{BR}(Z\to \mu^\pm\tau^\mp)$ & \quad$< 5.3\times 10^{-6}$\quad (ATLAS~\cite{ATLAS:2021bdj}) & \quad $\mathcal O (10^{-10})$\quad (FCC-ee~\cite{Abada:2019lih})\\
    \hline  
    $\mathrm{BR}(H\to e^\pm\mu^\mp)$ & \quad$< 6.1\times 10^{-5}$ (PDG~\cite{ParticleDataGroup:2024cfk})  & $<1.2\times10^{-5}$ (FCC-ee~\cite{Qin:2017aju})\\
    $\mathrm{BR}(H\to e^\pm\tau^\mp)$ & \quad$< 4.7\times 10^{-3}$ (PDG~\cite{ParticleDataGroup:2024cfk})  &$< 1.6\times 10^{-4}$ (FCC-ee~\cite{Qin:2017aju})\\
    $\mathrm{BR}(H\to \mu^\pm\tau^\mp)$ & \quad$< 2.5\times 10^{-3}$\quad(PDG~\cite{ParticleDataGroup:2024cfk})  & $<1.4\times 10^{-4}$ (FCC-ee~\cite{Qin:2017aju})\\
    \hline
    \end{tabular}}
    \caption{Current experimental bounds and future sensitivities for cLFV $Z$ and $H$ decays.}
    \label{tab:cLFV-Z-H-decays}
\end{table}
\renewcommand{\arraystretch}{1.}

\begin{figure}[h!]
    \centering
        \centering
        \begin{tikzpicture}
        \begin{feynman}
        \vertex (a) at (0,0) {\(Z\)};
        \vertex (b) at (1,0);
        \vertex (c) at (2,-1);
        \vertex (d) at (2,1);
        \vertex (e) at (3,-1) {\(\ell_\beta\)};
        \vertex (f) at (3,1) {\(\ell_\alpha\)};
        \diagram* {
        (a) -- [boson] (b),
        (b) -- [plain, edge label'=\( \chi_{_i}\)] (c),
        (b) -- [plain, edge label=\( \chi_{_j}\)] (d),
        (c) -- [scalar, edge label'=\( \eta^\pm\)] (d),
        (d) -- [fermion] (f),
        (e) -- [fermion] (c)
        };
        \end{feynman}
        \end{tikzpicture}
    \hfill
        \centering
        \begin{tikzpicture}
        \begin{feynman}
        \vertex (a) at (0,0) {\(Z\)};
        \vertex (b) at (1,0);
        \vertex (c) at (2,-1);
        \vertex (d) at (2,1);
        \vertex (e) at (3,-1) {\(\ell_\beta\)};
        \vertex (f) at (3,1) {\(\ell_\alpha\)};
        \diagram* {
        (a) -- [boson] (b),
        (b) -- [scalar, edge label'=\( \phi_{_k}\)] (c),
        (b) -- [scalar, edge label=\( \phi_l\)] (d),
        (c) -- [fermion, edge label'=\( \psi^\pm\)] (d),
        (d) -- [fermion] (f),
        (e) -- [fermion] (c)
        };
        \end{feynman}
        \end{tikzpicture}
    \hfill
        \centering
        \begin{tikzpicture}
        \begin{feynman}
        \vertex (a) at (0,0) {\(Z\)};
        \vertex (b) at (1,0);
        \vertex (c) at (2,-1);
        \vertex (d) at (2,1);
        \vertex (e) at (3,-1) {\(\ell_\beta\)};
        \vertex (f) at (3,1) {\(\ell_\alpha\)};
        \diagram* {
        (a) -- [boson] (b),
        (c) -- [fermion, edge label=\( \psi^+\)] (b),
        (b) -- [fermion, edge label=\( \psi^-\)] (d),
        (c) -- [scalar, edge label'=\( \phi_{_k}\)] (d),
        (d) -- [fermion] (f),
        (e) -- [fermion] (c)
        };
        \end{feynman}
        \end{tikzpicture}
    \hfill
        \centering
        \begin{tikzpicture}
        \begin{feynman}
        \vertex (a) at (0,0) {\(Z\)};
        \vertex (b) at (1,0);
        \vertex (c) at (2,-1);
        \vertex (d) at (2,1);
        \vertex (e) at (3,-1) {\(\ell_\beta\)};
        \vertex (f) at (3,1) {\(\ell_\alpha\)};
        \diagram* {
        (a) -- [boson] (b),
        (c) -- [scalar, edge label=\( \eta^+\)] (b),
        (b) -- [scalar, edge label=\( \eta^-\)] (d),
        (c) -- [plain, edge label'=\( \chi_{_i}\)] (d),
        (d) -- [fermion] (f),
        (e) -- [fermion] (c)
        };
        \end{feynman}
        \end{tikzpicture}
\mbox{  \hspace*{19mm}
            (a) \hspace*{38mm}
            (b) \hspace*{36mm}
            (c) \hspace*{36mm}
            (d) \hspace*{25mm}
    }
    \caption{Feynman diagrams of the NP contributions to $Z \rightarrow \ell_{\alpha}\ell_{\beta}$, with $i, j=1-4$ and $k, l=1-3$ (vertex corrections).}
    \label{fig:Zlldiag}
\end{figure}
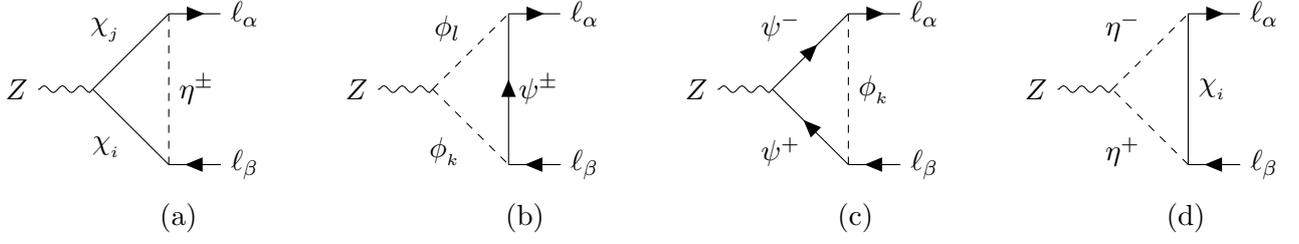

\begin{figure}[h!]
       \centering
        \begin{tikzpicture}
        \begin{feynman}
        \vertex (a) at (0,0) {\(H\)};
        \vertex (b) at (1,0);
        \vertex (c) at (2,-1);
        \vertex (d) at (2,1);
        \vertex (e) at (3,-1) {\(\ell_\beta\)};
        \vertex (f) at (3,1) {\(\ell_\alpha\)};
        \diagram* {
        (a) -- [scalar] (b),
        (b) -- [plain, edge label'=\( \chi_{_i}\)] (c),
        (b) -- [plain, edge label=\( \chi_{_j}\)] (d),
        (c) -- [scalar, edge label'=\( \eta^\pm\)] (d),
        (d) -- [fermion] (f),
        (e) -- [fermion] (c)
        };
        \end{feynman}
        \end{tikzpicture}
        \hspace*{20mm}
        \centering
        \begin{tikzpicture}
        \begin{feynman}
        \vertex (a) at (0,0) {\(H\)};
        \vertex (b) at (1,0);
        \vertex (c) at (2,-1);
        \vertex (d) at (2,1);
        \vertex (e) at (3,-1) {\(\ell_\beta\)};
        \vertex (f) at (3,1) {\(\ell_\alpha\)};
        \diagram* {
        (a) -- [scalar] (b),
        (b) -- [scalar, edge label'=\( \phi_{_k}\)] (c),
        (b) -- [scalar, edge label=\( \phi_{_l}\)] (d),
        (c) -- [fermion, edge label'=\( \psi^\pm\)] (d),
        (d) -- [fermion] (f),
        (e) -- [fermion] (c)
        };
        \end{feynman}
        \end{tikzpicture}
          \hspace*{20mm}
        \centering
        \begin{tikzpicture}
        \begin{feynman}
        \vertex (a) at (0,0) {\(H\)};
        \vertex (b) at (1,0);
        \vertex (c) at (2,-1);
        \vertex (d) at (2,1);
        \vertex (e) at (3,-1) {\(\ell_\beta\)};
        \vertex (f) at (3,1) {\(\ell_\alpha\)};
        \diagram* {
        (a) -- [scalar] (b),
        (c) -- [scalar, edge label=\( \eta^+\)] (b),
        (b) -- [scalar, edge label=\( \eta^-\)] (d),
        (c) -- [plain, edge label'=\( \chi_{_i}\)] (d),
        (d) -- [fermion] (f),
        (e) -- [fermion] (c)
        };
        \end{feynman}
        \end{tikzpicture}
    \mbox{  \hspace*{27mm}
            (a) \hspace*{52mm}
            (b) \hspace*{52mm}
            (c) \hspace*{45mm}
    }
    \caption{Feynman diagrams of the contributions to $H \rightarrow \ell_{\alpha}\ell_{\beta}$, with $i,j=1-4$ and $k,l=1-3$ (vertex corrections).}
    \label{fig:Hlldiag}
\end{figure}
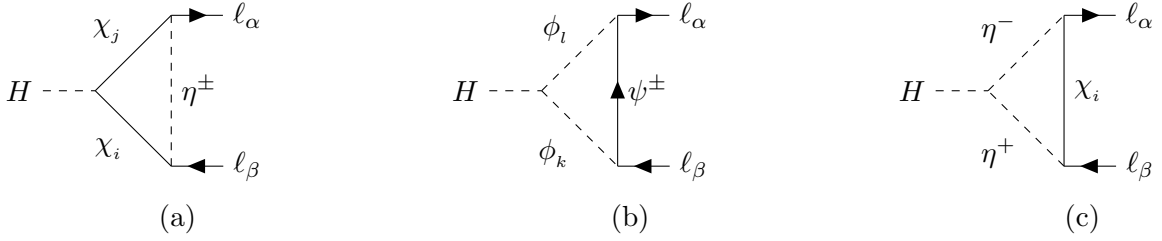

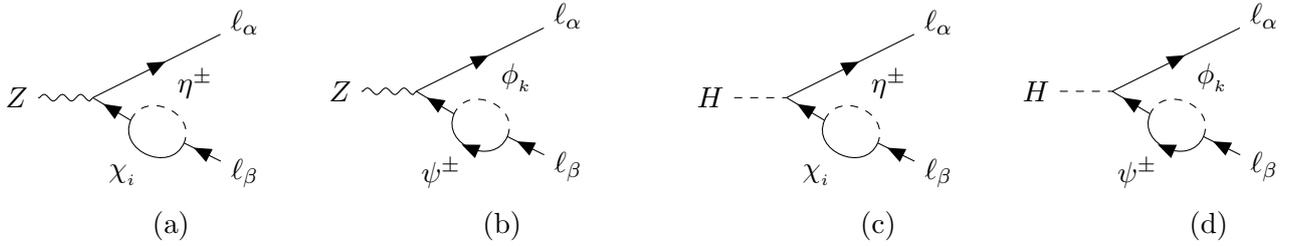
\begin{figure}[h!]
    \centering
        \centering
        \begin{tikzpicture}
        \begin{feynman}
        \vertex (a) at (0,0) {\(Z\)};
        \vertex (b) at (1,0);
        \vertex (c) at (3,1) {\(\ell_\alpha\)};
        \vertex (d) at (1.5,-0.3);
        \vertex (e) at (2.2,-0.6);
        \vertex (f) at (3,-1) {\(\ell_\beta\)};
        \diagram* {
        (a) -- [boson] (b),
        (b) -- [fermion] (c),
        (d) -- [plain, half right, edge label'=\( \chi_{_i}\)] (e),
        (d) -- [scalar, half left, edge label=\( \eta^\pm\)] (e),
        (d) -- [fermion] (b),
        (f) -- [fermion] (e)
        };
        \end{feynman}
        \end{tikzpicture}
    \hfill
        \centering
        \begin{tikzpicture}
        \begin{feynman}
        \vertex (a) at (0,0) {\(Z\)};
        \vertex (b) at (1,0);
        \vertex (c) at (3,1) {\(\ell_\alpha\)};
        \vertex (d) at (1.5,-0.3);
        \vertex (e) at (2.2,-0.6);
        \vertex (f) at (3,-1) {\(\ell_\beta\)};
        \diagram* {
        (a) -- [boson] (b),
        (b) -- [fermion] (c),
        (e) -- [fermion, half left, edge label=\( \psi^\pm\)] (d),
        (e) -- [scalar, half right, edge label'=\( \phi_{_k}\)] (d),
        (d) -- [fermion] (b),
        (f) -- [fermion] (e)
        };
        \end{feynman}
        \end{tikzpicture}
\hspace*{10mm}
        \begin{tikzpicture}
        \begin{feynman}
        \vertex (a) at (0,0) {\(H\)};
        \vertex (b) at (1,0);
        \vertex (c) at (3,1) {\(\ell_\alpha\)};
        \vertex (d) at (1.5,-0.3);
        \vertex (e) at (2.2,-0.6);
        \vertex (f) at (3,-1) {\(\ell_\beta\)};
        \diagram* {
        (a) -- [scalar] (b),
        (b) -- [fermion] (c),
        (d) -- [plain, half right, edge label'=\( \chi_{_i}\)] (e),
        (d) -- [scalar, half left, edge label=\( \eta^\pm\)] (e),
        (d) -- [fermion] (b),
        (f) -- [fermion] (e)
        };
        \end{feynman}
        \end{tikzpicture}
    \hfill
        \centering
        \begin{tikzpicture}
        \begin{feynman}
        \vertex (a) at (0,0) {\(H\)};
        \vertex (b) at (1,0);
        \vertex (c) at (3,1) {\(\ell_\alpha\)};
        \vertex (d) at (1.5,-0.3);
        \vertex (e) at (2.2,-0.6);
        \vertex (f) at (3,-1) {\(\ell_\beta\)};
        \diagram* {
        (a) -- [scalar] (b),
        (b) -- [fermion] (c),
        (e) -- [fermion, half left, edge label=\( \psi^\pm\)] (d),
        (e) -- [scalar, half right, edge label'=\( \phi_{_k}\)] (d),
        (d) -- [fermion] (b),
        (f) -- [fermion] (e)
        };
        \end{feynman}
        \end{tikzpicture}
\mbox{  \hspace*{18mm}
            (a) \hspace*{36mm}
            (b) \hspace*{42mm}
            (c) \hspace*{36mm}
            (d) \hspace*{25mm}
    }
    \caption{cLFV external leg corrections for $Z, H \rightarrow \ell_{\alpha}\ell_{\beta}$, with $i=1-4$ and $k=1-3$ (we have omitted the analogous corrections in the $\ell_\beta$ lines).}
    \label{fig:ZHlegdiag:cLFV}
\end{figure}

The cLFV decays receive contributions from several diagrams, as depicted in Figs.~\ref{fig:Zlldiag} and~\ref{fig:Hlldiag} (for the NP corrections to the vertex), and Fig.~\ref{fig:ZHlegdiag:cLFV} (cLFV corrections to the external lepton lines).
The expressions for the branching ratios can be written as
\begin{align}\label{eq:ZBr}
    \text{BR}(Z \to \bar{\ell}_\alpha \ell_\beta) &= \frac{\lambda^{1/2}(M_Z^2,m_{\ell_\alpha}^2,m_{\ell_\beta}^2)}{192 \pi\, M_Z^5\, \Gamma_Z} \left[ \left( |F^{\alpha \beta}_L|^2 + |F^{\alpha \beta}_R|^2 \right) \left( M_Z^2 - m_{\ell_\alpha}^2 - m_{\ell_\beta}^2\right) \lambda(M_Z^2,m_{\ell_\alpha}^2,m_{\ell_\beta
    }^2) \nonumber \right.\\
    &- \left. 4 \left( |F^{\alpha \beta}_{SL}|^2 + |F^{\alpha \beta}_{SR}|^2\right) \left( \lambda(M_Z^2,m_{\ell_\alpha}^2,m_{\ell_\beta}^2) - 3 M_Z^2 \left( M_Z^2 - m_{\ell_\alpha}^2 - m_{\ell_\beta}^2\right) \right) \nonumber \right. \\
    &- \left. 4 \left( m_{\ell_\alpha}\, m_{\ell_\beta} \,\text{Re} \left(F^{\alpha \beta}_L (F^{\alpha \beta}_R)^* \right) + m_{\ell_\beta}\, \text{Re}  \left( F^{\alpha \beta}_R (F^{\alpha \beta}_{SL})^* + F^{\alpha \beta}_L (F^{\alpha \beta}_{SR})^*\right) \right. \right. \nonumber \\
    &+ \left. \left. m_{\ell_\alpha} \,\text{Re} \left( F^{\alpha \beta}_R (F^{\alpha \beta}_{SR})^* + F^{\alpha \beta}_L (F^{\alpha \beta}_{SL})^*\right) \right)\lambda(M_Z^2,m_{\ell_\alpha}^2,m_{\ell_\beta}^2) \right. \nonumber\\
    &+ \left. 48 \,M_Z^2 \,m_{\ell_\alpha} \,m_{\ell_\beta} \text{Re} \left( F^{\alpha \beta}_{SL} (F^{\alpha \beta}_{SR})^*\right)\right]\,,
\end{align}
and
\begin{align}\label{eq:HBr}
    \text{BR}(H \to \bar{\ell}_\alpha\ell_\beta) &= \frac{\lambda^{1/2}(M_H^2,m_{\ell_\alpha}^2,m_{\ell_\beta}^2)}{16 \pi \,M_H^3\, \Gamma_H} \left[ \left( |G^{\alpha \beta}_L|^2 + |G^{\alpha \beta}_R|^2\right) (M_H^2 - m_{\ell_\alpha}^2 - m_{\ell_\beta}^2) \right. \nonumber\\
    &- \left. 4 m_{\ell_\alpha}\, m_{\ell_\beta} \,\text{Re} \left( G^{\alpha \beta}_L (G^{\alpha \beta}_R)^*\right)\right],
\end{align}
with $M_{Z,H}$ and $\Gamma_{Z,H}$ the boson masses and total decay widths, and $\lambda$ the Källén function,
\begin{equation}\label{eq:Kallen}
    \lambda(x,y,z) \equiv x^2+y^2+z^2-2x\,y-2x\,z-2y\,z\,.
\end{equation}
In Eqs.~(\ref{eq:ZBr},\ref{eq:HBr}) above, $F^{\alpha \beta}$ and $G^{\alpha \beta}$ respectively denote the form factors entering the $Z$ and $H$ decay amplitudes, which one can generically cast as 
\begin{align}\label{eq:ZHTensDec}
        F^{\alpha \beta}_X &= i \left( 
        F^{\alpha \beta}_{X\,2\chi} +  
        F^{\alpha \beta}_{X\,2\phi} + 
        F^{\alpha \beta}_{X\,1\phi} + 
        F^{\alpha \beta}_{X\,1\chi} +
        F^{\alpha \beta}_{X\,x\ell} +
        \delta^{\alpha\beta} F^{\beta}_{X\, CT} +
        \delta^{\alpha\beta} R^{\beta}_{X}\right) \,, \nonumber \\
        G^{\alpha \beta}_X &= i \left( 
        G^{\alpha \beta}_{X\,2\chi} +  
        G^{\alpha\beta}_{X\,2\phi} + 
        G^{\alpha \beta}_{X\,1\chi} + 
        G^{\alpha \beta}_{X\,x\ell} +
        \delta^{\alpha\beta} G^{\beta}_{X\, CT} +
        \delta^{\alpha\beta} S^{\beta}_{X}\right) \,,
\end{align}
in which for $G$, $X=L,R$ while for $F$, $X=SL,SR,L,R$. 
Let us also notice, that even though not relevant for the cLFV decays (with $\alpha\neq \beta$), the tree level $Z(H) \ell_\alpha \ell_\alpha$ couplings, respectively denoted $R (S)^\alpha_X$, are given by $R^\alpha_{SL} = -i e (s_w^2 - c_w^2)/(2 c_w s_w)$, $R^\alpha_{SR} = -i e s_w/c_w$, $S^\alpha_L = S^\alpha_R = -i m_{\ell_{_\alpha}}/v$. Moreover, the (diagonal) terms, i.e. $F^{\beta}_{X\, CT}$ and $G^{\beta}_{X\, CT}$, denote the counter-terms arising from the renormalisation of the $Z(H)\ell_\alpha\ell_\beta$ interaction.

In the above, we have introduced the abbreviated notation $n\phi \,(n\chi)$ for the diagrams containing $n$ neutral NP scalar (fermion) fields in the loop and $x\ell$ for the corrections to the external legs. For example, for $Z\to \ell_\alpha \ell_\beta$ decays, the contributions to $F$ correspond to diagrams (a)-(d) of Fig.~\ref{fig:Zlldiag} and diagrams (a)-(b) of Fig.~\ref{fig:ZHlegdiag:cLFV}.
The complete expressions for the form factors have been fully computed, and are given in Appendix~\ref{sec:FF}. 

\subsection{Electroweak precision observables}\label{sec:EWPO}
As mentioned, we conducted a full evaluation of the NP contributions of the present ``T1-2-A'' variant to several EW observables. A full renormalisation of the interaction vertices under scrutiny was carried out, as mentioned below. 

In Table~\ref{tab:obs:EW} we summarise both SM predictions and current experimental status for flavour conserving and invisible decays which we will subsequently address. Concerning future prospects, the projections of the FCC Collaboration suggest that in view of the impressive increase in statistics, uncertainties may be reduced by at least one order of magnitude (or even more than two in certain cases)~\cite{FCC:2018evy,FCC:2018byv}; here we adopt a more conservative approach, and project a reduction of the uncertainties by a factor 4.
\renewcommand{\arraystretch}{1.3}
\begin{table}[h!]
    \centering
    \hspace*{-2mm}{\small\begin{tabular}{|c|c|c|}
    \hline
    Observable & Exp. measurement & SM prediction  \\
    \hline
    $\Gamma(Z\to e^+e^-)$ & $83.91\pm0.12\:\mathrm{MeV}$ (LEP~\cite{ALEPH:2005ab}) & $83.965\pm0.016\:\mathrm{MeV}$~\cite{Freitas:2014hra}\\
    $\Gamma(Z\to \mu^+\mu^-)$ & $83.99\pm0.18\:\mathrm{MeV}$ (LEP~\cite{ALEPH:2005ab}) & $83.965\pm0.016\:\mathrm{MeV}$~\cite{Freitas:2014hra}\\
    $\Gamma(Z\to \tau^+\tau^-)$ & $84.08\pm0.22\:\mathrm{MeV}$ (LEP~\cite{ALEPH:2005ab}) & $83.775\pm0.016\:\mathrm{MeV}$~\cite{Freitas:2014hra}\\
    \hline
    $\Gamma(Z\to\mathrm{inv.})$ & $499.0 \pm 1.5\:\mathrm{MeV}$ (PDG~\cite{ParticleDataGroup:2024cfk})& $501.45\pm 0.05\:\mathrm{MeV}$~\cite{Freitas:2014hra}\\
    \hline
    $R_{\mu e}(Z\to\ell\ell)$ & $1.0001\pm 0.0024$ (PDG~\cite{ParticleDataGroup:2024cfk}) & $1.0$~\cite{Freitas:2014hra}\\
    $R_{\tau e}(Z\to\ell\ell)$ & $1.0020\pm0.0032$ (PDG~\cite{ParticleDataGroup:2024cfk}) & $0.9977$~\cite{Freitas:2014hra}\\
    $R_{\tau \mu}(Z\to\ell\ell)$ & $1.0010\pm 0.0026$ (PDG~\cite{ParticleDataGroup:2024cfk}) & $0.9977$~\cite{Freitas:2014hra}\\
    \hline
    $R_{\tau \mu}(H\to\ell\ell)$ & $230\pm 146$ (PDG~\cite{ParticleDataGroup:2024cfk}) & $288$~\cite{LHCHiggsCrossSectionWorkingGroup:2016ypw}\\
    \hline
    $\mathrm{BR}(H\to\tau^+\tau^-)$ & $0.06_{-0.007}^{+0.008}$ (PDG~\cite{ParticleDataGroup:2024cfk}) & $0.0624\pm0.0035$~\cite{Denner:2011mq}\\
    $\mathrm{BR}(H\to\mu^+\mu^-)$ & $(2.6 \pm 1.3)\times 10^{-4}$ (PDG~\cite{ParticleDataGroup:2024cfk}) & $(2.17 \pm0.13)\times 10^{-4}$~\cite{Denner:2011mq}\\
    \hline
    \end{tabular}}
    \caption{Experimental  values and SM predictions  for several LFUV and EW observables. All uncertainties are given at 68\% C.L. (the parametric uncertainties are negligible for the SM predictions of the universality ratios).}
    \label{tab:obs:EW}
\end{table}
\renewcommand{\arraystretch}{1.}

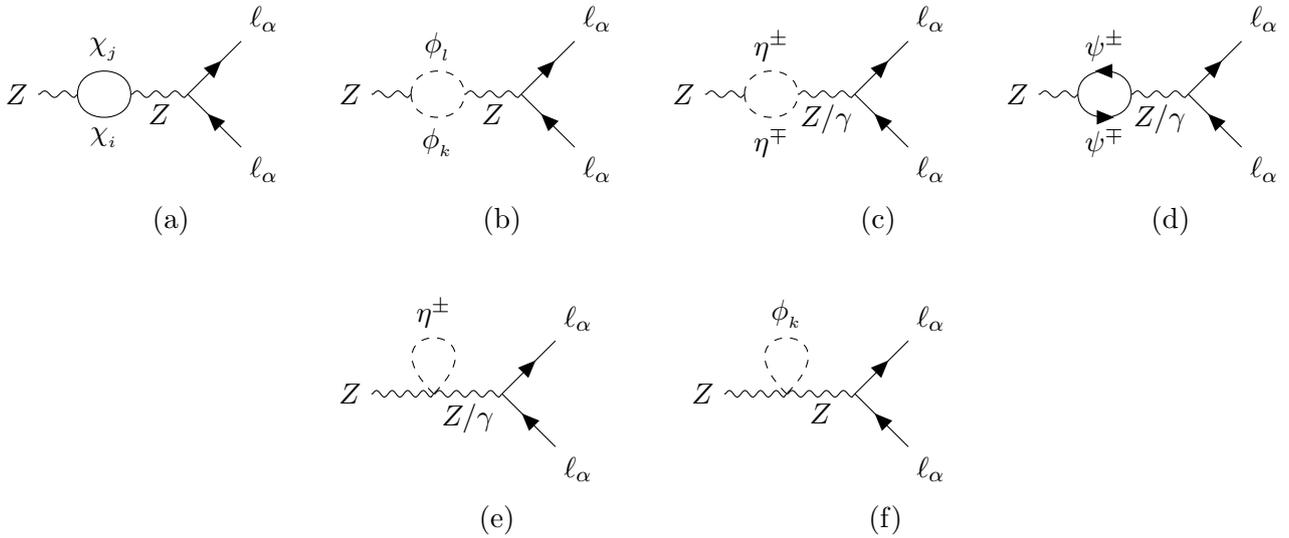
\begin{figure}[h!]
    \centering
        \centering
        \begin{tikzpicture}
        \begin{feynman}
        \vertex (a) at (0,0) {\(Z\)};
        \vertex (b) at (0.8,0);
        \vertex (c) at (1.5,0);
        \vertex (d) at (2.25,0);
        \vertex (e) at (3.25,1) {\(\ell_\alpha\)};
        \vertex (f) at (3.25,-1) {\(\ell_\alpha\)};
        \diagram* {
        (a) -- [boson] (b),
        (c) -- [boson, edge label'=\( Z\)] (d),
        (b) -- [plain, half right, edge label'=\( \chi_{_i}\)] (c),
        (b) -- [plain, half left, edge label=\( \chi_{_j}\)] (c),
        (d) -- [fermion] (e),
        (f) -- [fermion] (d)
        };
        \end{feynman}
        \end{tikzpicture}
    \hfill
        \centering
        \begin{tikzpicture}
        \begin{feynman}
        \vertex (a) at (0,0) {\(Z\)};
        \vertex (b) at (0.8,0);
        \vertex (c) at (1.5,0);
        \vertex (d) at (2.25,0);
        \vertex (e) at (3.25,1) {\(\ell_\alpha\)};
        \vertex (f) at (3.25,-1) {\(\ell_\alpha\)};
        \diagram* {
        (a) -- [boson] (b),
        (c) -- [boson, edge label'=\( Z\)] (d),
        (b) -- [scalar, half right, edge label'=\( \phi_{_k}\)] (c),
        (b) -- [scalar, half left, edge label=\( \phi_{_l}\)] (c),
        (d) -- [fermion] (e),
        (f) -- [fermion] (d)
        };
        \end{feynman}
        \end{tikzpicture}
    \hfill
        \centering
        \begin{tikzpicture}
        \begin{feynman}
        \vertex (a) at (0,0) {\(Z\)};
        \vertex (b) at (0.8,0);
        \vertex (c) at (1.5,0);
        \vertex (d) at (2.25,0);
        \vertex (e) at (3.25,1) {\(\ell_\alpha\)};
        \vertex (f) at (3.25,-1) {\(\ell_\alpha\)};
        \diagram* {
        (a) -- [boson] (b),
        (c) -- [boson, edge label'=\( Z/\gamma\)] (d),
        (b) -- [scalar, half right, edge label'=\( \eta^\mp\)] (c),
        (b) -- [scalar, half left, edge label=\( \eta^\pm\)] (c),
        (d) -- [fermion] (e),
        (f) -- [fermion] (d)
        };
        \end{feynman}
        \end{tikzpicture}
    \hfill
        \centering
        \begin{tikzpicture}
        \begin{feynman}
        \vertex (a) at (0,0) {\(Z\)};
        \vertex (b) at (0.8,0);
        \vertex (c) at (1.5,0);
        \vertex (d) at (2.25,0);
        \vertex (e) at (3.25,1) {\(\ell_\alpha\)};
        \vertex (f) at (3.25,-1) {\(\ell_\alpha\)};
        \diagram* {
        (a) -- [boson] (b),
        (c) -- [boson, edge label'=\( Z/\gamma\)] (d),
        (b) -- [fermion, half right, edge label'=\( \psi^\mp\)] (c),
        (c) -- [fermion, half right, edge label'=\( \psi^\pm\)] (b),
        (d) -- [fermion] (e),
        (f) -- [fermion] (d)
        };
        \end{feynman}
        \end{tikzpicture}
\mbox{  \hspace*{18mm}
            (a) \hspace*{36mm}
            (b) \hspace*{42mm}
            (c) \hspace*{31mm}
            (d) \hspace*{27mm}
    }
\vspace*{3mm}
\\
        \centering
        \begin{tikzpicture}
        \begin{feynman}
        \vertex (a) at (0,0) {\(Z\)};
        \vertex (b) at (1.1,0);
        \vertex (d) at (2,0);
        \vertex (e) at (3,1) {\(\ell_\alpha\)};
        \vertex (f) at (3,-1) {\(\ell_\alpha\)};
        \diagram* {
        (a) -- [boson] (b),
        (b) -- [boson, edge label'=\( Z/\gamma\)] (d),
        b -- [scalar, loop, min distance=1.4cm, in=45, out=135, edge label=\( \eta^\pm\)] b,
        (d) -- [fermion] (e),
        (f) -- [fermion] (d)
        };
        \end{feynman}
        \end{tikzpicture}
 \hspace*{8mm}
        \centering
        \begin{tikzpicture}
        \begin{feynman}
        \vertex (a) at (0,0) {\(Z\)};
        \vertex (b) at (1.1,0);
        \vertex (d) at (2,0);
        \vertex (e) at (3,1) {\(\ell_\alpha\)};
        \vertex (f) at (3,-1) {\(\ell_\alpha\)};
        \diagram* {
        (a) -- [boson] (b),
        (b) -- [boson, edge label'=\( Z\)] (d),
        b -- [scalar, loop, min distance=1.4cm, in=45, out=135, edge label=\( \phi_{_k}\)] b,
        (d) -- [fermion] (e),
        (f) -- [fermion] (d)
        };
        \end{feynman}
        \end{tikzpicture}
\mbox{  \hspace*{61mm}
            (e) \hspace*{44mm}
            (f) \hspace*{64mm}
    }
    \caption{Flavour conserving external leg corrections: Feynman diagrams for $Z \rightarrow \ell_{\alpha}\ell_{\alpha}$, with $i,j=1-4$ and $k,l=1-3$.}
    \label{fig:Zlegdiag}
\end{figure}

\begin{figure}[h!]
    \centering
        \centering
        \begin{tikzpicture}
        \begin{feynman}
        \vertex (a) at (0,0) {\(H\)};
        \vertex (b) at (0.8,0);
        \vertex (c) at (1.5,0);
        \vertex (d) at (2,0);
        \vertex (e) at (3,1) {\(\ell_\alpha\)};
        \vertex (f) at (3,-1) {\(\ell_\alpha\)};
        \diagram* {
        (a) -- [scalar] (b),
        (c) -- [scalar] (d),
        (b) -- [plain, half right, edge label'=\( \chi_{_i}\)] (c),
        (b) -- [plain, half left, edge label=\( \chi_{_j}\)] (c),
        (d) -- [fermion] (e),
        (f) -- [fermion] (d)
        };
        \end{feynman}
        \end{tikzpicture}
    \hfill
        \centering
        \begin{tikzpicture}
        \begin{feynman}
        \vertex (a) at (0,0) {\(H\)};
        \vertex (b) at (0.8,0);
        \vertex (c) at (1.5,0);
        \vertex (d) at (2,0);
        \vertex (e) at (3,1) {\(\ell_\alpha\)};
        \vertex (f) at (3,-1) {\(\ell_\alpha\)};
        \diagram* {
        (a) -- [scalar] (b),
        (c) -- [scalar] (d),
        (b) -- [scalar, half right, edge label'=\( \phi_{_k}\)] (c),
        (b) -- [scalar, half left, edge label=\( \phi_{_l}\)] (c),
        (d) -- [fermion] (e),
        (f) -- [fermion] (d)
        };
        \end{feynman}
        \end{tikzpicture}
    \hfill
        \centering
        \begin{tikzpicture}
        \begin{feynman}
        \vertex (a) at (0,0) {\(H\)};
        \vertex (b) at (0.8,0);
        \vertex (c) at (1.5,0);
        \vertex (d) at (2,0);
        \vertex (e) at (3,1) {\(\ell_\alpha\)};
        \vertex (f) at (3,-1) {\(\ell_\alpha\)};
        \diagram* {
        (a) -- [scalar] (b),
        (c) -- [scalar] (d),
        (b) -- [scalar, half right, edge label'=\( \eta^\mp\)] (c),
        (b) -- [scalar, half left, edge label=\( \eta^\pm\)] (c),
        (d) -- [fermion] (e),
        (f) -- [fermion] (d)
        };
        \end{feynman}
        \end{tikzpicture}
\mbox{ \hspace*{10mm}
            (a) \hspace*{60mm}
            (b) \hspace*{60mm}
            (c) \hspace*{60mm}
    }
\vspace*{3mm}
\\
    \centering
        \begin{tikzpicture}
        \begin{feynman}
        \vertex (a) at (0,0) {\(H\)};
        \vertex (b) at (1.1,0);
        \vertex (d) at (2,0);
        \vertex (e) at (3,1) {\(\ell_\alpha\)};
        \vertex (f) at (3,-1) {\(\ell_\alpha\)};
        \diagram* {
        (a) -- [scalar] (b),
        (b) -- [scalar] (d),
        b -- [scalar, loop, min distance=1.4cm, in=45, out=135, edge label=\( \eta^\pm\)] b,
        (d) -- [fermion] (e),
        (f) -- [fermion] (d)
        };
        \end{feynman}
        \end{tikzpicture}
 \hspace*{25mm}
        \centering
        \begin{tikzpicture}
        \begin{feynman}
        \vertex (a) at (0,0) {\(H\)};
        \vertex (b) at (1.1,0);
        \vertex (d) at (2,0);
        \vertex (e) at (3,1) {\(\ell_\alpha\)};
        \vertex (f) at (3,-1) {\(\ell_\alpha\)};
        \diagram* {
        (a) -- [scalar] (b),
        (b) -- [scalar] (d),
        b -- [scalar, loop, min distance=1.4cm, in=45, out=135, edge label=\( \phi_{_k}\)] b,
        (d) -- [fermion] (e),
        (f) -- [fermion] (d)
        };
        \end{feynman}
        \end{tikzpicture}
\mbox{  \hspace*{45mm}
            (e) \hspace*{56mm}
            (f) \hspace*{60mm}
    }
    \caption{Flavour conserving external leg corrections: Feynman diagrams for $H \rightarrow \ell_{\alpha}\ell_{\alpha}$, with $i,j=1-4$ and $k,l=1-3$.}
    \label{fig:Hlegdiag}
\end{figure}
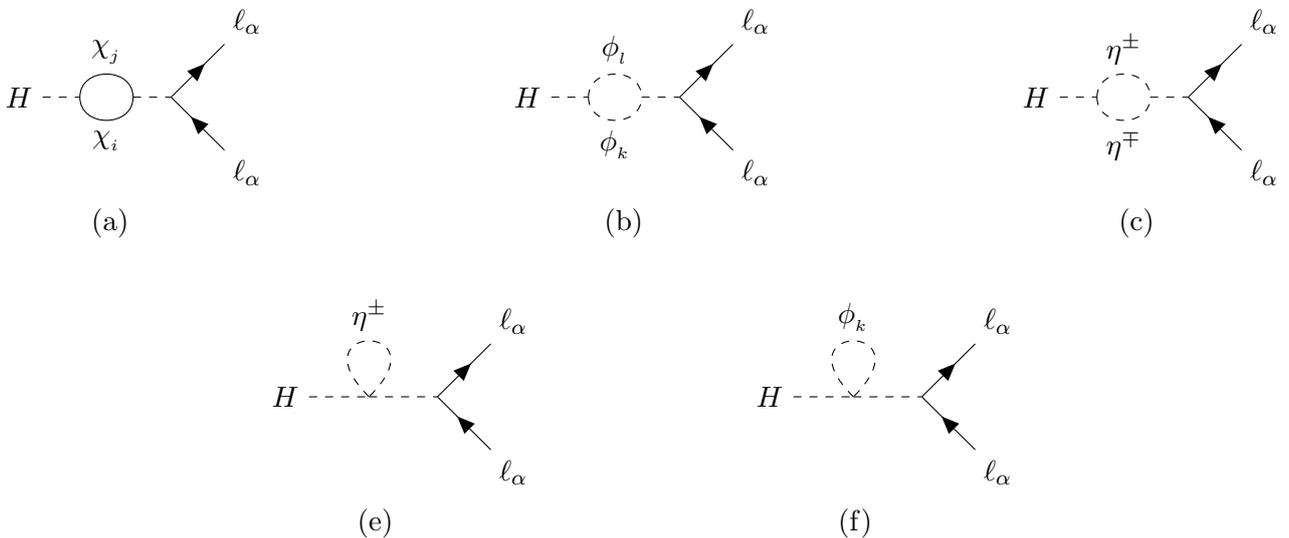

\paragraph{Flavour conserving leptonic $Z$ and $H$ decays} 
Concerning the flavour conserving leptonic $Z$ and Higgs decays, there are  NP corrections to the vertices (trivially obtained from the digrams depicted in Figures~\ref{fig:Zlldiag} and~\ref{fig:Hlldiag}, by taking same flavoured final states, i.e. $\ell_\alpha = \ell_\beta$); further contributions arise from corrections to the boson lines, and these diagrams are collected in Figs.~\ref{fig:Zlegdiag} and~\ref{fig:Hlegdiag}. 

The branching ratios are given in full analogy to the ones of Eqs.~(\ref{eq:ZBr},~\ref{eq:HBr}), and the associated form factor decomposition follows Eqs.~(\ref{eq:ZHTensDec}) (for $\ell_\alpha = \ell_\beta$). Appendix~\ref{sec:FF} contains the full expressions of the form factors.
Notice that now one must include the corresponding counter-terms; the description of the renormalisation procedure for the $Z$ and Higgs leptonic interactions is presented in 
Appendix~\ref{app:renormalisation}.

In addition to complying with current experimental measurements, the above flavour conserving decay rates further allow to test the SM paradigm of lepton flavour universality (only broken by finite lepton mass effects). 
The presence of the NP can also contribute to flavour conserving $Z$ bosons in a non-universal manner, and thus potentially induce the violation of lepton flavour universality of (effective) $Z$-boson couplings. One can construct the ratios of decay widths
\begin{equation}\label{eq:RZab}
    R^Z_{\alpha\beta}\, = \, \dfrac{\Gamma(Z \to \ell_\alpha^+\ell_\alpha^-)}{\Gamma(Z \to \ell_\beta^+\ell_\beta^-)}\,, \quad \text{with } \alpha \neq \beta \, =\, e, \, \mu, \, \tau\,,
\end{equation}
(which have the advantage of allowing the cancellation of QED corrections in the theoretical predictions). The full contributions arising in the present scotogenic construction can then be confronted to the current experimental bounds.

As done for $Z$ decays, one can also study the ``T1-2-A'' contributions to dimuon and ditau Higgs decays, for which current LHC measurements are becoming very precise. In addition to studying the individual branching ratios, one can also construct the ratios of rates, 
\begin{equation}\label{eq:RHalphabeta:def}
    R^H_{\alpha \beta} \,= \,\frac{\Gamma(H\to \ell_\alpha \ell_\alpha)}{\Gamma(H\to \ell_\beta \ell_\beta)}\,,\quad \alpha=\mu,\quad\text{and}\quad \beta =\tau\,,
\end{equation}
which is a theoretically very clean observable:
in the context of the SM, and at leading order, it translates that only the lepton masses (i.e. the charged lepton Yukawa couplings) do violate LFU, $R^H_{\alpha \beta}|_\text{SM} \approx \mathcal{O}(m_\alpha^2/m_\beta^2)$. 

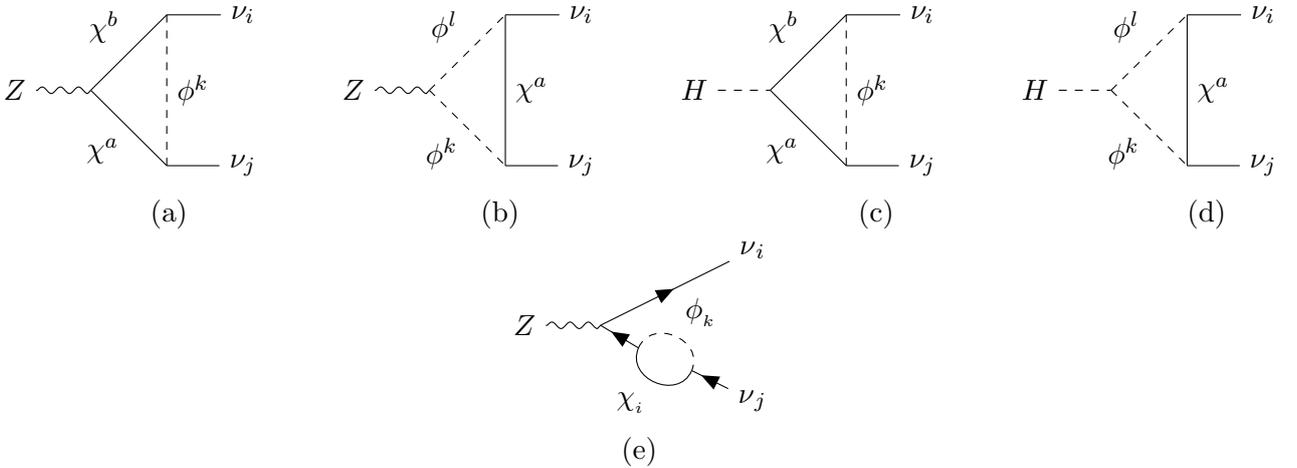
\begin{figure}[h!]
    \centering
        \begin{tikzpicture}
        \begin{feynman}
        \vertex (a) at (0,0) {\(Z\)};
        \vertex (b) at (1,0);
        \vertex (c) at (2,-1);
        \vertex (d) at (2,1);
        \vertex (e) at (3,-1) {\(\nu_j\)};
        \vertex (f) at (3,1) {\(\nu_i\)};
        \diagram* {
        (a) -- [boson] (b),
        (b) -- [plain, edge label'=\( \chi^a\)] (c),
        (b) -- [plain, edge label=\( \chi^b\)] (d),
        (c) -- [scalar, edge label'=\( \phi^k\)] (d),
        (d) -- [plain] (f),
        (c) -- [plain] (e)
        };
        \end{feynman}
        \end{tikzpicture}
    \hfill
    \centering
        \begin{tikzpicture}
        \begin{feynman}
        \vertex (a) at (0,0) {\(Z\)};
        \vertex (b) at (1,0);
        \vertex (c) at (2,-1);
        \vertex (d) at (2,1);
        \vertex (e) at (3,-1) {\(\nu_j\)};
        \vertex (f) at (3,1) {\(\nu_i\)};
        \diagram* {
        (a) -- [boson] (b),
        (b) -- [scalar, edge label'=\( \phi^k\)] (c),
        (b) -- [scalar, edge label=\( \phi^l\)] (d),
        (c) -- [plain, edge label'=\( \chi^a\)] (d),
        (d) -- [plain] (f),
        (c) -- [plain] (e)
        };
        \end{feynman}
        \end{tikzpicture}
    \hfill
    \centering
        \begin{tikzpicture}
        \begin{feynman}
        \vertex (a) at (0,0) {\(H\)};
        \vertex (b) at (1,0);
        \vertex (c) at (2,-1);
        \vertex (d) at (2,1);
        \vertex (e) at (3,-1) {\(\nu_j\)};
        \vertex (f) at (3,1) {\(\nu_i\)};
        \diagram* {
        (a) -- [scalar] (b),
        (b) -- [plain, edge label'=\( \chi^a\)] (c),
        (b) -- [plain, edge label=\( \chi^b\)] (d),
        (c) -- [scalar, edge label'=\( \phi^k\)] (d),
        (d) -- [plain] (f),
        (c) -- [plain] (e)
        };
        \end{feynman}
        \end{tikzpicture}
    \hfill
    \centering
        \begin{tikzpicture}
        \begin{feynman}
        \vertex (a) at (0,0) {\(H\)};
        \vertex (b) at (1,0);
        \vertex (c) at (2,-1);
        \vertex (d) at (2,1);
        \vertex (e) at (3,-1) {\(\nu_j\)};
        \vertex (f) at (3,1) {\(\nu_i\)};
        \diagram* {
        (a) -- [scalar] (b),
        (b) -- [scalar, edge label'=\( \phi^k\)] (c),
        (b) -- [scalar, edge label=\( \phi^l\)] (d),
        (c) -- [plain, edge label'=\( \chi^a\)] (d),
        (d) -- [plain] (f),
        (c) -- [plain] (e)
        };
        \end{feynman}
        \end{tikzpicture}
\mbox{  \hspace*{18mm}
            (a) \hspace*{36mm}
            (b) \hspace*{42mm}
            (c) \hspace*{36mm}
            (d) \hspace*{25mm}
    }
\\
    \centering
        \begin{tikzpicture}
        \begin{feynman}
        \vertex (a) at (0,0) {\(Z\)};
        \vertex (b) at (1,0);
        \vertex (c) at (3,1) {\(\nu{_i}\)};
        \vertex (d) at (1.5,-0.3);
        \vertex (e) at (2.2,-0.6);
        \vertex (f) at (3,-1) {\(\nu{_j}\)};
        \diagram* {
        (a) -- [boson] (b),
        (b) -- [fermion] (c),
        (d) -- [plain, half right, edge label'=\( \chi_{_i}\)] (e),
        (d) -- [scalar, half left, edge label=\( \phi_{_k}\)] (e),
        (d) -- [fermion] (b),
        (f) -- [fermion] (e)
        };
        \end{feynman}
        \end{tikzpicture}
\mbox{  \hspace*{80mm}
        (e) \hspace*{130mm}
    }
    \caption{Additional invisible decay diagrams: $Z, H \rightarrow \nu_i\nu_j$, with $a,b=1-4$ and $k,l=1-3$ (plus similar contributions in the $\nu_j$ lines for the external leg corrections).}
    \label{fig:ZHlegdiag}
\end{figure}

\paragraph{Invisible $Z$ and Higgs decays}
In the present model, all the ``invisible'' NP particles turn out to be heavier than the EW scale, hence only neutrinos compose the final state of the invisible decays. The contributions of the new states strictly occur through virtual corrections: the invisible widths thus receive diagonal and non-diagonal contributions ($\nu_i \nu_j$ with $i=j$ or $i\neq j$).
In addition to the contributions already present for the charged leptonic decay modes (now strictly imposing $\alpha=\beta$) - i.e. corresponding to the diagrams that were presented in Figs.~\ref{fig:Zlegdiag} and~\ref{fig:Hlegdiag}, further modifications\footnote{Notice that now the final states comprise $\nu_i \nu_j$ final states instead of $\ell_\alpha \ell_\beta$.} concern only the triangle diagrams and corrections to the external lepton leg, as summarised in Fig.~\ref{fig:ZHlegdiag}.

The branching ratios are given in full analogy to the ones of Eqs.~(\ref{eq:ZBr},~\ref{eq:HBr}), noticing that one implicitly replaces $(\alpha, \beta) \to (i,j)$, with sums carried over all combinations of physical neutrinos in the final states.
The form factor decomposition is similar to that previously presented in Eqs.~(\ref{eq:ZHTensDec}). 
For the diagonal contributions, the tree-level couplings are given by $R^i_{SL} = - R^i_{SR} = i e/(2 c_w s_w)$ (with $S^i_L = S^i_R = 0$), and again, one must now include the corresponding counter-terms, see Appendices~\ref{sec:FF} and~\ref{app:renormalisation} for the form factors and details on the renormalisation procedure.

\paragraph{Oblique parameters} 
Finally, let us briefly comment on these observables, which are in general expressed through six quantities ($S$, $T$, $U$, $V$, $W$ and $X$)~\cite{Peskin:1990zt,Peskin:1991sw,Grimus:2008nb,Hagedorn:2018spx,Ahriche:2022bpx,AbuSiam:2025voc}. Under the assumption that the NP scale is considerably larger than the EW one (the so-called linear approximation in momentum), $V$, $W$ and $X$ vanish, and one is left with new contributions to $S$, $T$, $U$ parameters. 
The details of the computation are given in Appendix~\ref{app:Oblique}.
The current EW fit for the latter parameters (cf.~\cite{ParticleDataGroup:2024cfk}) leads to the following ($1\sigma$) intervals 
\begin{align}
        \Delta S &= -0.04 \pm 0.10\,, \nonumber \\ 
        \Delta T &= 0.01 \pm 0.12\,, \nonumber \\
        \Delta U &= -0.01 \pm 0.09\,.
\end{align}
Preliminary prospects for the sensitivity of FCC-ee to the $S$ and $T$ parameters have also been very recently presented~\cite{EPPSSU2026:EW}.

\section{Numerical results and discussion}\label{sec:numeric}

As already presented, in this class of models flavoured interactions are weighed by combinations of entries of the $\mathcal{G}$ matrix (itself written in terms of $g_\psi^\alpha$ and $g_{F_{1,2}}^\alpha$), which in turn encodes neutrino oscillation data through the modified Casas-Ibarra parametrisation.
Since usual Markov-chain MonteCarlo techniques do not allow to successfully explore the parameter space\footnote{A recent approach based on machine learning techniques has been deployed in~\cite{deSouza:2025uxb}, leading to an efficient scan of the phenomenologically motivated regions in parameter space. This novel approach allowed the authors to explore in more detail the parameter space, and to enrich the dark matter candidates, while preserving the driving assumptions of~\cite{Alvarez:2023dzz}.}, we have adopted the proposed technique of~\cite{Alvarez:2023dzz}, a hybrid approach to ultimately parametrise $\mathcal{G}$. This is detailed in Appendix~\ref{app:neutrinomass}, in which we also briefly comment on the regimes for certain entries of $\mathcal{G}$.

We have implemented the model's Lagrangian in FeynRules~\cite{Alloul:2013bka}, further relying on the packages FeynArts~\cite{Hahn:2000kx}, FeynCalc~\cite{Shtabovenko:2023idz,Shtabovenko:2020gxv,Shtabovenko:2016sxi,Mertig:1990an}, and LoopTools~\cite{Hahn:1998yk} to derive the fundamental interactions, and to facilitate the computation of the analytic expressions for the different observables. We have also implemented the model in SARAH~\cite{Staub:2015kfa,Staub:2013tta} to generate the SPheno~\cite{Porod:2003um,Porod:2011nf} spectrum file, which allowed cross-checking the predictions for a subset of observables.
As already mentioned, all relevant DM quantities (relic density, interaction cross-sections, constraints from direct and indirect detection) are obtained via a link to micrOMEGAs~\cite{Alguero:2023zol}.

In addition to the regimes for the entries of $\mathcal{G}$ (see Appendix~\ref{app:neutrinomass}), throughout the numerical analysis, the scanned ranges of the different parameters are summarised in Table~\ref{table:parameters}. 
\begin{table}[h!]
\centering\renewcommand{\arraystretch}{1.8} 
\begin{tabular}{cc|cc}
\hline
\hline
\textbf{Parameter} & \textbf{Range} & \textbf{Parameter} & \textbf{Range} \\ 
\hline
$m_{\nu_{1}}$ & $\left[ 10^{-19}, 10^{-10} \right]$ GeV & $M_{S}^{2}, M_{\eta}^{2}$ & $\left[ 5 \times 10^{5}, 5 \times 10^{6} \right]$ GeV \\
$\lambda_{4S}, \lambda_{4\eta}$ & $\left[ 10^{-7}, 1 \right]$ & $M_{1}, M_{2}$ & $\left[ 100, 20000 \right]$ GeV \\ 
$\lambda_{S\eta}, \lambda_{S}$ & $\pm\left[ 10^{-3}, 1 \right]$ & $M_{\psi}$ & $\left[ 700, 2000 \right]$ GeV \\
$\lambda_{\eta}, \lambda_{\eta^{\prime}}, \lambda_{\eta^{\prime\prime}}$ & $\pm\left[ 10^{-3}, 1 \right]$ & $y_{11, 12, 21, 22}$ & $\pm\left[ 10^{-10}, 10^{-4} \right]$ \\
$\alpha$ & $\pm\left[ 10, 10^{4} \right]$ GeV & $\log_{10}(r_e)$ & $\mathcal{N}(4.0,0.7)$\\
$\log_{10}(r_\mu )$ & $\mathcal{N}(0.0,0.3)$ & $\log_{10}(r_\tau)$ & $\mathcal{N}(3.0,0.7)$
\\
\hline
\hline
\end{tabular}
\caption{Input parameters for the random scan. Here, we have introduced 
$r_{\alpha} = | g_{\psi}^{\alpha}|/ | g_{R}^{\alpha}|$  
(cf. discussion in Appendix~\ref{app:neutrinomass}; moreover, $\mathcal{N}(\mu,\sigma)$ denotes a normal distribution.} 
\label{table:parameters}
\renewcommand{\arraystretch}{1} 
\end{table}

As emphasised before, we explore thoroughly the phenomenology of the model, considering the impact of specific requirements (other than neutrino mass generation) on certain distinctive features; this is for instance the case of relaxing the assumptions on sizeable NP contributions to $(g-2)_\mu$.

Although we will not enter in a detailed discussion (as most findings do corroborate that of previous analyses), we nevertheless briefly discuss some points concerning the NP spectrum and viable DM candidates. Regarding neutrino mass generation, here we have focused on the case of a normal ordering of the neutrino spectrum, for which NuFit~\cite{Esteban:2024eli} data was then used to determine the relevant couplings. 
We have nevertheless verified that starting from an inverted ordering one would be led to similar phenomenological implications (concerning favoured regimes and results for the observables) which would thus not affect the conclusions drawn in this section.

For the points subsequently deemed phenomenologically valid (including having a viable dark matter relic density), the spectrum of the NP states lies in the following intervals 
\begin{align}
 &   280~\text{GeV} \leq M_{\phi_1} \leq 2200~\text{GeV}\,, \quad 
    720~\text{GeV} \leq M_{\phi_2} \leq 2250~\text{GeV}\,, \nonumber \\
  &  680~\text{GeV} \leq M_{A^0} \leq 2250~\text{GeV}\,, \quad 
    670~\text{GeV} \leq M_{\eta^\pm} \leq 2240~\text{GeV}\,, \nonumber \\
 &    240~\text{GeV} \leq M_{\chi_{_1}} \leq 1990~\text{GeV}\,, \quad 
390~\text{GeV} \leq M_{\chi_{_2}} \leq 2240~\text{GeV}\,, \quad 
700~\text{GeV} \leq M_{\chi_{_{3,4}}} \leq 20~\text{TeV}\,.
\end{align}

Finally, and concerning the nature of the dark matter candidate, and in addition to (phenomenologically viable) scalar and fermion candidates, our scan of the parameter space further revealed that one could also have pseudoscalar candidates. These subdominant regimes had escaped the analysis of~\cite{Alvarez:2023dzz} (but were identified in the recent study of~\cite{deSouza:2025uxb}).  
CP-odd scalar dark matter candidates emerge for intermediate values of the scalar trilinear coupling $\alpha \lesssim \mathcal{O}(1~\text{TeV})$, and are not associated with distinctive features in what concerns the flavour and precision observables here studied. It is also worth noticing that direct DM detection bounds are the source of important constraints.

\subsection{Charged lepton flavour violation: pure leptonic decays}\label{sec:results:cLFV}
\begin{figure}[h!]
\centering
\includegraphics[width=0.45 \textwidth] {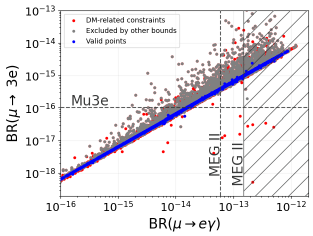}
\hspace*{10mm}
\includegraphics[width=0.45 \textwidth] {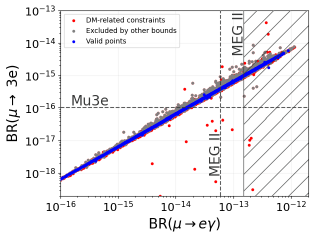}
\\
\includegraphics[width=0.45 \textwidth]{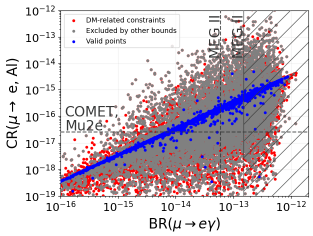}
\hspace*{10mm}
\includegraphics[width=0.45 \textwidth]{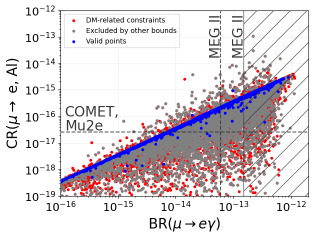}
\caption{Predicted rates for $\mu \to 3e$ (top) and $\mu -e$ conversion in nuclei (bottom), both versus BR($\mu \to e \gamma$). On the left panels, all displayed points lead to a SM-like $(g-2)_\mu$ (i.e. $\Delta a_\mu \approx 1.5\sigma$), while the right panels exhibit a significant NP contribution ($\Delta a_\mu \approx 4.2\sigma$).  
Red points denote exclusion due to conflict with dark matter constraints, grey points correspond to the violation of at least one phenomenological bound (flavour, EWPO) other than those under study; blue points are viable under all constraints but those under consideration (in this case BR($\mu \to e\gamma$)). Full (dashed) lines correspond to current bounds (future sensitivity), with hatched areas being already excluded.} 
\label{fig:Mu3e:Mu-e:mueg}
\end{figure}

A first analysis of the parameter space of this scotogenic variant naturally addresses the prospects for cLFV observables\footnote{As discussed in detail upon the description of how the model's parameter space is surveyed (Appendix~\ref{app:neutrinomass}), certain radiative cLFV decays are a preliminary input in the procedure to reconstruct the matrix of couplings, $\mathcal{G}$.}.
In Fig.~\ref{fig:Mu3e:Mu-e:mueg}, we thus present the prospects for flavour violation in the muon sector: on the top row, we display the rates for $\mu \to 3e$ versus BR($\mu \to e \gamma$), while on the bottom row rates for  $\mu -e$ conversion in Aluminium nuclei again versus BR($\mu \to e \gamma$). As aforementioned, and to better assess the impact of saturating (or not) the previously existing tension in the muon anomalous magnetic moment, left and right panels respectively feature a SM-like prediction to $(g-2)_\mu$ or alternatively, significant NP contributions to account for a $4.2 \sigma$ deviation. Here (and as it will be done throughout our analysis), we explicitly display in red regimes ruled out due to dark matter related constraints (relic density and direct detection experiments), and in grey, points excluded due to the violation of at least one phenomenological bound, other than regarding the observable under study. 
Finally, viable points are represented in blue (including those explicitly in conflict with the depicted observables). We recall that relevant cLFV bounds and future sensitivities were collected in Table~\ref{tab:cLFV_current_future_sensitivity}.

As can be readily seen, and once both phenomenological and DM-related constraints are imposed, a correlated behaviour between three-body and the radiative muon cLFV decays becomes apparent, in agreement with what had been identified\footnote{Such correlations between cLFV observables do occur in other scotogenic model realisations, see for instance~\cite{Toma:2013zsa}.} in previous studies of this scotogenic variant~\cite{Alvarez:2023dzz}. 
The valid points (i.e. lying on a blue band)
are associated with dominant photon-penguin contributions to the $\mu \to 3 e$ decays; albeit strongly disfavoured throughout the explored parameter space, one can also have contributions from $Z$-penguins.

The comparison with the projections of the model's parameter space obtained upon aiming at explaining a significant tension in $\Delta a_\mu$ are displayed for completeness on the right panel: although the correlation obtained for the viable points is not altered, it is nevertheless interesting to verify how enhancing the flavour-conserving dipole contribution (i.e. the NP contributions to the muon anomalous magnetic moment) in turn induces a clear dominance of the cLFV dipole contribution to the $\mu \to 3 e$ decays, with other contributions being clearly subdominant in this case. For a SM-like $(g-2)_\mu$ (cf. left panel), $Z$-penguin exchanges to $\mu \to 3 e$ decays can play a role, leading to a potential loss of correlation (albeit for regimes excluded due to other cLFV bounds).  

In both cases, it is important to highlight that one can easily have contributions for both cLFV observables which lie within future sensitivity for both MEGII and Mu3e: future observations of these decays can thus help falsifying the model, or then support its viability. For cLFV predictions beyond MEGII reach, part of the explored parameter space can still be probed via searches for $\mu \to 3 e$ decays.

The bottom row of Fig.~\ref{fig:Mu3e:Mu-e:mueg} offers the predictions for the neutrinoless muon-electron conversion, again presented versus the radiative muon decay, for both SM- and NP-like scenarios of $\Delta a_\mu$. For the case of a $4.2 \sigma$ tension in the muon anomalous magnetic moment one again recovers a clear correlation between both observables, a direct consequence of dominant dipole contributions to $\mu-e$ conversion\footnote{Notice that in view of the NP interactions and symmetries, there are no new couplings to quarks, and hence no box contributions to muon-electron conversion in nuclei.}; $Z$-penguin and anapole contributions can become sizeable and destructively interfere leading to the (mostly) excluded points below the distinctive correlation line. Once one relaxes the requirement of a large NP contribution to $(g-2)_\mu$, and in addition to the destructive interference mentioned above (which is now more pronounced), $Z$-penguin contributions to $\mu-e$ conversion can become orders of magnitude larger than those of the dipole (as much as 1000 larger), and one observes a significant spread around the experimentally allowed blue band, which is also significantly thicker than in the previous case. 
In all cases,  muon-electron conversion in nuclei offers excellent prospects for tests of this scotogenic variant.

\begin{figure}[h!]
\centering
\includegraphics[width=0.45\textwidth]
{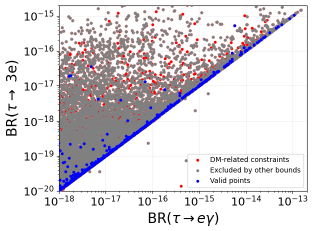}
\hspace*{10mm}
\includegraphics[width=0.45 \textwidth]
{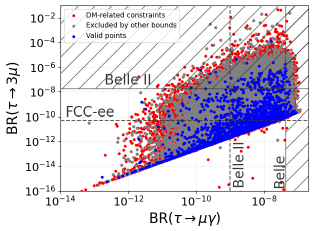}
\caption{Prospects for cLFV tau-lepton decays: on the left, BR($\tau \to 3 e$) vs. BR($\tau \to e \gamma$), and on the right
BR($\tau \to 3 \mu$) vs. BR($\tau \to \mu \gamma$). Line and colour code as in Fig.~\ref{fig:Mu3e:Mu-e:mueg}.}
\label{fig:tau3mu:tau-mugamma}
\end{figure}

Concerning cLFV tau lepton decays, our results are summarised in Fig.~\ref{fig:tau3mu:tau-mugamma}, in which we display $\tau \to 3 e$ and $\tau \to 3 \mu$, respectively versus $\tau \to e \gamma$ and $\tau \to \mu \gamma$. 
As expected, we have verified that considering distinct regimes for $\Delta a_\mu$ leads to little effects on $\tau$ cLFV decay rates. Hence we have only displayed the currently favoured SM-like scenario. 
As can be immediately seen from both plots, the observed patterns for tau cLFV leptonic decays are quite different from those encountered for muon decays; strikingly (and for both viable and phenomenologically excluded points), one has a significant spread in what concerns the predictions of the 3-body decays; notice however that no destructive interferences occur between the distinct contributions to $\tau \to 3\ell$. 

While the dominant contributions to the tau 3-body decays still indeed arise in most cases from the dipole operators (corresponding to a very saturated ``correlation'' line, almost invisible to the naked eye), for $\tau \to 3 \mu$ there are important contributions from box diagrams, which are at the source of the largest rates. Furthermore, and while for cLFV decays in the $\tau-e$ the predictions are clearly below any future experimental sensitivity, $\tau \to \mu$ decays (both radiative and three-body) are well within future experimental sensitivity; in fact, and for extensive regions in the model's parameter space,  $\tau \to 3\mu$ decays are one of the most constraining observables.

The results here presented are in good agreement with the subset presented~\cite{Alvarez:2023dzz}; notice however that relaxing the requirement on $(g-2)_\mu$ allows the presence of distinct regimes for leptonic cLFV decays (although most already phenomenologically excluded).

\subsection{Lepton flavour violating $Z$ and $H$ decays}

\begin{figure}[h!]
\centering
\includegraphics[width=0.45 \textwidth]{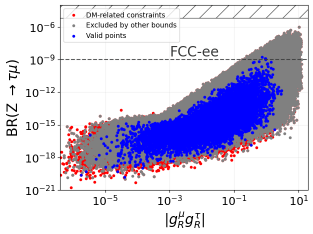}
\hspace*{10mm}
\includegraphics[width=0.45 \textwidth]
{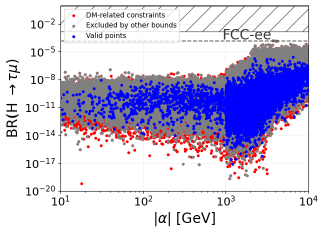}
\caption{On the left, BR($Z \to \tau \mu$)
as a function of the right-handed charged lepton couplings to $\eta$ and $\Psi_1$, $g_R^{\mu,\tau}$ (cf. Eq.~(\ref{eq:lagrangian:fermion})).
On the right panel, BR($H \to \tau \mu$) versus 
the triple scalar coupling, $\alpha$ (in~GeV), see Eq.~(\ref{eq:Vscalar}).
Line and colour code as in Fig.~\ref{fig:Mu3e:Mu-e:mueg}.
}
\label{fig:ZTauMu:HTauMu}
\end{figure}

We now turn our attention to cLFV decays of neutral SM bosons. In view of the findings regarding cLFV in the $\tau-\mu$ sector, one could in principle expect sizable contributions to the $Z, H \to \tau \mu$ decays. 
We present the results of our study (relying on independent computations) in Fig.~\ref{fig:ZTauMu:HTauMu}. The relevant cLFV bounds and future sensitivities have been summarised in Table~\ref{tab:cLFV_current_future_sensitivity}.
We have chosen to illustrate the behaviour of the BR($Z\to \tau \mu$) as a function of two particularly driving quantities: the couplings of $\eta$ and $\Psi_1$ to the second and third generation right-handed charged leptons, $g_R^{\mu,\tau}$, taking their product for simplicity (we recall that there is a significant hierarchy for the $g_R^{\alpha}$ couplings, which accounts for the range displayed). 

As visible from the left panel of Fig.~\ref{fig:ZTauMu:HTauMu}, large values of 
BR($Z\to \tau \mu$) are indeed possible, in association to sizeable values of the couplings, $|g_R^{\mu}\, g_R^{\tau}| \geq 0.1$, as would be expected. Nevertheless, recall that $Z$-penguin contributions to $\tau \to 3\mu$ were responsible for very large values of this observable, some regimes even already excluded by current Belle II bounds. Ultimately, this precludes ``observable'' values for $Z\to \mu \tau$ decays\footnote{From a close inspection of Fig.~\ref{fig:tau3mu:tau-mugamma} notice that while having sizeable values of BR($Z\to \tau\mu$) - within FCC-ee reach - is technically possible, such regimes are statistically disfavoured.}.
 
Regarding flavour violating Higgs decays, we summarise our results on the right panel of Fig.~\ref{fig:tau3mu:tau-mugamma}, in which we display the associated rates as a function of the NP trilinear scalar coupling, $\alpha$, which plays a driving role for this observable\footnote{The apparent effect associated with a greater density of points for $\alpha \gtrsim 1$~TeV is simply an artifact of the scanning technique.}. Although many distinct contributions are present (cf. diagrams of Fig.~\ref{fig:Hlldiag} and~\ref{fig:ZHlegdiag}), vertices involving  $\alpha$ (which is also a key ingredient in the scalar spectrum) offer dominant contributions.
Despite potential predictions very close to future experimental sensitivity, the feasibility of observing $H\to \tau\mu$ decays is again hampered by conflicts with the current bounds on $\tau \to 3 \mu$. 

\subsection{Electroweak precision observables}
In view of the potential of future lepton colliders in what concerns electroweak precision, addressing the impact of a given NP model regarding the latter becomes important. 
We have thus taken special care to address EWPO in this scotogenic variant, as well as any potential impact regarding fundamental tests of SM paradigms (such as universality of lepton flavour interactions).
As detailed in Section~\ref{sec:EWPO}, in our study we have carried a full computation of all the observables; in particular, the renormalisation of the distinct quantities (taking into account all relevant higher order corrections) has been done, with the details given in the appendices.

\paragraph{Invisible $\pmb{Z}$ and Higgs decays}
The invisible $Z$ width has been a very strong constraints for models with an extended and/or modified neutral lepton sector. In what follows, we consider how the NP contributions to the $Z\to \nu \nu$ decays can interfere with the SM ones.
We again emphasise that all the expressions have been derived from a first principle approach, and that a full renormalisation procedure has been implemented. 
The numerical results are obtained for points associated with a SM-like anomalous magnetic moment of the muon.
\begin{figure}[h!]
\centering
\includegraphics[width=.6\textwidth]{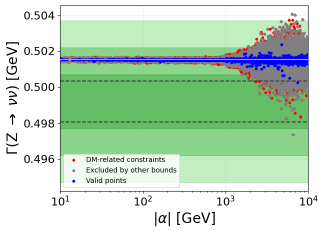}
\caption{Prospects for $\Gamma$($Z\to \nu\nu$) as a function of $|\alpha|$. 
Point-colour scheme as in Fig.~\ref{fig:Mu3e:Mu-e:mueg}. From darker to lighter, the green bands denote current bounds at 
$1\sigma$, $2\sigma$ and $3\sigma$.
The full orange line corresponds to the SM 2-loop prediction~\cite{Dubovyk:2018rlg}, while dashed grey lines denote future sensitivity at FCC-ee, centred on the current experimental central value.}
\label{fig:Zinv}
\end{figure}
Figure~\ref{fig:Zinv} illustrates our findings for $\Gamma$($Z\to \nu\nu$), displayed as a function of $|\alpha|$. 

As can be seen from the numerical results, the scotogenic variant under study can be at the origin of sizeable contributions to the invisible $Z$ width, and with the advent of a precision era at FCC-ee, important regimes in parameter space can be probed. For regimes of very large trilinear couplings (i.e. $|\alpha| \gtrsim 1$~TeV), the invisible $Z$ width becomes very sensitive to the latter, and the NP contributions exhibit a significant departure from the SM expectation (but are excluded due to violating other constraints).
Albeit statistically disfavoured, certain points could even explain significant tensions between the SM prediction and observation. 

\begin{figure}[h!]
\centering
\includegraphics[width=0.6\textwidth]{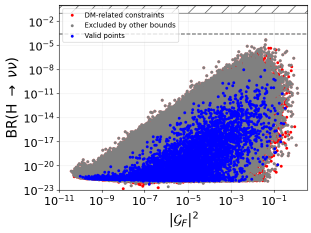}
\caption{Invisible Higgs decays: BR($H\to \text{inv.}$) versus $|\mathcal{G}_F|^2={\sum_{a \beta}|\mathcal{G}_{F_{a \beta}}|^2}$. 
Line and colour code as in Fig.~\ref{fig:Mu3e:Mu-e:mueg}.
}
\label{fig:Hinv}
\end{figure}
For completeness, in Fig.~\ref{fig:Hinv} we display the predictions to the invisible Higgs decays (corresponding to $H \to \nu\nu$), presenting them as a function of a sum over the entries of the lower $2 \times 3$ submatrix of the generalised neutrino coupling matrix, $\mathcal{G}_F$ (see Eq.~(\ref{eq:Gmatrix})), as such quantities prove convenient to encode in an effective way the different couplings to the Higgs. While clearly different from the SM case (strictly massless neutrinos), the NP contributions lie beyond experimental sensitivity.

\paragraph{Lepton flavour universality:  flavour conserving $\pmb{Z}$ and Higgs decays}
We finally address the decays of the $Z$ and the Higgs to a pair of same-flavoured leptons. Such rates - especially $Z \to \ell \ell$ - further allow to test the SM paradigm of flavour-universality in gauge-lepton interactions. (For the Higgs, this allows testing deviations from the ratio of charged lepton masses). 
Once more we highlight that in our computation we have taken into account higher order (1-loop) effects, and carried out a full (analytical) renormalisation of the interactions, as detailed in the appendices. 

\begin{figure}[h!]
\centering
\includegraphics[width=0.45 \textwidth]{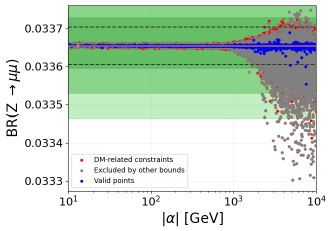}
\hspace*{10mm}
\includegraphics[width=0.45 \textwidth]
{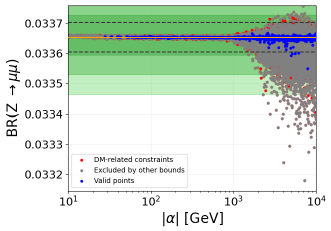}
\caption{BR($Z \to \mu \mu$) as a function of the scalar trilinear coupling $\alpha$. Line and colour code as in Fig.~\ref{fig:Zinv}. 
The left panel (right) corresponds to a SM-like (NP-like) $(g-2)_\mu$.
}
\label{fig:ZMuMu}
\end{figure}
In the panels of Fig.~\ref{fig:ZMuMu} we present our results for BR($Z\to \mu\mu$), shown here as a function of the most relevant parameter - the trilinear scalar coupling. Although for large values of $\alpha \gtrsim 1$~TeV one can have sizeable contributions to this decay, they are in general disfavoured due to conflict with other bounds (in particular $H \to \ell_\alpha \ell_\alpha$); nevertheless, a small subset of points can significantly deviate from the current measured value (or then from the SM expectation, corresponding to a 2-loop evaluation~\cite{Dubovyk:2018rlg}). Such deviations could be in principle testable at FCC-ee. 
Although we do not present it here, let us mention that a similar behaviour is found for $Z\to ee$ and $Z\to \tau \tau$ decays. At the end of this section we will address the prospects for the LFUV sensitive ratio, $R^Z_{\mu \tau}$. 

\begin{figure}[h!]
\centering
\includegraphics[width=0.45 \textwidth]{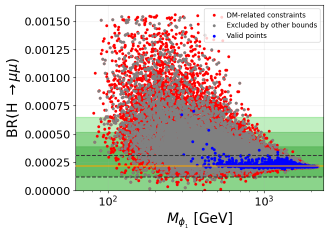}
\hspace*{10mm}
\includegraphics[width=0.45 \textwidth]
{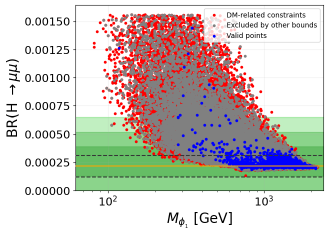}
\caption{BR($H \to \mu \mu$) as a function of the lightest scalar mass, $M_{\phi_1}$, for a SM-like $(g-2)_\mu$ (left) and a significant tension ($\Delta a_\mu = 4.2 \sigma$, on the right).
Line and colour code as in Fig.~\ref{fig:Zinv}, with the full line denoting the SM prediction~\cite{LHCHiggsCrossSectionWorkingGroup:2016ypw}.}
\label{fig:HMuMu}
\end{figure}

In Fig.~\ref{fig:HMuMu} we present our projections for flavour conserving Higgs decays, in particular $H\to \mu\mu$; just like done in previous subsections, we consider the effect that saturating a significant tension in $\Delta a_\mu$ might have on this observable. 
Firstly let us notice that although lying beyond the displayed range, one can have even larger contributions corresponding to huge deviations from the SM expectations, which are clearly excluded; secondly, and contrary to what was done in previous studies~\cite{Alvarez:2023dzz}, we allow for regimes in which $M_{\phi_1}$ can be potentially light. Although not exclusively, such regimes can occur in association with cancellations driven by large values of the trilinear scalar coupling, $\alpha$ (above 1~TeV), leading to conflict with other observables. 
However, it is interesting to verify that in this scotogenic variant, the contributions to $H\to \mu\mu$ decays can be extremely large (in certain cases, albeit statistically less meaningful, beyond $3\sigma$ from the SM prediction). The comparison of left and right panels of Fig.~\ref{fig:HMuMu} further reveals the effect of relaxing the enhancement of the muon dipole contributions, with a considerable larger spread of phenomenologically allowed points for a NP-like $(g-2)_\mu$. 

We have also investigated the prospects for $H\to \tau\tau$ decays, with very similar findings: the flavour conserving Higgs to tau decays also offer the possibility of large deviations from the SM, although less sensitive to $\Delta a_\mu$.

\begin{figure}[h!]
\centering
\includegraphics[width=0.45\textwidth]{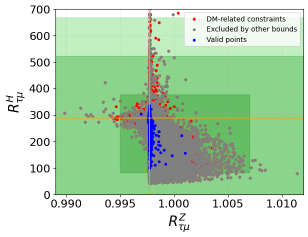}
\hspace*{10mm}
\includegraphics[width=.45\textwidth]{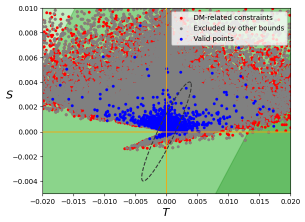}
\caption{On the left panel, $R^Z_{\tau \mu}$ versus $R^H_{\tau \mu}$; on the right, prospects for the $S$ and $T$ parameters.
The full orange line corresponds to the SM 2-loop prediction~\cite{Dubovyk:2018rlg}, while dashed grey lines denote future sensitivity at FCC-ee, centred on the current experimental central value.
Line and colour code as in Fig.~\ref{fig:Zinv}; on the right panel, the preliminary sensitivity of FCC-ee~\cite{EPPSSU2026:EW} is denoted by the dashed contour.}
\label{fig:RHZTaMu:oblique}
\end{figure}

We finally consider the ratios of flavour conserving decays, $R_Z^{\tau \mu}$ and $R_H^{\tau \mu}$, which are sensitive to new sources of lepton flavour universality violation (in addition to those present in the SM, i.e. the Yukawa couplings). For simplicity, we summarise in a single view $R^Z_{\tau \mu}$ and $R^H_{\tau \mu}$, as displayed in 
the left panel of Fig.~\ref{fig:RHZTaMu:oblique} (for a SM-like $(g-2)_\mu$). 
We thus recover a global picture in agreement with the discussion so far: while large deviations from the SM expectations could in principle be present, such regimes are mostly excluded due to conflict with DM and /or phenomenological constraints. 
In view of this, it is thus clear that deviations from the SM expectation in the individual decay rates offer better prospects to study the implications of the ``T1-2-A'' scotogenic variant.

\paragraph{Oblique parameters}
Finally, we have also computed the NP contributions to the $S$, $T$ and $U$ parameters; our numerical results confirm that these play a non-constraining role on the ``T1-2-A'' parameter space\footnote{Our computation, relying on the expressions detailed in Appendix~\ref{app:renormalisation} lead to numerical results which - despite being in agreement with observation and SM expectation - reveal a certain tension with respect to those of SPheno~\cite{Porod:2003um,Porod:2011nf} regarding the $U$ parameter.}. However, the preliminary expected precision of a future FCC-ee~\cite{EPPSSU2026:EW} might lead to a significant reduction in the associated uncertainties, so that the current scotogenic realisation might be efficiently probed via its predictions for the oblique $S$ and $T$ parameters, as visible by the constraining impact of the dashed contour depicted in the right panel of Fig.~\ref{fig:RHZTaMu:oblique}.


\section{Conclusions}\label{sec:conclusion}
Scotogenic models offer an appealing and natural connection between the mechanism of neutrino mass generation and an explanation for the observed dark matter relic density. In recent years, numerous variants and realisations have been explored: lepton (flavoured) observables have been identified as powerful probes of these NP constructions.

In view of its potential to generate light neutrino masses, put forward viable DM candidates and explain the observed baryon asymmetry of the Universe, the ``T1-2-A'' variant has been under intense scrutiny in recent years. 
Its impact for cLFV transitions and decays, and also the potential to saturate the formerly existing tension in the anomalous magnetic moment of the muon, have rendered it the object of several studies~\cite{Alvarez:2023dzz}; novel machine learning techniques have also been employed to efficiently scan a highly non-trivial parameter space (and address indirect detection at the LHC)~\cite{deSouza:2025uxb}. 

In this work we have revisited this class of ``T1-2-A'' scotogenic variant, conducting a thorough study of its phenomenological implications for numerous charged lepton flavour observables, as well as electroweak (precision) tests. We have independently derived the full analytical expressions for all the cLFV rates (which included pure leptonic decays, neutrinoless conversion in nuclei, and cLFV decays of Higgs and $Z$ bosons). 
In view of the excellent prospects of the next generation of lepton colliders, we have also addressed several EW observables (among them invisible $Z$ and Higgs decays, and oblique parameters), as well as flavour conserving leptonic decays, $Z(H) \to \ell \ell$, fully carrying out a renormalisation procedure - which is presented in detail for future studies. For all the latter observables, we have done a full computation of all relevant one-loop contributions to the decay widths without any simplifying approximations.

In our analysis, we have relaxed certain driving assumptions of previous studies (see, e.g.~\cite{Alvarez:2023dzz}), in particular in what concerns explaining the BAU from leptogenesis; furthermore, we have addressed the impact of the evolution concerning the anomalous magnetic moment of the muon (which is now in good agreement with the SM prediction), in particular the impact of the latter on the expected patterns for muon cLFV decays and transitions. Our findings suggest that one is still led to correlations between the observables - albeit not as strong (due to a reduced dominance of the dipole contributions); the cLFV patterns in the muon sector thus remain excellent means to probe and falsify these constructions. 

Concerning the prospects for cLFV $Z$ and Higgs decays, one can have very large contributions, but these are however excluded due to conflict with other observables, in particular cLFV $\tau \to 3\mu$ decays, and are thus in general beyond future sensitivity.
Invisible $Z$ decays could receive sizeable contributions, in association with very large values of the scalar trilinear coupling ($\alpha \geq 1$~TeV), but these are also precluded due to conflict with cLFV bounds. It is important to mention that such large regimes of $\alpha$ were favoured to explain the observed BAU (as done in~\cite{Alvarez:2023dzz}). 

Other than the leptonic cLFV processes, the most promising observables to look for deviations from the SM are perhaps dilepton Higgs decays, for which one can have contributions beyond $2-3\sigma$ of current measurements. In the future, the preliminary expected sensitivity of the FCC-ee might also allow probing (end excluding) significant parts of the model's parameter space, which would be otherwise phenomenologically viable.

This class of scotogenic models clearly offers a rich canvas for NP studies; its extremely rich phenomenology should allow for signals to be observed in the coming cLFV low-energy experiments, or at colliders via di-muon or di-tau Higgs decays. Moreover, it might prove interesting to explore peculiar decay chains, leading to signatures that can be searched for at the LHC (or possibly at future lepton colliders).

\section*{Acknowledgements}
The authors are grateful to M.~Sarazin and B.~Herrmann for useful discussions. We are also indebted to A.~Goudelis, J.~Kriewald and E.~Pinsard for valuable advice.
This project has received support from the IN2P3 (CNRS) Master Project, ``Hunting for Heavy Neutral Leptons'' (12-PH-0100).

\appendix

\section{Neutrino mass generation in the ``T1-2-A'' scotogenic model }\label{app:neutrinomass}

As a consequence of the $Z_{2}$ symmetry introduced to stabilise the potential DM candidate, neutrino masses are realised at the one-loop level. 
In what follows, we compute the contributions to the neutrino mass matrix, $\mathcal{M}_{\nu}$, arising from the diagrams already depicted in Fig.~\ref{fig:NMassInt}.
We then present a modified Casas-Ibarra parametrisation allowing to successfully accommodate oscillation data, and finally describe the procedure allowing to fully reconstruct $\mathcal{M}_{\nu}$, relying on an interplay of oscillation data and charged lepton flavour violation limits.

\subsection{Higher-order contributions to neutrino masses}
As mentioned in Section~\ref{sec:model}, after EWSB, and starting from the interactions present in the fermion Lagrangian of Eq.~(\ref{eq:lagrangian:fermion}), the neutrino mass matrix can be cast in terms of a ``coupling" matrix
$\mathcal{G}$, and of $\mathcal{M}_{L}$, the latter encoding the relevant information regarding the new massive fields propagating in the loop~\cite{Alvarez:2023dzz}, which we rewrite here for convenience
\begin{equation}\label{eq:Gmatrix-App}
    \mathcal{M}_{\nu} = \mathcal{G}^T \mathcal{M}_{L} \mathcal{G}, \quad \text{where} \quad \mathcal{G} = \begin{pmatrix}
        g_{\psi}^{e} & g_{\psi}^{\mu} & g_{\psi}^{\tau} \\[5pt]
        g_{F_{1}}^{e} & g_{F_{1}}^{\mu} & g_{F_{1}}^{\tau} \\[5pt]
        g_{F_{2}}^{e} & g_{F_{2}}^{\mu} & g_{F_{2}}^{\tau}
    \end{pmatrix}\,,
\end{equation}
with $\mathcal{M}_{L}$ ($3 \times 3$ symmetric matrix) written in terms of the $\chi$ mixing matrix $U_{\chi}$ (see Eqs.~(\ref{eq:chi:M:Uchi})) and of the scalar mixing matrix $U_{\phi}$ (cf. Eq.~(\ref{eq:UPhi:def})),
\begin{align}
    \left( \mathcal{M}_{L} \right)_{11} &= \sum_{ik} b_{ik} \,( U_{\chi}^* )_{4i}^{2} \,( U_{\phi} )_{1k}^{2}\,, \nonumber\\
    \left( \mathcal{M}_{L} \right)_{22} &= \frac{1}{2} \sum_{ik} b_{ik} \,( U_{\chi}^* )_{1i}^{2} \left[ ( U_{\phi} )_{2k}^{2} - (U_{\phi} )_{3k}^{2} \right] \,,\nonumber\\
    \left( \mathcal{M}_{L} \right)_{33} &= \frac{1}{2} \sum_{ik} b_{ik} \,( U_{\chi}^* )_{2i}^{2} \left[ ( U_{\phi} )_{2k}^{2} - (U_{\phi} )_{3k}^{2} \right] \,,\nonumber\\
    \left( \mathcal{M}_{L} \right)_{12} = \left( \mathcal{M}_{L} \right)_{21} &= \frac{1}{\sqrt{2}} \sum_{ik} b_{ik} \,( U_{\chi}^* )_{1i} \,( U_{\chi}^* )_{4i} \,( U_{\phi} )_{1k} \,( U_{\phi} )_{2k} \,,\nonumber\\
    \left( \mathcal{M}_{L} \right)_{13} = \left( \mathcal{M}_{L} \right)_{31} &= \frac{1}{\sqrt{2}} \sum_{ik} b_{ik}\, ( U_{\chi}^* )_{2i} \,( U_{\chi}^* )_{4i} \,( U_{\phi} )_{1k}\, ( U_{\phi} )_{2k} \,,\nonumber\\
    \left( \mathcal{M}_{L} \right)_{23} = \left( \mathcal{M}_{L} \right)_{32} &= \frac{1}{2} \sum_{ik} b_{ik} \,( U_{\chi}^* )_{2i} \,( U_{\chi}^* )_{1i} \left[ ( U_{\phi} )_{2k}^{2} - U_{\phi} )_{3k}^{2} \right]\,,
\end{align}
in which $i = 1, 2, 3, 4$ and $k = 1, 2, 3$. Finally the loop function $b_{ik}$ is given by~\cite{Alvarez:2023dzz}:
\begin{equation}
    b_{ik} \,= \,b(m_{\chi_{_i}^{0}},m_{\phi_{_k}^{0}}) \,= \,
    \frac{1}{16\pi^2} \,\frac{m_{\chi_{_i}^{0}}}{m_{\phi_{_k}^{0}}^{2}-m_{\chi_{_i}^{0}}^{2}} \left[ m_{\chi_{_i}^{0}}^{2} \ln m_{\chi_{_i}^{0}}^{2} - m_{\phi_{_k}^{0}}^{2} \ln m_{\phi_{_k}^{0}}^{2} \right]\,.
\end{equation}
Using a modified Casas-Ibarra parametrisation~\cite{Casas:2001sr,Basso:2012voo}, one can then encode neutrino oscillation data in $g_{\psi}$ and $g_{F}$ (entering in $\mathcal{G}$),
\begin{equation}
    \mathcal{G} \,= U_{L} \,D_{L}^{-1/2}\, R\, D_{\nu}^{1/2} \,U_{\text{PMNS}}^{*}\,,
    \label{eqn:casas_ibarra}
\end{equation}
in which
\begin{equation}
    D_{L} \,= \,U_{L}^{T} \,\mathcal{M}_{L} \,U_{L}\,,
\end{equation}
and $D_{\nu}$ is the diagonal matrix of neutrino mass eigenvalues; the unitary $3 \times 3$ $U_{\text{PMNS}}$ matrix encodes leptonic mixing; as usually done, the remaining degrees of freedom can be expressed through the $R$ mixing matrix:
\begin{equation}
    R = \begin{pmatrix}
        c_3 & -s_3 & 0\\
        s_3 & c_3 & 0\\
        0 & 0 & 1
    \end{pmatrix}\begin{pmatrix}
        c_1 & 0 & -s_1\\
        0 & 1 & 0\\
        s_1 & 0 & c_1
    \end{pmatrix}\begin{pmatrix}
        1 & 0 & 0\\
        0 & c_2 & -s_2\\
        0 & s_2 & c_2
    \end{pmatrix}\,,
 \label{eqn:R_matrix}
\end{equation}
where $s_{i} = \sin\theta_{i}$ and $c_{i} = \cos\theta_{i}$, with three complex mixing angles $\theta_{1,2,3}$. As usual, the $R$ matrix is paramount to lepton flavour contributions (in our case including not only cLFV transitions but also magnetic moments).
 
\subsection{Full reconstruction of $\mathcal{M}_{\nu}$: determining $\mathcal{G}$}
As visible from Eq.~(\ref{eqn:casas_ibarra}), 
$\mathcal{G}$ calls upon elements from two distinct sectors: neutrino oscillation data and generalised loop elements involving the new scalar and fermion fields, as manifest from Fig.~\ref{fig:NMassInt}. Data from NuFit~\cite{Esteban:2024eli}, allows defining both $D_{\nu}$ and $U_{\text{PMNS}}$; for a given choice of parameters of the extended scalar and fermion sectors, one can also determine $U_{L}$ and $D_{L}$. At this stage, the full determination of $\mathcal{G}$ relies on the three complex angles of the $R$ matrix. Such an inverse problem cannot be solved exactly (neither analytically nor numerically). A hybrid approach\footnote{As already mentioned in the main body of the manuscript, machine learning techniques have been recently used to carry out a more efficient can of the parameter space~\cite{deSouza:2025uxb}.}, as proposed in~\cite{Alvarez:2023dzz}, relies on an iterative procedure, which allows preliminary approximate ``guestimates'' relying on the external input of a set of observables - cLFV radiative decays and the muon anomalous magnetic moment.

As can be inferred from the discussion in Sections~\ref{sec:AMM} and~\ref{sec:cLFV:lepton}, two classes of diagrams generically contribute to the radiative leptonic processes; however, those featuring charged scalars and neutral fermions (i.e. diagram (b) in Fig.~\ref{fig:radiative:AMM:cLFV}) are typically subdominant due to the hierarchy between $\alpha$ and $y_{ij}$. Neglecting the latter, one easily verifies that the magnetic moment and the radiative cLFV rates can be cast (as a first approximation) in terms of $g_\psi^\ell$ and $g_R^\ell$ (with $\ell=e, \mu, \tau$), see Eqs.~(\ref{eqn:wilson_cRij}, \ref{eq:GammaLR}). One further parametrises (again, as first approximation) $g_\psi^\ell$ in terms of $g_R^\ell$ as 
\begin{equation}
\left| g_{\psi}^{\alpha}\right| \, 
= r_{\alpha} \, \left| g_{R}^{\alpha} \right|\,,
\end{equation}
in which the ratios $r_{\alpha}$ correspond to arbitrary normal distributions\footnote{Although the distributions are indeed arbitrary, we have chosen the following values in order to maximise the number of (final) outputs: $\log_{10}(r_{e}) = \mathcal{N}( 4.0, \ 0.7)$, $\log_{10}(r_{\mu}) = \mathcal{N}( 0.0, \ 0.3)$ and $\log_{10}(r_{\tau}) = \mathcal{N}( 3.0, \ 0.7)$.}. (The argument of $r_\alpha$ is also randomly chosen.) 

One subsequently generates values for the cLFV rates ($\text{BR}(\mu \to e \gamma)$, $\text{BR}(\tau \to \mu \gamma)$) following a skewed random law with a maximum close to the experimental limit at $95\%$ C.L., as well as a value for $\Delta a_\mu$; for the latter we take a random distribution around both benchmark cases (i.e. a $4.2 \sigma$ tension or a fair $1.5 \sigma$ agreement with the SM). Relying on the already taken values of the spectrum and interactions of the extended scalar and fermion sectors, this allows to infer a first solution for $g_R^\ell$ (and hence to  $g_\psi^\ell$).
Finally, and relying on the yet unconstrained complex angles of the $R$ matrix, one can now reconstruct $\mathcal{G}$: while $\theta_3$ is taken as real and randomly generated, $\theta_{1,2}$ are determined by iteratively solving (a subset of) the equations for the first line of $\mathcal{G}$. The thus reconstructed matrix of couplings is then used as input for the computation of the above mentioned observables, and the procedure repeated until stable entries for $\mathcal{G}$ are encountered (i.e. for which the output rates of the observables are in good agreement with the input values). Moreover, it is required that all couplings obey the perturbativity limit, $ |g | < \sqrt{4\pi}$. 

\begin{figure}[h!]
    \centering
    \includegraphics[width=0.45 \textwidth] {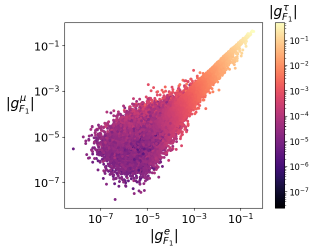}
    \hspace*{10mm}
    \includegraphics[width=0.45 \textwidth] {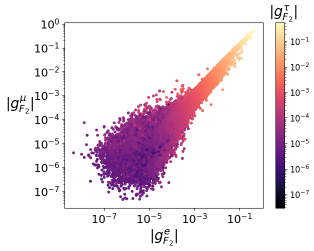}
    \\
    \includegraphics[width=0.45 \textwidth]{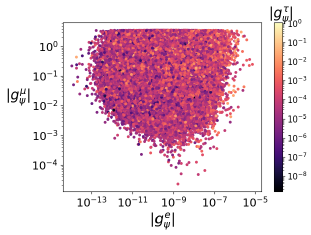}
    \hspace*{10mm}
    \includegraphics[width=0.45 \textwidth]{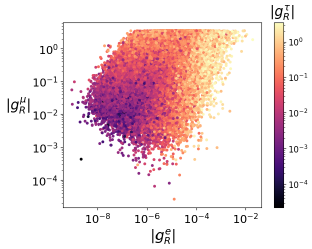}
    \caption{Distributions of the absolute values of the components of the Yukawa-like couplings $g_{F_1}^\alpha$ (upper left), $g_{F_2}^\alpha$ (upper right), $g_{\psi}^\alpha$ (lower left), $g_{R}^\alpha$ (lower right) obtained from a random scan over the extended scalar and fermion sectors. The plots put forward the hierarchy between the different flavours/components enforced by the neutrino mass fit, constraints from $\mu \to e \gamma$ and $\tau \to \mu \gamma$ as well as the different accommodations of $(g-2)_\mu$.}\label{fig:Gmatrix}
\end{figure}

For completeness, we display in Fig.~\ref{fig:Gmatrix} our ranges for the (absolute) values of the different couplings entering $\mathcal{G}$, in fair agreement (although not identical) to those obtained in the study of~\cite{Alvarez:2023dzz}.

\section{Form Factors}\label{sec:FF}
We present here the derived expressions of the form factors relevant for the calculation of several observables of interest, including cLFV transitions and decays (including cLFV $Z$ and Higgs decays), and electroweak observables, in particular invisible and lepton flavour conserving $Z$ and Higgs decays. In order to derive the amplitudes of interest, we made extensive use of FeynRules~\cite{Alloul:2013bka} for the implementation of Feynman rules, FeynArts~\cite{Hahn:2000kx}, as well as FeynCalc~\cite{Shtabovenko:2023idz,Shtabovenko:2020gxv,Shtabovenko:2016sxi,Mertig:1990an} for the Dirac algebra, and finally LoopTools~\cite{Hahn:1998yk} for the numerical evaluation of the Passarino-Veltman functions.  

\subsection{Leptonic cLFV transitions and decays}
In what concerns the cLFV observables whose rates were presented in Section~\ref{sec:cLFV:lepton}, here we provide the detailed expressions for the form factors relevant for the computation of the three-body decays and of the muon-electron conversion.
Working in the limit of negligibly small lepton masses (i.e. $m_{\ell_\alpha}^2/M_{Z}^2 \ll 1$), one can approximate the anapole contributions as
\begin{align}\label{eq:K1}
    K_1^L & = \frac{1}{576 \pi^2} \left( \sum_i \left[ \frac{ \left( \Gamma_L^{\beta i} \right)^* \Gamma_L^{\alpha i}}{M_{\chi_{_i}}^2} f_{1,\chi} \left( \frac{M_{\eta^\pm}^2}{M_{\chi_{_i}}^2} \right)\right] + \sum_k \left[ \frac{ \left( \Gamma_L^{\beta k} \right)^* \Gamma_L^{\alpha k}}{M_{\phi_{_k}}^2} f_{1,\phi} \left( \frac{M_\psi^2}{M_{\phi_{_k}}^2} \right)\right] \right), \nonumber\\
    K_1^R & = K_1^L \quad (L \rightarrow R)\,,
\end{align}
where $i = 1, 4$ and $k = 1, 3$. The associated loop functions are given by
\begin{align}\label{eq:K1_loop}
       f_{1,\chi}(x) &= \frac{(x-1)(2x^2-7x+11)-6\ln(x)}{
       (x-1)^{4}}\, , \quad \forall x \neq 1\, ;
       \quad 
        f_{1,\chi}(1) = \frac{3}{2} \,,\nonumber \\
       f_{1,\phi}(x) &= \frac{(1-x)(7x^2-29x+16)-6(3x-2)\ln(x)}{
       (x-1)^{4}}\, , \quad 
       \forall x \neq 1\, ;
       \quad 
        f_{1,\phi}(1) = \frac{9}{2}\,.
\end{align}
We recall that the dipole form factors (i.e. $K^L_2 = 2 c_R/m_{\ell_{_\alpha}}$) were already given in Eq.~(\ref{eqn:wilson_cRij}). 
The $A$ form factors encode all contributions associated with 4-lepton interactions; for simplicity, we perform the following decomposition
\begin{align}
    A_{LL}^S &= B_{LL}^S, \nonumber \\
    A_{LR}^S &= B_{LR}^S, \nonumber \\
    A_{LL}^V &= B_{LL}^V + i  
   R_{SL}^\beta \,\frac{F_L}{M_Z^2}, \nonumber \\
    A_{LR}^V &= B_{LR}^V + i R_{SL}^\beta\,\frac{F_R}{M_Z^2}, \nonumber \\
    A_{LL}^T &= B_{LL}^T, \nonumber \\
    A^{S,V,T}_{RR} &= A^{S,V,T}_{LL}(L \leftrightarrow R), \nonumber \\
    A^{S,V,T}_{RL} &= A^{S,V,T}_{LR}(L \leftrightarrow R),
\end{align}
with $R_{SL}^\alpha = -ie(s_w^2 - c_w^2)/(2 c_w s_w)$ and $R_{SR}^\alpha = -ies_w/c_w$ the tree-level lepton couplings to the $Z$; in the above, $B$ corresponds to the box contributions and $F$ to those of the $Z$-penguin mediated diagrams\footnote{Due to their clearly subdominant role, we do not include here the expressions for the scalar (Higgs) penguins; we nevertheless included them in all the numerical studies.}. 
The box form factors are given by
\begin{align}\label{eq:Boxes}
    B^S_{LL} &= \frac{1}{32 \pi^2} \left( \sum_{ij} \left[ M_{\chi_{_i}} M_{\chi_{_j}} \left(\Gamma^{\beta j}_R\right)^* \Gamma^{\alpha i}_L \left( \left(\Gamma^{\beta j}_R\right)^* \Gamma^{\beta i}_L - 2 \left( \Gamma^{\beta i}_R\right)^* \Gamma^{\beta j}_L \right) \operatorname{D_0} \left( M_{\eta^\pm}^2, M_{\eta^\pm}^2, M_{\chi_{_i}}^2, M_{\chi_{_j}}^2 \right)\right] \right. \nonumber\\
    &- \left. 2 \sum_{kl} \left[ M_\psi^2 \left(\Gamma^{\beta l}_R\right)^* \Gamma^{\alpha k}_L \left( \Gamma^{\beta l}_R\right)^* \Gamma^{\beta k}_L \operatorname{D_0} \left( M_\psi^2, M_\psi^2, M_{\phi_{_l}}^2, M_{\phi_{_k}}^2 \right)\right]\right),
    \\
    B^S_{LR} &= \frac{1}{16 \pi^2} \left( \sum_{ij} \left[\left(\Gamma^{\beta j}_L\right)^* \Gamma^{\alpha i}_L \left(2 \left(\Gamma^{\beta j}_R\right)^* \operatorname{D_{0 0}}(M_{\eta^\pm}^2,M_{\eta^\pm}^2,M_{\chi_{_i}}^2,M_{\chi_{_j}}^2) \Gamma^{\beta i}_R \right. \right. \right. \nonumber\\
    &- \left. \left. \left.\left(\Gamma^{\beta i}_R\right)^* \operatorname{D_0}(M_{\eta^\pm}^2,M_{\chi_{_j}}^2,M_{\eta^\pm}^2,M_{\chi_{_i}}^2) M_{\chi_{_i}} M_{\chi_{_j}} \Gamma^{\beta j}_R\right)\right] \right. \nonumber\\
    &- \left. \sum_{kl} \left[M_\psi^2 |\Gamma^{\beta l}_R|^2 \left(\Gamma^{\beta k}_L\right)^* \operatorname{D_0}(M_{\phi_{_l}}^2,M_\psi^2,M_{\phi_{_k}}^2,M_\psi^2) \Gamma^{\alpha k}_L\right]\right),
    \\
    B^V_{LL} &= \frac{1}{32 \pi^2} \left( \sum_{ij} \left[\Gamma^{\alpha i}_L \left(-2 |\Gamma^{\beta j}_L|^2 \left(\Gamma^{\beta i}_L\right)^* \operatorname{D_{0 0}}(M_{\eta^\pm}^2,M_{\eta^\pm}^2,M_{\chi_{_i}}^2,M_{\chi_{_j}}^2) \right. \right. \right. \nonumber \\
    &+ \left. \left. \left. \left(\Gamma^{\beta j}_L\right)^* \left(\Gamma^{\beta j}_L\right)^* \operatorname{D_0}(M_{\chi_{_i}}^2,M_{\eta^\pm}^2,M_{\chi_{_j}}^2,M_{\eta^\pm}^2) M_{\chi_{_i}} M_{\chi_{_j}} \Gamma^{\beta i}_L\right)\right] \right. \nonumber\\
    &- 2 \left. \sum_{kl} \left[|\Gamma^{\beta l}_L|^2 \left(\Gamma^{\beta k}_L\right)^* \operatorname{D_{0 0}}(M_\psi^2,M_\psi^2,M_{\phi_{_l}}^2,M_{\phi_{_k}}^2) \Gamma^{\alpha k}_L\right]\right),
    \\
    B^V_{LR} &= -\frac{1}{16 \pi^2} \left( \sum_{ij} \left[\left(\Gamma^{\beta j}_R\right)^* \operatorname{D_{0 0}}(M_{\eta^\pm}^2,M_{\eta^\pm}^2,M_{\chi_{_i}}^2,M_{\chi_{_j}}^2) \Gamma^{\alpha i}_L \left(\left(\Gamma^{\beta j}_L\right)^* \Gamma^{\beta i}_R+\left(\Gamma^{\beta i}_L\right)^* \Gamma^{\beta j}_R\right)\right] \right. \nonumber\\
    &+ \left. \sum_{kl} \left[\left(\Gamma^{\beta l}_L\right)^* \left(\Gamma^{\beta k}_R\right)^* \operatorname{D_{0 0}}(M_\psi^2,M_\psi^2,M_{\phi_{_l}}^2,M_{\phi_{_k}}^2) \Gamma^{\alpha k}_L \Gamma^{\beta l}_R\right]\right),
    \\
    B^T_{LL} &= -\frac{1}{128 \pi^2} \sum_{ij} \left[ \left(\Gamma^{\beta j}_R\right)^* \left(\Gamma^{\beta j}_R\right)^* \operatorname{D_0}(M_{\eta^\pm}^2,M_{\eta^\pm}^2,M_{\chi_{_i}}^2,M_{\chi_{_j}}^2) M_{\chi_{_i}} M_{\chi_{_j}} \Gamma^{\alpha i}_L \Gamma^{\beta i}_L\right],
    \\
    B_{RR}^{S,V,T} &= B^{S,V,T}_{LL}(L \leftrightarrow R),
    \\
    B_{RL}^{S,V,T} &= B^{S,V,T}_{LR}(L \leftrightarrow R).
\end{align}
The $Z$ penguin form factors can be written as
\begin{align}
    F_L &= \frac{e}{64 \pi^2 c_w s_w} \left( \sum_i \left[ (s_w^2-c_w^2) \Gamma_L^{\alpha i} \left(\Gamma_L^{\beta i}\right)^* \left(\operatorname{B_0}(M_{\eta^\pm}^2,M_{\chi_{_i}}^2)-4 \operatorname{C_{0 0}}(M_{\eta^\pm}^2,M_{\eta^\pm}^2,M_{\chi_{_i}}^2) \right. \right. \right. \nonumber\\
    &+ \left. \left. \left. \left(M_{\chi_{_i}}^2-M_{\eta^\pm}^2\right) \operatorname{B_0^\prime}(M_{\eta^\pm}^2,M_{\chi_{_i}}^2)\right) \right. \right. \nonumber \\
    &+ \left. \left. \sum_j \left[4 c_w s_w \Gamma_L^{\alpha i} \left(\Gamma_L^{\beta j}\right)^* \left(\Gamma_{Z,L}^{ij} \left(\operatorname{B_0}(M_{\eta^\pm}^2,M_{\chi_{_j}}^2)-2 \operatorname{C_{0 0}}(M_{\eta^\pm}^2,M_{\chi_{_i}}^2,M_{\chi_{_j}}^2)\right)\right. \right. \right. \right. \nonumber\\
    &+ \left. \left. \left. \left.M_{\chi_{_i}} \operatorname{C_0}(M_{\chi_{_j}}^2,M_{\chi_{_i}}^2,M_{\eta^\pm}^2) \left(M_{\chi_{_j}} \left(\Gamma_{Z,L}^{ij}\right)^*+M_{\chi_{_i}} \Gamma_{Z,L}^{ij}\right)\right)\right]\right] \right. \nonumber \\
    &+ \left. \sum_k \left[ 2 \left(\Gamma_L^{\beta k}\right)^* \left(\Gamma_L^{\alpha k} \left(c_w^2 \operatorname{B_0}(M_\psi^2,M_{\phi_{_k}}^2)+2 \left(s_w^2-c_w^2\right) \operatorname{C_{0 0}}(M_\psi^2,M_\psi^2,M_{\phi_{_k}}^2) \right. \right. \right. \right. \nonumber\\
    &+ \left. \left. \left. \left. s_w^2 \left(M_\psi^2-M_{\phi_{_k}}^2\right) \operatorname{B_0^\prime}(M_\psi^2,M_{\phi_{_k}}^2)\right)+2 i U_\phi^{2 k} \Gamma_L^{\alpha 3} \operatorname{C_{0 0}}(M_{A^0}^2,M_\psi^2,M_{\phi_{_k}}^2)\right) \right. \right. \nonumber\\
    &- \left. \left. 4 i U_\phi^{2 k} \left(\Gamma_L^{\beta 3}\right)^* \Gamma_L^{\alpha k} \operatorname{C_{0 0}}(M_{A^0}^2,M_\psi^2,M_{\phi_{_k}}^2) \right] \right),\\
    F_R &= - \left( F_L \right)^*\,.
\end{align}
We have also introduced for convenience the following shortened notation for the Passarino-Veltman functions:
$\operatorname{C_i}(m_1^2,m_2^2,m_3^2) \equiv \operatorname{C_i}(0,0,0,m_1^2,m_2^2,m_3^2)$,  $\operatorname{B_0}(m_1^2,m_2^2) \equiv \operatorname{B_0}(0,m_1^2,m_2^2)$ and $\operatorname{B_0^\prime}(m_1^2,m_2^2) \equiv \operatorname{B_0^\prime}(0,m_1^2,m_2^2)$, and 
$\operatorname{D_i} (m_1^2, m_2^2, m_3^2, m_4^2) \equiv \operatorname{D_i} (0,0,0,0,0,0,m_1^2, m_2^2, m_3^2, m_4^2)$.

\subsection{cLFV $\pmb{Z}$ decays: $\pmb{Z \rightarrow \ell^\pm_{\alpha}\ell^\mp_{\beta}}$}\label{sec:ZcLdec}
We recall that the relevant diagrams were presented in Fig.~\ref{fig:Zlldiag}; in order to have the appendix self-consistent, we again provide the expressions for the generalised coupling matrices, $\Gamma^{\alpha}_{L,R}$, first defined in Eq.~(\ref{eq:GammaLR}) (see Section~\ref{sec:observable}),
\begin{align}
    \Gamma_L^{\alpha k}  &=  - g_\psi^\alpha\, U_\phi^{1 k}\,, \nonumber \\
    \Gamma_R^{\alpha k} & = \frac{\left(g_R^\alpha \right)^*}{\sqrt{2}} 
 \left( U_\phi^{2 k} + i U_\phi^{3 k} \right)\,, \nonumber \\ 
 \Gamma_L^{\alpha i} & =  g_{F_1}^\alpha \left(U_\chi^{1 i} \right)^* + g_{F_2}^\alpha \left(U_\chi^{2 i} \right)^* \,, \nonumber \\
    \Gamma_R^{\alpha i} & = \left(g_R^\alpha \right)^* 
 U_\chi^{3 i}\,. \nonumber
\end{align}
To render the expressions more compact and easy to grasp, we further introduce the following quantities (which encode the $Z$ couplings to the NP neutral scalars and neutral fermions):
\begin{align}\label{eq:ZVert}
    \Gamma_{Z,L}^{i j} & = U_\chi^{3 i}\, ( U_\chi^{3 d })^* - U_\chi^{4 i} \,( U_\chi^{4 d })^* \,,\\
    \Gamma_{Z,R}^{i j} & = - ( \Gamma_{Z,L}^{i j})^* \,.
\end{align}
The relevant form factors entering in the computation of the widths (and branching ratios) - see Section~\ref{sec:cLFV:Z-H} - are given below.
\begin{align}\label{eq:ZllFF1}
    F_{SL}^{2 \chi, \alpha\beta} & = \frac{i e}{32 \pi^2 c_w s_w} \sum_{i j} \left(\Gamma^{i j}_{Z, L} \left(\Gamma^{\beta i}_L \left(\Gamma^{\alpha j}_L\right)^* \left(\operatorname{B_0}(m_{\ell_\beta}^2,M_{\eta^\pm}^2,M_{\chi_{_i}}^2)+M_{\chi_{_j}}^2 \operatorname{C_0}(M_Z^2,m_{\ell_\alpha}^2,m_{\ell_\beta}^2,M_{\chi_{_i}}^2,M_{\chi_{_j}}^2,M_{\eta^\pm}^2) \right. \right. \right. \nonumber\\
    &- \left. \left. \left. 2 \operatorname{C_{0 0}}(M_Z^2,m_{\ell_\alpha}^2,m_{\ell_\beta}^2,M_{\chi_{_i}}^2,M_{\chi_{_j}}^2,M_{\eta^\pm}^2)+m_{\ell_\alpha}^2 \operatorname{C_1}(m_{\ell_\alpha}^2,m_{\ell_\beta}^2,M_Z^2,M_{\chi_{_j}}^2,M_{\eta^\pm}^2,M_{\chi_{_i}}^2)\right. \right. \right. \nonumber\\
    &+ \left. \left. \left.M_Z^2 \operatorname{C_1}(M_Z^2,m_{\ell_\beta}^2,m_{\ell_\alpha}^2,M_{\chi_{_j}}^2,M_{\chi_{_i}}^2,M_{\eta^\pm}^2)\right)+\Gamma^{\beta i}_L \left(\Gamma^{\alpha j}_R\right)^* m_{\ell_\alpha} M_{\chi_{_j}} \operatorname{C_1}(M_Z^2,m_{\ell_\alpha}^2,m_{\ell_\beta}^2,M_{\chi_{_i}}^2,M_{\chi_{_j}}^2,M_{\eta^\pm}^2)\right. \right. \nonumber\\
    &+ \left. \left.\Gamma^{\beta i}_R \left(\Gamma^{\alpha j}_L\right)^* m_{\ell_\beta} M_{\chi_{_i}} \operatorname{C_1}(M_Z^2,m_{\ell_\beta}^2,m_{\ell_\alpha}^2,M_{\chi_{_j}}^2,M_{\chi_{_i}}^2,M_{\eta^\pm}^2)\right)\right. \nonumber\\
    &- \left.\left(\Gamma^{i j}_{Z, L}\right)^* \left(\Gamma^{\beta i}_L \left(\Gamma^{\alpha j}_R\right)^* m_{\ell_\alpha} M_{\chi_{_i}} \left(\operatorname{C_1}(M_Z^2,m_{\ell_\beta}^2,m_{\ell_\alpha}^2,M_{\chi_{_j}}^2,M_{\chi_{_i}}^2,M_{\eta^\pm}^2)\right. \right. \right. \nonumber\\
    &+ \left. \left. \left.\operatorname{C_1}(m_{\ell_\beta}^2,m_{\ell_\alpha}^2,M_Z^2,M_{\chi_{_j}}^2,M_{\eta^\pm}^2,M_{\chi_{_i}}^2)\right)-\Gamma^{\beta i}_R \left(\Gamma^{\alpha j}_L\right)^* m_{\ell_\beta} M_{\chi_{_j}} \left(\operatorname{C_0}(M_Z^2,m_{\ell_\alpha}^2,m_{\ell_\beta}^2,M_{\chi_{_i}}^2,M_{\chi_{_j}}^2,M_{\eta^\pm}^2)\right. \right. \right. \nonumber\\
    &+ \left. \left. \left.\operatorname{C_1}(M_Z^2,m_{\ell_\alpha}^2,m_{\ell_\beta}^2,M_{\chi_{_j}}^2,M_{\chi_{_i}}^2,M_{\eta^\pm}^2)\right)+\Gamma^{\beta i}_R \left(\Gamma^{\alpha j}_R\right)^* m_{\ell_\beta} m_{\ell_\alpha} \operatorname{C_1}(m_{\ell_\alpha}^2,m_{\ell_\beta}^2,M_Z^2,M_{\chi_{_j}}^2,M_{\eta^\pm}^2,M_{\chi_{_i}}^2)\right. \right. \nonumber\\
    &- \left. \left.\Gamma^{\beta i}_L \left(\Gamma^{\alpha j}_L\right)^* M_{\chi_{_i}} M_{\chi_{_j}} \operatorname{C_0}(M_Z^2,m_{\ell_\alpha}^2,m_{\ell_\beta}^2,M_{\chi_{_i}}^2,M_{\chi_{_j}}^2,M_{\eta^\pm}^2)\right)\right)\,,
\\
    F_{SL}^{2 S, \alpha\beta} & = - \sum_k \frac{e U_\phi^{2 k}}{16 \pi^2 c_w s_w} \Gamma^{\beta 3}_L \left(\Gamma^{\alpha k}_L\right)^* \left(\operatorname{C_{0 0}}(M_Z^2,m_{\ell_\alpha}^2,m_{\ell_\beta}^2,M_{\phi_{_k}}^2,M_{A^0}^2,M_\psi^2) \right. \nonumber\\
    &+ \left. \operatorname{C_{0 0}}(M_Z^2,m_{\ell_\alpha}^2,m_{\ell_\beta}^2,M_{A^0}^2,M_{\phi_{_k}}^2,M_\psi^2)\right)\,,
\\
    F_{SL}^{1 \chi, \alpha\beta} & = \frac{i e \left( c_w^2 - s_w^2 \right)}{16 \pi^2 c_w s_w} \sum_i \Gamma^{\beta i}_L \left(\Gamma^{\alpha i}_L\right)^* \operatorname{C_{0 0}}(M_Z^2,m_{\ell_\alpha}^2,m_{\ell_\beta}^2,M_{\eta^\pm}^2,M_{\eta^\pm}^2,M_{\chi_{_i}}^2)\,,
\\
    F_{SL}^{1 S, \alpha\beta} & = \frac{i e \left( c_w^2 - s_w^2 \right)}{32 \pi^2 c_w s_w} \sum_k \left(\Gamma^{\beta k}_L \left(\Gamma^{\alpha k}_L\right)^* \left(\operatorname{B_0}(m_{\ell_\beta}^2,M_\psi^2,M_{\phi_{_k}}^2)-2 \operatorname{C_{0 0}}(M_Z^2,m_{\ell_\alpha}^2,m_{\ell_\beta}^2,M_\psi^2,M_\psi^2,M_{\phi_{_k}}^2)\right. \right. \nonumber\\
    &+ \left. \left. M_Z^2 \operatorname{C_1}(M_Z^2,m_{\ell_\beta}^2,m_{\ell_\alpha}^2,M_\psi^2,M_\psi^2,M_{\phi_{_k}}^2)+m_{\ell_\alpha}^2 \operatorname{C_1}(m_{\ell_\alpha}^2,m_{\ell_\beta}^2,M_Z^2,M_\psi^2,M_{\phi_{_k}}^2,M_\psi^2)\right)\right. \nonumber\\
    &- \left.\Gamma^{\beta k}_L \left(\Gamma^{\alpha k}_R\right)^* M_\psi m_{\ell_\alpha} \operatorname{C_0}(M_Z^2,m_{\ell_\alpha}^2,m_{\ell_\beta}^2,M_\psi^2,M_\psi^2,M_{\phi_{_k}}^2)\right. \nonumber\\
    &- \left.\Gamma^{\beta k}_R \left(\Gamma^{\alpha k}_L\right)^* M_\psi m_{\ell_\beta} \operatorname{C_0}(M_Z^2,m_{\ell_\alpha}^2,m_{\ell_\beta}^2,M_\psi^2,M_\psi^2,M_{\phi_{_k}}^2) \right. \nonumber\\
    &+ \left.\Gamma^{\beta k}_R \left(\Gamma^{\alpha k}_R\right)^* m_{\ell_\alpha} m_{\ell_\beta} \operatorname{C_1}(m_{\ell_\alpha}^2,m_{\ell_\beta}^2,M_Z^2,M_\psi^2,M_{\phi_{_k}}^2,M_\psi^2)\right)\,,
\\
F_{SR} & = F_{SL} (L \leftrightarrow R)\,.
\end{align}
The $L/R$ form factors are given by:
\begin{align}\label{eq:ZllFF2}
    F_{L}^{2 \chi, \alpha\beta} & = \frac{-i e}{16 \pi^2c_w s_w} \sum_{i j} \left(\Gamma^{i j}_{Z, L} \left(\Gamma^{\beta i}_L \left(\Gamma^{\alpha j}_R\right)^* M_{\chi_{_j}} \operatorname{C_1}(M_Z^2,m_{\ell_\beta}^2,m_{\ell_\alpha}^2,M_{\chi_{_i}}^2,M_{\chi_{_j}}^2,M_{\eta^\pm}^2)\right. \right. \nonumber\\
    &- \left. \left.\Gamma^{\beta i}_L \left(\Gamma^{\alpha j}_L\right)^* m_{\ell_\alpha} \operatorname{C_{1 2}}(M_Z^2,m_{\ell_\alpha}^2,m_{\ell_\beta}^2,M_{\chi_{_i}}^2,M_{\chi_{_j}}^2,M_{\eta^\pm}^2)\right)\right. \nonumber\\
    &- \left.\left(\Gamma^{i j}_{Z, L}\right)^* \left(\Gamma^{\beta i}_L \left(\Gamma^{\alpha j}_R\right)^* M_{\chi_{_i}} \operatorname{C_1}(M_Z^2,m_{\ell_\beta}^2,m_{\ell_\alpha}^2,M_{\chi_{_j}}^2,M_{\chi_{_i}}^2,M_{\eta^\pm}^2) \right. \right. \nonumber\\
    &- \left. \left. \Gamma^{\beta i}_R \left(\Gamma^{\alpha j}_R\right)^* m_{\ell_\beta} \operatorname{C_{1 2}}(M_Z^2,m_{\ell_\beta}^2,m_{\ell_\alpha}^2,M_{\chi_{_j}}^2,M_{\chi_{_i}}^2,M_{\eta^\pm}^2)\right)\right)\,,
\\
    F_{L}^{2 S, \alpha\beta} & = \sum_k \frac{e U_\phi^{2 k}}{16 \pi^2 c_w s_w} \left(\Gamma^{\beta 3}_R \left(\Gamma^{\alpha k}_R\right)^* m_{\ell_\beta} \left(\operatorname{C_{1 2}}(M_Z^2,m_{\ell_\beta}^2,m_{\ell_\alpha}^2,M_{\phi_{_k}}^2,M_{A^0}^2,M_\psi^2) \right. \right. \nonumber\\
    &+ \left. \left. \operatorname{C_{1 2}}(M_Z^2,m_{\ell_\beta}^2,m_{\ell_\alpha}^2,M_{A^0}^2,M_{\phi_{_k}}^2,M_\psi^2)\right) \right. \nonumber\\
    &+ \left. \Gamma^{\beta k}_L \left(\Gamma^{\alpha 3}_R\right)^* M_\psi \operatorname{C_1}(m_{\ell_\alpha}^2,m_{\ell_\beta}^2,M_Z^2,M_{A^0}^2,M_\psi^2,M_{\phi_{_k}}^2)\right)\,,
\\
    F_{L}^{1 \chi, \alpha\beta} & = -\frac{i e \left( c_w^2 - s_w^2 \right)}{16 \pi^2 c_w s_w} \sum_i \left(\Gamma^{\beta i}_L \left(\Gamma^{\alpha i}_L\right)^* m_{\ell_\alpha} \operatorname{C_{1 2}}(M_Z^2,m_{\ell_\alpha}^2,m_{\ell_\beta}^2,M_{\eta^\pm}^2,M_{\eta^\pm}^2,M_{\chi_{_i}}^2) \right. \nonumber\\
    &+ \left. \Gamma^{\beta i}_R \left(\Gamma^{\alpha i}_R\right)^* m_{\ell_\beta} \operatorname{C_{1 2}}(M_Z^2,m_{\ell_\beta}^2,m_{\ell_\alpha}^2,M_{\eta^\pm}^2,M_{\eta^\pm}^2,M_{\chi_{_i}}^2) \right. \nonumber\\
    &- \left. \Gamma^{\beta i}_L \left(\Gamma^{\alpha i}_R\right)^* M_{\chi_{_i}} \operatorname{C_1}(m_{\ell_\alpha}^2,m_{\ell_\beta}^2,M_Z^2,M_{\eta^\pm}^2,M_{\chi_{_i}}^2,M_{\eta^\pm}^2)\right)\,,
\\
    F_{L}^{1 S, \alpha\beta} & = \frac{i e \left( c_w^2 - s_w^2 \right)}{16 \pi^2 c_w s_w} \sum_k \left(\Gamma^{\beta k}_L \left(\Gamma^{\alpha k}_R\right)^* M_\psi \left(\operatorname{C_0}(M_Z^2,m_{\ell_\alpha}^2,m_{\ell_\beta}^2,M_\psi^2,M_\psi^2,M_{\phi_{_k}}^2) \right. \right. \nonumber\\
    &+ \left. \left. \operatorname{C_1}(m_{\ell_\alpha}^2,m_{\ell_\beta}^2,M_Z^2,M_\psi^2,M_{\phi_{_k}}^2,M_\psi^2)\right)-\Gamma^{\beta k}_L \left(\Gamma^{\alpha k}_L\right)^* m_{\ell_\alpha} \operatorname{C_{1 2}}(M_Z^2,m_{\ell_\alpha}^2,m_{\ell_\beta}^2,M_\psi^2,M_\psi^2,M_{\phi_{_k}}^2) \right. \nonumber\\
    &+ \left. \Gamma^{\beta k}_R \left(\Gamma^{\alpha k}_R\right)^* m_{\ell_\beta} \operatorname{C_{1 2}}(M_Z^2,m_{\ell_\beta}^2,m_{\ell_\alpha}^2,M_\psi^2,M_\psi^2,M_{\phi_{_k}}^2)\right)\,,
\\
F_{R} & = - {F_{L}}^* (\alpha \leftrightarrow \beta)\,.
\end{align}

\subsection{$\pmb{Z}$ decays into massive neutrinos}\label{sec:ZInvdec}
In the physical basis, the Lagrangian contains the following term: $\mathcal{L} \supset \overline{\nu_i}(\Gamma_L^{i c k} P_L + \Gamma_R^{i c k} P_R)\phi_k \chi_c$. Notice that 
due to the Majorana nature of the fermions one has $\Gamma_L^{i c k} = {\Gamma_R^{i c k}}^* = \Gamma_{c k}^{i}$. 
The interaction vertex can be cast as follows
\begin{equation}\label{eq:HNLVert}
   \Gamma_{b k}^{i} =  \sum_{\alpha=1}^3 {U_\text{PMNS}^{\alpha i}}^* \left( {U_\chi^{4 b}}^* g_\psi^\alpha U_\phi^{1 k} + \frac{\left( {U_\chi^{1 b}}^* g_{F_1}^\alpha + {U_\chi^{2 b}}^* g_{F_2}^\alpha \right) \left( U_\phi^{2 k} - i U_\phi^{3 k} \right)}{\sqrt{2}}\right)\,.
\end{equation}
This leads to the following form factors:
\begin{align}\label{eq:ZInvFFSL}
    F_{SL}^{2 \chi, i j} & = \sum_{a b k} \frac{i e}{32 \pi^2 c_w s_w} \Gamma^{a b}_{Z, L} \left(-\Gamma^j_{a k} \left(\Gamma^i_{b k}\right)^* \left(\operatorname{B_0}(m_{\nu_{_i}}^2,M_{\phi_{_k}}^2,M_{\chi_{_b}}^2)+M_{\chi_{_a}}^2 \operatorname{C_0}(M_Z^2,m_{\nu_{_j}}^2,m_{\nu_{_i}}^2,M_{\chi_{_b}}^2,M_{\chi_{_a}}^2,M_{\phi_{_k}}^2) \right. \right. \nonumber\\
    &- \left. \left. 2 \operatorname{C_{0 0}}(M_Z^2,m_{\nu_{_j}}^2,m_{\nu_{_i}}^2,M_{\chi_{_b}}^2,M_{\chi_{_a}}^2,M_{\phi_{_k}}^2)+M_Z^2 \operatorname{C_1}(M_Z^2,m_{\nu_{_i}}^2,m_{\nu_{_j}}^2,M_{\chi_{_a}}^2,M_{\chi_{_b}}^2,M_{\phi_{_k}}^2) \right. \right. \nonumber\\
    &+ \left. \left. m_{\nu_{_j}}^2 \operatorname{C_1}(m_{\nu_{_j}}^2,m_{\nu_{_i}}^2,M_Z^2,M_{\chi_{_a}}^2,M_{\phi_{_k}}^2,M_{\chi_{_b}}^2)\right)-\Gamma^j_{b k} \left(\Gamma^i_{a k}\right)^* M_{\chi_{_a}} M_{\chi_{_b}} \operatorname{C_0}(M_Z^2,m_{\nu_{_j}}^2,m_{\nu_{_i}}^2,M_{\chi_{_a}}^2,M_{\chi_{_b}}^2,M_{\phi_{_k}}^2) \right. \nonumber\\
    &+ \left. \Gamma^i_{a k} \left(\Gamma^j_{b k}\right)^* m_{\nu_{_j}} m_{\nu_{_i}} \operatorname{C_1}(m_{\nu_{_j}}^2,m_{\nu_{_i}}^2,M_Z^2,M_{\chi_{_b}}^2,M_{\phi_{_k}}^2,M_{\chi_{_a}}^2) \right. \nonumber\\
    &+ \left. \left(\Gamma^i_{a k}\right)^* \left(\Gamma^j_{b k}\right)^* m_{\nu_{_j}} M_{\chi_{_a}} \left(\operatorname{C_0}(M_Z^2,m_{\nu_{_j}}^2,m_{\nu_{_i}}^2,M_{\chi_{_b}}^2,M_{\chi_{_a}}^2,M_{\phi_{_k}}^2) + \operatorname{C_1}(M_Z^2,m_{\nu_{_i}}^2,m_{\nu_{_j}}^2,M_{\chi_{_a}}^2,M_{\chi_{_b}}^2,M_{\phi_{_k}}^2) \right. \right. \nonumber\\
    &+ \left. \left. \operatorname{C_1}(M_Z^2,m_{\nu_{_i}}^2,m_{\nu_{_j}}^2,M_{\chi_{_b}}^2,M_{\chi_{_a}}^2,M_{\phi_{_k}}^2)+\operatorname{C_1}(m_{\nu_{_j}}^2,m_{\nu_{_i}}^2,M_Z^2,M_{\chi_{_a}}^2,M_{\phi_{_k}}^2,M_{\chi_{_b}}^2) \right. \right. \nonumber\\
    &+ \left. \left. \operatorname{C_1}(m_{\nu_{_j}}^2,m_{\nu_{_i}}^2,M_Z^2,M_{\chi_{_b}}^2,M_{\phi_{_k}}^2,M_{\chi_{_a}}^2)\right)-\Gamma^i_{a k} \Gamma^j_{b k} m_{\nu_{_i}} M_{\chi_{_b}} \left(\operatorname{C_0}(M_Z^2,m_{\nu_{_j}}^2,m_{\nu_{_i}}^2,M_{\chi_{_a}}^2,M_{\chi_{_b}}^2,M_{\phi_{_k}}^2) \right. \right. \nonumber\\
    &+ \left. \left. \operatorname{C_1}(M_Z^2,m_{\nu_{_i}}^2,m_{\nu_{_j}}^2,M_{\chi_{_a}}^2,M_{\chi_{_b}}^2,M_{\phi_{_k}}^2)+\operatorname{C_1}(M_Z^2,m_{\nu_{_i}}^2,m_{\nu_{_j}}^2,M_{\chi_{_b}}^2,M_{\chi_{_a}}^2,M_{\phi_{_k}}^2)\right)\right) \, ,
    \\
    F_{SL}^{1 \chi, i j} & = \sum_{a k} \frac{e U_\Phi^{2 k}}{16 \pi^2 c_w s_w} \left(\Gamma^j_{a k} \left(\Gamma^i_{a 3}\right)^* \operatorname{C_{0 0}}(M_Z^2,m_{\nu_{_j}}^2,m_{\nu_{_i}}^2,M_{A^0}^2,M_{\phi_{_k}}^2,M_{\chi_{_a}}^2) \right. \nonumber\\
    &- \left. \Gamma^j_{a 3} \left(\Gamma^i_{a k}\right)^* \operatorname{C_{0 0}}(M_Z^2,m_{\nu_{_j}}^2,m_{\nu_{_i}}^2,M_{\phi_{_k}}^2,M_{A^0}^2,M_{\chi_{_a}}^2)\right) \, ,
    \\
    F_{SR} & = {F_{SL}}^* \, ,
\end{align}
in which $a, b= 1-4$,  and $k=1-3$. Let us also mention that the $1 \chi$ form factor vanishes for $k=3$. 
Regarding the $L/R$ form factors, one has
\begin{align}\label{eq:ZInvFFL}
    F_{L}^{2 \chi, i j} & = \sum_{a b k} \frac{-i e}{16 \pi^2 c_w s_w} \Gamma^{i j}_{Z, L} \left(M_{\chi_{_b}} \Gamma^i_{a k} \Gamma^j_{b k} \left(\operatorname{C_1}(M_Z^2,m_{\nu_{_i}}^2,m_{\nu_{_j}}^2,M_{\chi_{_a}}^2,M_{\chi_{_b}}^2,M_{\phi_{_k}}^2) \right. \right. \nonumber\\
    &- \left. \left. \operatorname{C_1}(M_Z^2,m_{\nu_{_j}}^2,m_{\nu_{_i}}^2,M_{\chi_{_a}}^2,M_{\chi_{_b}}^2,M_{\phi_{_k}}^2)\right)+m_{\nu_{_j}} \Gamma^i_{a k} \left(\Gamma^j_{b k}\right)^* \operatorname{C_{1 2}}(M_Z^2,m_{\nu_{_j}}^2,m_{\nu_{_i}}^2,M_{\chi_{_a}}^2,M_{\chi_{_b}}^2,M_{\phi_{_k}}^2) \right. \nonumber\\
    &- \left. m_{\nu_{_i}} \Gamma^j_{a k} \left(\Gamma^i_{b k}\right)^* \operatorname{C_{1 2}}(M_Z^2,m_{\nu_{_i}}^2,m_{\nu_{_j}}^2,M_{\chi_{_a}}^2,M_{\chi_{_b}}^2,M_{\phi_{_k}}^2)\right) \, ,
    \\
    F_{L}^{1 \chi, i j} & = \sum_{a k} \frac{-i e U_\Phi^{2 k}}{16 \pi^2 c_w s_w} \left(\Gamma^j_{a 3} \left(M_{\chi_{_a}} \Gamma^i_{a 3} \left(\operatorname{C_1}(m_{\nu_{_j}}^2,m_{\nu_{_i}}^2,M_Z^2,M_{\phi_{_k}}^2,M_{\chi_{_a}}^2,M_{A^0}^2) \right. \right. \right. \nonumber\\
    &- \left. \left. \left. \operatorname{C_1}(m_{\nu_{_j}}^2,m_{\nu_{_i}}^2,M_Z^2,M_{A^0}^2,M_{\chi_{_a}}^2,M_{\phi_{_k}}^2)\right)+m_{\nu_{_i}} \left(\Gamma^i_{a k}\right)^* \left(\operatorname{C_{1 2}}(M_Z^2,m_{\nu_{_i}}^2,m_{\nu_{_j}}^2,M_{A^0}^2,M_{\phi_{_k}}^2,M_{\chi_{_a}}^2) \right. \right. \right. \nonumber\\
    &+ \left. \left. \left. \operatorname{C_{1 2}}(M_Z^2,m_{\nu_{_i}}^2,m_{\nu_{_j}}^2,M_{\phi_{_k}}^2,M_{A^0}^2,M_{\chi_{_a}}^2)\right)\right)+m_{\nu_{_j}} \Gamma^i_{a k} \left(\Gamma^j_{a 3}\right)^* \left(\operatorname{C_{1 2}}(M_Z^2,m_{\nu_{_i}}^2,m_{\nu_{_j}}^2,M_{A^0}^2,M_{\phi_{_k}}^2,M_{\chi_{_a}}^2) \right. \right. \nonumber\\
    &+ \left. \left. \operatorname{C_{1 2}}(M_Z^2,m_{\nu_{_i}}^2,m_{\nu_{_j}}^2,M_{\phi_{_k}}^2,M_{A^0}^2,M_{\chi_{_a}}^2)\right)\right) \, ,
    \\
    F_R & = - F_L^* \, .
\end{align}

\subsection{Higgs decays into charged leptons
}\label{sec:HcLdec}
Let us first introduce the quantities $ \Gamma_{H (L,R)}$, which encode the Higgs couplings to the new neutral fields, 
\begin{align}\label{eq:HVert}
    \Gamma_{H,L}^{i j} & = \sum_{a b} \frac{y_{a b}}{\sqrt{2}} \,\left( U_\chi^{a + 2, i}\, U_\chi^{b, j} + U_\chi^{a + 2, i}\, U_\chi^{b, j} \right)\,, \\
    \Gamma_{H,R}^{i j} & = ( \Gamma_{H,L}^{i j})^*\,,\\
    \Gamma_H^{k l} & = - \left( \left[ v \,\lambda_\eta^{\prime \prime} \,
    \left( U_\phi^{2 k} + i U_\phi^{3 k}\right) \left( U_\phi^{2 l} + i U_\phi^{3 l}\right) + 2\, \alpha \,U_\phi^{1 k} U_\phi^{2 l} + \text{c.c.}(k \leftrightarrow l)\right] \right. \nonumber \\
    & + \left. v 
    \left[ \lambda_S \,U_\phi^{1 k} \,U_\phi^{1 l} + \left( \lambda_\eta + \lambda_\eta^\prime \right) \left( U_\phi^{2 k} + i U_\phi^{3 k}\right) \left( U_\phi^{2 l} - i U_\phi^{3 l}\right) + (k \leftrightarrow l)\right]\right)\,.
\end{align}
The form factors are
\begin{align}\label{eq:HllFF}
    F_{L}^{2 \chi, \alpha\beta} & = -\frac{i}{16 \pi^2} \sum_{i j} \left(\Gamma^{i j}_{H, L} \left(\Gamma^{\beta i}_L \left(\Gamma^{\alpha j}_R\right)^* \left(\operatorname{B_0}(m_{\ell_\beta}^2,M_{\eta^\pm}^2,M_{\chi_{_i}}^2)+M_{\chi_{_j}}^2 \operatorname{C_0}(M_H^2,m_{\ell_\alpha}^2,m_{\ell_\beta}^2,M_{\chi_{_i}}^2,M_{\chi_{_j}}^2,M_{\eta^\pm}^2) \right. \right. \right. \nonumber\\
    &+ \left. \left. \left. m_{\ell_\alpha}^2 \operatorname{C_1}(m_{\ell_\alpha}^2,m_{\ell_\beta}^2,M_H^2,M_{\chi_{_j}}^2,M_{\eta^\pm}^2,M_{\chi_{_i}}^2)+M_H^2 \operatorname{C_1}(M_H^2,m_{\ell_\beta}^2,m_{\ell_\alpha}^2,M_{\chi_{_j}}^2,M_{\chi_{_i}}^2,M_{\eta^\pm}^2)\right) \right. \right. \nonumber\\
    &+ \left. \left. \Gamma^{\beta i}_L \left(\Gamma^{\alpha j}_L\right)^* m_{\ell_\alpha} M_{\chi_{_j}} \operatorname{C_1}(M_H^2,m_{\ell_\alpha}^2,m_{\ell_\beta}^2,M_{\chi_{_i}}^2,M_{\chi_{_j}}^2,M_{\eta^\pm}^2) \right. \right. \nonumber\\
    &+ \left. \left. \Gamma^{\beta i}_R \left(\Gamma^{\alpha j}_R\right)^* m_{\ell_\beta} M_{\chi_{_i}} \operatorname{C_1}(M_H^2,m_{\ell_\beta}^2,m_{\ell_\alpha}^2,M_{\chi_{_j}}^2,M_{\chi_{_i}}^2,M_{\eta^\pm}^2)\right) \right. \nonumber\\
    &+ \left. \Gamma^{i j}_{H, R} \left(\Gamma^{\beta i}_L \left(\Gamma^{\alpha j}_R\right)^* M_{\chi_{_i}} M_{\chi_{_j}} \operatorname{C_0}(M_H^2,m_{\ell_\alpha}^2,m_{\ell_\beta}^2,M_{\chi_{_i}}^2,M_{\chi_{_j}}^2,M_{\eta^\pm}^2) \right. \right. \nonumber\\
    &- \left. \left. \Gamma^{\beta i}_R \left(\Gamma^{\alpha j}_L\right)^* m_{\ell_\alpha} m_{\ell_\beta} \operatorname{C_1}(m_{\ell_\alpha}^2,m_{\ell_\beta}^2,M_H^2,M_{\chi_{_j}}^2,M_{\eta^\pm}^2,M_{\chi_{_i}}^2) \right. \right. \nonumber\\
    &- \left. \left. \Gamma^{\beta i}_L \left(\Gamma^{\alpha j}_L\right)^* m_{\ell_\alpha} M_{\chi_{_i}} \left(\operatorname{C_1}(M_H^2,m_{\ell_\beta}^2,m_{\ell_\alpha}^2,M_{\chi_{_j}}^2,M_{\chi_{_i}}^2,M_{\eta^\pm}^2)+\operatorname{C_1}(m_{\ell_\alpha}^2,m_{\ell_\beta}^2,M_H^2,M_{\chi_{_j}}^2,M_{\eta^\pm}^2,M_{\chi_{_i}}^2)\right) \right. \right. \nonumber\\
    &+ \left. \left. \Gamma^{\beta i}_R \left(\Gamma^{\alpha j}_R\right)^* m_{\ell_\beta} M_{\chi_{_j}} \left(\operatorname{C_0}(M_H^2,m_{\ell_\alpha}^2,m_{\ell_\beta}^2,M_{\chi_{_i}}^2,M_{\chi_{_j}}^2,M_{\eta^\pm}^2) \right. \right. \right. \nonumber\\
    &+ \left. \left. \left. \operatorname{C_1}(M_H^2,m_{\ell_\beta}^2,m_{\ell_\alpha}^2,M_{\chi_{_j}}^2,M_{\chi_{_i}}^2,M_{\eta^\pm}^2)\right)\right)\right)\,,
\\
    F_{L}^{1 \chi, \alpha\beta} & = \frac{i v \lambda_\eta}{16 \pi^2} \sum_i  \left(\Gamma^{\beta i}_L \left(\Gamma^{\alpha i}_L\right)^* m_{\ell_\alpha} \operatorname{C_1}(M_H^2,m_{\ell_\alpha}^2,m_{\ell_\beta}^2,M_{\eta^\pm}^2,M_{\eta^\pm}^2,M_{\chi_{_i}}^2) \right. \nonumber\\
    &- \left. \Gamma^{\beta i}_L \left(\Gamma^{\alpha i}_R\right)^* M_{\chi_{_i}} \operatorname{C_0}(M_H^2,m_{\ell_\alpha}^2,m_{\ell_\beta}^2,M_{\eta^\pm}^2,M_{\eta^\pm}^2,M_{\chi_{_i}}^2) \right. \nonumber\\
    &+ \left. \Gamma^{\beta i}_R \left(\Gamma^{\alpha i}_R\right)^* m_{\ell_\beta} \operatorname{C_1}(M_H^2,m_{\ell_\beta}^2,m_{\ell_\alpha}^2,M_{\eta^\pm}^2,M_{\eta^\pm}^2,M_{\chi_{_i}}^2)\right)\,,
\\
    F_{L}^{2 S, \alpha\beta} & = - \frac{i}{32 \pi^2} \sum_{k l} \Gamma^{k l}_H\left(\Gamma^{\beta k}_L \left(\Gamma^{\alpha l}_L\right)^* m_{\ell_\alpha} \operatorname{C_1}(M_H^2,m_{\ell_\alpha}^2,m_{\ell_\beta}^2,M_{\phi_{_k}}^2,M_{\phi_{_l}}^2,M_\psi^2) \right. \nonumber\\
    &- \left. \Gamma^{\beta k}_L \left(\Gamma^{\alpha l}_R\right)^* M_\psi \operatorname{C_0}(M_H^2,m_{\ell_\alpha}^2,m_{\ell_\beta}^2,M_{\phi_{_k}}^2,M_{\phi_{_l}}^2,M_\psi^2) \right. \nonumber\\
    &+ \left. \Gamma^{\beta k}_R \left(\Gamma^{\alpha l}_R\right)^* m_{\ell_\beta} \operatorname{C_1}(M_H^2,m_{\ell_\beta}^2,m_{\ell_\alpha}^2,M_{\phi_{_l}}^2,M_{\phi_{_k}}^2,M_\psi^2)\right)\,,
\\
F_{R} & = F_{L} (L \leftrightarrow R)\,.
\end{align}

\subsection{$\pmb{H}$ decays into neutrinos}\label{sec:HInvdec}
In view of its similarities with $Z \to \nu_i \nu_j$, one can use elements already introduced in Eq.~(\ref{eq:HNLVert}), and readily derive the associated form factors
\begin{align}\label{eq:HInvFF}
    F_{L}^{2 \chi, i j} & = \sum_{a b k} \frac{-i}{16 \pi^2} \left(\left(\Gamma^j_{b k}\right)^* m_{\nu_{_j}} \left(\left(\operatorname{C_1}(M_H^2,m_{\nu_{_i}}^2,m_{\nu_{_j}}^2,M_{\chi_{_b}}^2,M_{\chi_{_a}}^2,M_{\phi_{_k}}^2) \right. \right. \right. \nonumber\\
    & + \left. \left. \left. \operatorname{C_1}(m_{\nu_{_j}}^2,m_{\nu_{_i}}^2,M_H^2,M_{\chi_{_b}}^2,M_{\phi_{_k}}^2,M_{\chi_{_a}}^2)\right) \Gamma^{a b}_{H, L} M_{\chi_{_b}} \Gamma^i_{a k} \right. \right. \nonumber\\
    & + \left. \left. \left(\operatorname{C_1}(M_H^2,m_{\nu_{_i}}^2,m_{\nu_{_j}}^2,M_{\chi_{_b}}^2,M_{\chi_{_a}}^2,M_{\phi_{_k}}^2) M_{\chi_{_a}} \Gamma^i_{a k} \right. \right. \right. \nonumber\\
    & + \left. \left. \left. \operatorname{C_1}(m_{\nu_{_j}}^2,m_{\nu_{_i}}^2,M_H^2,M_{\chi_{_b}}^2,M_{\phi_{_k}}^2,M_{\chi_{_a}}^2) \left(\left(\Gamma^i_{a k}\right)^* m_{\nu_{_i}}+M_{\chi_{_a}} \Gamma^i_{a k}\right)\right) \left(\Gamma^{a b}_{H, L}\right)^*\right) \right. \nonumber\\
    & - \left. \operatorname{C_0}(M_H^2,m_{\nu_{_j}}^2,m_{\nu_{_i}}^2,M_{\chi_{_a}}^2,M_{\chi_{_b}}^2,M_{\phi_{_k}}^2) M_{\chi_{_b}} \left(\Gamma^{a b}_{H, L} \Gamma^i_{a k} \left(-\left(\left(\Gamma^j_{b k}\right)^* m_{\nu_{_j}}\right)+M_{\chi_{_b}} \Gamma^j_{b k}\right)\right. \right. \nonumber\\
    & + \left. \left.\left(\left(\Gamma^i_{a k}\right)^* m_{\nu_{_i}}+M_{\chi_{_a}} \Gamma^i_{a k}\right) \Gamma^j_{b k} \left(\Gamma^{a b}_{H, L}\right)^*\right)-\Gamma^j_{b k} \left(\Gamma^{a b}_{H, L} \left(\operatorname{B_0}(m_{\nu_{_i}}^2,M_{\phi_{_k}}^2,M_{\chi_{_a}}^2) \right. \right. \right. \nonumber\\
    & + \left. \left. \left. M_H^2 \operatorname{C_1}(M_H^2,m_{\nu_{_i}}^2,m_{\nu_{_j}}^2,M_{\chi_{_b}}^2,M_{\chi_{_a}}^2,M_{\phi_{_k}}^2)+\operatorname{C_1}(m_{\nu_{_j}}^2,m_{\nu_{_i}}^2,M_H^2,M_{\chi_{_b}}^2,M_{\phi_{_k}}^2,M_{\chi_{_a}}^2) m_{\nu_{_j}}^2\right) \Gamma^i_{a k} \right. \right. \nonumber\\
    & + \left. \left. \operatorname{C_1}(M_H^2,m_{\nu_{_i}}^2,m_{\nu_{_j}}^2,M_{\chi_{_b}}^2,M_{\chi_{_a}}^2,M_{\phi_{_k}}^2) \left(\Gamma^i_{a k}\right)^* m_{\nu_{_i}} \left(\Gamma^{a b}_{H, L} M_{\chi_{_a}}+M_{\chi_{_b}} \left(\Gamma^{a b}_{H, L}\right)^*\right)\right)\right) \, ,
    \\
    F_{L}^{1 \chi, i j} & = \sum_{a k l} \frac{i}{32 \pi^2} \Gamma^{k l}_H \left(\operatorname{C_1}(M_H^2,m_{\nu_{_i}}^2,m_{\nu_{_j}}^2,M_{\phi_{_l}}^2,M_{\phi_{_k}}^2,M_{\chi_{_a}}^2) \left(\Gamma^i_{a k}\right)^* m_{\nu_{_i}} \Gamma^j_{a l} \right. \nonumber\\
    & -\left. \left(\left(\operatorname{C_1}(M_H^2,m_{\nu_{_i}}^2,m_{\nu_{_j}}^2,M_{\phi_{_l}}^2,M_{\phi_{_k}}^2,M_{\chi_{_a}}^2)+\operatorname{C_1}(m_{\nu_{_j}}^2,m_{\nu_{_i}}^2,M_H^2,M_{\phi_{_l}}^2,M_{\chi_{_a}}^2,M_{\phi_{_k}}^2)\right) \left(\Gamma^j_{a l}\right)^* m_{\nu_{_j}} \right. \right. \nonumber\\
    & + \left. \left. \operatorname{C_0}(M_H^2,m_{\nu_{_j}}^2,m_{\nu_{_i}}^2,M_{\phi_{_k}}^2,M_{\phi_{_l}}^2,M_{\chi_{_a}}^2) \left(\left(\Gamma^j_{a l}\right)^* m_{\nu_{_j}}+M_{\chi_{_a}} \Gamma^j_{a l}\right)\right) \Gamma^i_{a k}\right) \, ,
    \\
    F_{R} & = - {F_{L}}^* \, ,
\end{align}
in which $a,b = 1-4$ and $k,l=1-3$. We notice here that in $F_{L}^{2 \chi, \alpha\beta}$, the coefficient of the $B_0$ function vanishes when summed over $a,b,k$. (Since there are no tree-level $H \bar{\nu^c_i} \nu_i$ interactions in the Lagragian, these processes need not be renormalised.)

\section{Renormalisation procedure and results}\label{app:renormalisation}

In this appendix we describe the renormalisation procedure adopted throughout the study of the ``T1-2-A'' scotogenic variant. Below we also collect all the relevant expressions that were used in the derivation of the observables discussed in the phenomenological analysis.

As usually done, starting from the bare Lagrangian, we follow the expansion rules from~\cite{Denner:1991kt} in order to derive the counter-term Lagrangian (and the associated Feynman rules), which also includes field renormalisation quantities ($\delta_{XY}$ or $\delta^{XY}_{L/R}$). The latter are defined for convenience and allow a compact presentation of the otherwise lengthy analytical expressions (in particular regarding corrections to propagators).
In what follows we thus go over the renormalisation procedure of the neutral leptonic interaction, corrections to the massive lepton's propagators, as well as self-energies for the vector bosons.

\subsection{Renormalisation of the $\pmb{Z}$ and Higgs leptonic interactions}
We begin by considering the interactions with charged leptons ($Z\ell\ell$ and $H\ell\ell$), and subsequently address the case of $Z\nu\nu$ and $H\nu\nu$.
For the former, we adopt the on-shell renormalisation scheme as well as dimensional regularisation; we expand all relevant counter-terms to $m^2$ (or $m$ for fermions), coupling counter-terms (mainly the electric charge), and also field renormalisation quantities.
The quantities that are required to address the leptonic decays of the $Z$ and Higgs bosons discussed in Sections~\ref{sec:cLFV:Z-H} and~\ref{sec:EWPO}, have been introduced in Eqs.~(\ref{eq:ZHTensDec}). 
The counter-terms and other relevant form factors are thus given by the following set of definitions - beginning by those pertinent to flavour conserving charged lepton interactions: 
\begin{align}\label{eq:CTFFcLFC}
    F^{\alpha}_{SL \, CT} &= -\frac{i e}{4 c_w s_w^3} \left( \frac{\delta_{M_Z^2}}{M_Z^2}- \frac{\delta_{M_W^2}}{M_W^2} + 2 s_w^2 \left(s_w^2-c_w^2\right) \delta_e \right),
    \\
    F^{\alpha}_{SR \, CT} &= -\frac{i e}{2 c_w s_w} \left(\frac{\delta_{M_Z^2}}{M_Z^2}-\frac{\delta_{M_W^2}}{M_W^2} + 2 \frac{\delta_e}{s_w^2}\right),
    \\
    F^{\alpha}_{SL \, x \ell} &= \frac{i e}{4 c_w s_w} \left(
    \left(c_w^2-s_w^2\right) \left(\overline{\delta^\alpha_L} + \delta^\alpha_L + \delta_{Z Z} \right) + 2 c_w s_w \delta_{\gamma Z} \right),
    \\
    F^{\alpha}_{SR \, x \ell} &= -\frac{i e s_w}{2 c_w} \left( \overline{\delta^\alpha_R} + \delta^\alpha_R + \delta _{Z Z} - \frac{c_w}{s_w} \delta_{\gamma Z} \right),
    \\
    G^{\alpha}_{L\,CT} &= -\frac{i}{2 s_w^2 v} \left( c_w^2 m_{\ell_\alpha} \left( \frac{\delta_{M_Z^2}}{M_Z^2} - \left( 1 - \frac{s_w^2}{c_w^2}\right) \frac{\delta_{M_W^2}}{M_W^2} \right) + 2 s_w^2 \left(m_{\ell_\alpha} \delta_e - \delta_{m_{\ell_\alpha}} \right) \right),
    \\
    G^{\alpha}_{R\,CT} &= G^{\alpha}_{L\,CT},
    \\
    G^{\alpha}_{L \, x \ell} &= \frac{i m_{\ell_\alpha}}{2 v} \left( \delta_H + \overline{\delta^\alpha_R} + \delta^\alpha_L\right),
    \\
    G^{\alpha}_{R \, x \ell} &= G^{\alpha}_{L \, x \ell} (L \leftrightarrow R),
\end{align}
in which we recall that $\alpha$ denotes the leptonic flavour, and $s_w(c_w)$ the (co)sine of the weak angle. One must further define quantities entering the flavour-violating corrections, 
\begin{align}\label{eq:CTFFcLFV}
    F^{\alpha \beta}_{SR \,x\ell} &=- \frac{i e s_w}{2 c_w} \left(\delta^{\alpha \beta}_R + \overline{\delta^{\alpha \beta}_R } \right),
    \\
    F^{\alpha \beta}_{SL \,x\ell} &= \frac{i e }{4} \left(\frac{c_w}{s_w} - \frac{s_w}{c_w}\right) \left(\delta^{\alpha \beta}_L + \overline{\delta^{\alpha \beta}_L } \right),
    \\
    G^{\alpha \beta}_{L \,x\ell} &= \frac{i e}{4 c_w M_Z s_w}  \left( m_{\ell_\alpha} \delta^{\alpha \beta}_L + m_{\ell_\beta} \overline{\delta^{\alpha \beta}_R} \right),
    \\
    G^{\alpha \beta}_{R \,x\ell} &= G^{\alpha \beta}_{L \,x\ell}(L \leftrightarrow R)\,.
\end{align}
For $Z\nu_i\nu_i$ interactions, the relevant quantities can be cast as 
\begin{align}\label{CTFFInvdiag}
    F_{SL \,CT}^{i} & = \frac{i e}{4 c_w s_w} \left( \left( \frac{c_w^2}{s_w^2} - 1 \right) \left( \frac{\delta_{M_W^2}}{M_W^2} - \frac{\delta_{M_Z^2}}{M_Z^2}\right) + 2 \delta_e \right),\\
    F_{SR \,CT}^{i} & = \left( F_{SL \,CT}^{i} \right)^*,\\
    F_{SL \, x \ell}^{i} & = \frac{i e}{4 c_w s_w} \left( \overline{\delta^{i}_L} +  \delta^{i}_L  + \delta_{ZZ}\right),\\
    F_{SR \, x \ell}^{i} & = \left( F_{SL \, x \ell}^{i} \right)^*\,,
\end{align}
while for $Z\nu_i\nu_j$ one finds
\begin{align}\label{CTFFNInvoffdiag}
    F_{SL \, x \ell}^{i j} & = \frac{i e}{4 c_w s_w} \left( \overline{\delta^{i j}_L} +  \delta^{i j}_L\right),\\
    F_{SR \, x \ell}^{i j} & = \left( F_{SL \, x \ell}^{i j} \right)^*\,.
\end{align}
In the above, we have introduced the following renormalisation quantities: for the Higgs, $\delta_H$, and for the $Z$, $\gamma$ and $W$ boson (as well as associated counter-terms), $\delta_{M_{Z,W}^2}$, $\delta_{ZZ}$,  $\delta_{\gamma\gamma}$, $\delta_{\gamma Z}$ and $\delta_{Z \gamma}$. Likewise, the corrections to the charged lepton and neutrino propagators are respectively denoted $\delta_{L,R}^{\alpha (\beta)}$ and $\delta_{L}^{i(j)}$. The counter-terms and renormalised quantities (derived from the on-shell renormalisation equations) will be defined in the subsequent subsections.
Let us notice that the Ward identity allows to express the electric charge counter-term with respect to field renormalisation quantities~\cite{Denner:1991kt} as $\delta_e = -1/2( \delta_{\gamma \gamma} - s_w/c_w \delta_{Z \gamma} )$. 

\subsection{Correction to the lepton propagators}
\begin{figure}[h!]
    \centering
    \begin{subfigure}[b]{0.3\textwidth}
        \centering
        \begin{tikzpicture}
        \begin{feynman}
        \vertex (a) at (0,0) {\(\ell_\alpha\)};
        \vertex (b) at (1,0);
        \vertex (c) at (2,0);
        \vertex (d) at (3,0) {\(\ell_\beta\)};
        \diagram* {
        (a) -- [fermion] (b),
        (b) -- [plain, half right, edge label'=\( \chi_{_i}\)] (c),
        (b) -- [scalar, half left, edge label=\( \eta^\pm\)] (c),
        (c) -- [fermion] (d)
        };
        \end{feynman}
        \end{tikzpicture}
    \end{subfigure}
    \hspace*{10mm}
    \begin{subfigure}[b]{0.3\textwidth}
        \centering
        \begin{tikzpicture}
        \begin{feynman}
        \vertex (a) at (0,0) {\(\ell_\alpha\)};
        \vertex (b) at (1,0);
        \vertex (c) at (2,0);
        \vertex (d) at (3,0) {\(\ell_\beta\)};
        \diagram* {
        (a) -- [fermion] (b),
        (b) -- [fermion, half right, edge label'=\( \psi^\pm\)] (c),
        (b) -- [scalar, half left, edge label=\( \phi_{_k}\)] (c),
        (c) -- [fermion] (d)
        };
        \end{feynman}
        \end{tikzpicture}
    \end{subfigure}
    \caption{Corrections to the charged lepton propagators.}
    \label{fig:FSE}
\end{figure}
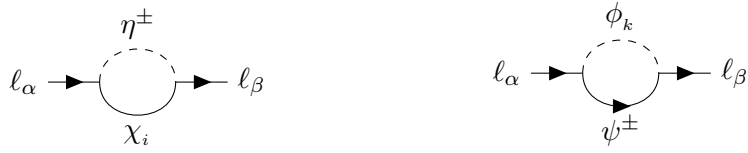

We first consider the case of charged leptons, and then that of the massive neutrinos.
\paragraph{Charged leptons}
The diagrams for the correction to the charged lepton propagators are depicted in Fig.~\ref{fig:FSE}.

The expressions for the off-diagonal fermion counter-terms
can be written as 
\begin{align}\label{eq:loffdiagCT}
    \delta_L^{\beta \alpha} & = \frac{1}{16 \pi^2 (m_{\ell_\alpha}^2 - m_{\ell_\beta}^2)} \left( \sum_i \left[\Gamma^{\alpha i}_L \left(\Gamma^{\beta i}_L\right)^* \left(\left(-M_{\chi_{_i}}^2-m_{\ell_\alpha}^2+M_{\eta^\pm}^2\right) \operatorname{B_0}(m_{\ell_\alpha}^2,M_{\eta^\pm}^2,M_{\chi_{_i}}^2)+\operatorname{A_0}(M_{\chi_{_i}}^2)\right. \right. \right. \nonumber\\
    & - \left. \left. \left.\operatorname{A_0}(M_{\eta^\pm}^2)\right)-\frac{m_{\ell_\beta}}{m_{\ell_\alpha}} \Gamma^{\alpha i}_R \left(\Gamma^{\beta i}_R\right)^* \left(\left(M_{\chi_{_i}}^2+m_{\ell_\alpha}^2-M_{\eta^\pm}^2\right) \operatorname{B_0}(m_{\ell_\alpha}^2,M_{\eta^\pm}^2,M_{\chi_{_i}}^2)-\operatorname{A_0}(M_{\chi_{_i}}^2)\right. \right. \right. \nonumber\\
    & + \left. \left. \left.\operatorname{A_0}(M_{\eta^\pm}^2)\right)-\Gamma^{\alpha i}_R \left(\Gamma^{\beta i}_L\right)^* 2 m_{\ell_\alpha} M_{\chi_{_i}} \operatorname{B_0}(m_{\ell_\alpha}^2,M_{\eta^\pm}^2,M_{\chi_{_i}}^2)-\Gamma^{\alpha i}_L \left(\Gamma^{\beta i}_R\right)^* 2 m_{\ell_\beta} M_{\chi_{_i}} \operatorname{B_0}(m_{\ell_\alpha}^2,M_{\eta^\pm}^2,M_{\chi_{_i}}^2)\right] \right. \nonumber\\
    & + \left. \sum_k \left[\Gamma^{\alpha k}_L \left(\Gamma^{\beta k}_L\right)^* \left(-\left(m_{\ell_\alpha}^2-M_{\phi_{_ k}}^2+M_\psi^2\right) \operatorname{B_0}(m_{\ell_\alpha}^2,M_\psi^2,M_{\phi_{_ k}}^2)-\operatorname{A_0}(M_{\phi_{_ k}}^2)+\operatorname{A_0}(M_\psi^2)\right)\right. \right. \nonumber\\
    & + \left. \left.\frac{m_{\ell_\beta}}{m_{\ell_\alpha}} \Gamma^{\alpha k}_R \left(\Gamma^{\beta k}_R\right)^* \left(-\left(m_{\ell_\alpha}^2-M_{\phi_{_ k}}^2+M_\psi^2\right) \operatorname{B_0}(m_{\ell_\alpha}^2,M_\psi^2,M_{\phi_{_ k}}^2)-\operatorname{A_0}(M_{\phi_{_ k}}^2)+\operatorname{A_0}(M_\psi^2)\right)\right. \right. \nonumber\\
    & - \left. \left.\Gamma^{\alpha k}_L \left(\Gamma^{\beta k}_R\right)^* 2 m_{\ell_\beta} M_\psi \operatorname{B_0}(m_{\ell_\alpha}^2,M_\psi^2,M_{\phi_{_ k}}^2)-\Gamma^{\alpha k}_R \left(\Gamma^{\beta k}_L\right)^* 2 m_{\ell_\alpha} M_\psi \operatorname{B_0}(m_{\ell_\alpha}^2,M_\psi^2,M_{\phi_{_ k}}^2)\right]\right)\,,
    \\
    \delta_R^{\beta \alpha} & = \frac{1}{16 \pi^2 (m_{\ell_\alpha}^2 - m_{\ell_\beta}^2)} \left( \sum_i \left[\Gamma^{\alpha i}_R \left(\Gamma^{\beta i}_R\right)^* \left(\left(-M_{\chi_{_i}}^2-m_{\ell_\alpha}^2+M_{\eta^\pm}^2\right) \operatorname{B_0}(m_{\ell_\alpha}^2,M_{\eta^\pm}^2,M_{\chi_{_i}}^2)+\operatorname{A_0}(M_{\chi_{_i}}^2)\right. \right. \right. \nonumber\\
    & - \left. \left. \left.\operatorname{A_0}(M_{\eta^\pm}^2)\right)-\frac{m_{\ell_\beta}}{m_{\ell_\alpha}} \Gamma^{\alpha i}_L \left(\Gamma^{\beta i}_L\right)^* \left(\left(M_{\chi_{_i}}^2+m_{\ell_\alpha}^2-M_{\eta^\pm}^2\right) \operatorname{B_0}(m_{\ell_\alpha}^2,M_{\eta^\pm}^2,M_{\chi_{_i}}^2)-\operatorname{A_0}(M_{\chi_{_i}}^2)\right. \right. \right. \nonumber\\
    & + \left. \left. \left.\operatorname{A_0}(M_{\eta^\pm}^2)\right)-\Gamma^{\alpha i}_L \left(\Gamma^{\beta i}_R\right)^* 2 m_{\ell_\alpha} M_{\chi_{_i}} \operatorname{B_0}(m_{\ell_\alpha}^2,M_{\eta^\pm}^2,M_{\chi_{_i}}^2)-\Gamma^{\alpha i}_R \left(\Gamma^{\beta i}_L\right)^* 2 m_{\ell_\beta} M_{\chi_{_i}} \operatorname{B_0}(m_{\ell_\alpha}^2,M_{\eta^\pm}^2,M_{\chi_{_i}}^2)\right] \right. \nonumber\\
    & + \left. \sum_k \left[\Gamma^{\alpha k}_R \left(\Gamma^{\beta k}_R\right)^* \left(-\left(m_{\ell_\alpha}^2-M_{\phi_{_ k}}^2+M_\psi^2\right) \operatorname{B_0}(m_{\ell_\alpha}^2,M_\psi^2,M_{\phi_{_ k}}^2)-\operatorname{A_0}(M_{\phi_{_ k}}^2)+\operatorname{A_0}(M_\psi^2)\right)\right. \right. \nonumber\\
    & + \left. \left.\frac{m_{\ell_\beta}}{m_{\ell_\alpha}} \Gamma^{\alpha k}_L \left(\Gamma^{\beta k}_L\right)^* \left(-\left(m_{\ell_\alpha}^2-M_{\phi_{_ k}}^2+M_\psi^2\right) \operatorname{B_0}(m_{\ell_\alpha}^2,M_\psi^2,M_{\phi_{_ k}}^2)-\operatorname{A_0}(M_{\phi_{_ k}}^2)+\operatorname{A_0}(M_\psi^2)\right)\right. \right. \nonumber\\
    & - \left. \left.\Gamma^{\alpha k}_R \left(\Gamma^{\beta k}_L\right)^* 2 m_{\ell_\beta} M_\psi \operatorname{B_0}(m_{\ell_\alpha}^2,M_\psi^2,M_{\phi_{_ k}}^2)-\Gamma^{\alpha k}_L \left(\Gamma^{\beta k}_R\right)^* 2 m_{\ell_\alpha} M_\psi \operatorname{B_0}(m_{\ell_\alpha}^2,M_\psi^2,M_{\phi_{_ k}}^2)\right]\right)\,;
\end{align}
for the self energies one has
\begin{align}\label{eq:ldiagCT}
    \delta_m^\alpha &= \frac{1}{64 \pi^2 m_{\ell_\alpha}} \left( \sum_i \left[\left(|\Gamma^{\alpha i}_R|^2+|\Gamma^{\alpha i}_L|^2\right) \left(\left(M_{\chi_{_i}}^2+m_{\ell_\alpha}^2-M_{\eta^\pm}^2\right) \operatorname{B_0}(m_{\ell_\alpha}^2,M_{\eta^\pm}^2,M_{\chi_{_i}}^2)\right. \right. \right. \nonumber \\
    &- \left. \left. \left.\operatorname{A_0}(M_{\chi_{_i}}^2)+\operatorname{A_0}(M_{\eta^\pm}^2)\right)+4 m_{\ell_\alpha} M_{\chi_{_i}} \text{Re}\left(\Gamma^{\alpha i}_L \left(\Gamma^{\alpha i}_R\right)^*\right) \operatorname{B_0}(m_{\ell_\alpha}^2,M_{\eta^\pm}^2,M_{\chi_{_i}}^2)\right] \right. \nonumber \\
    &+ \left. \sum_k \left[\left(|\Gamma^{\alpha k}_R|^2+|\Gamma^{\alpha k}_L|^2\right) \left(\left(M_\psi^2+m_{\ell_\alpha}^2-M_{\phi_{_ k}}^2\right) \operatorname{B_0}(m_{\ell_\alpha}^2,M_\psi^2,M_{\phi_{_ k}}^2)+\operatorname{A_0}(M_{\phi_{_ k}}^2)\right. \right. \right. \nonumber \\
    &- \left. \left. \left.\operatorname{A_0}(M_\psi^2)\right)+4 m_{\ell_\alpha} M_\psi \text{Re}\left(\Gamma^{\alpha k}_L \left(\Gamma^{\alpha k}_R\right)^*\right) \operatorname{B_0}(m_{\ell_\alpha}^2,M_\psi^2,M_{\phi_{_ k}}^2)\right]\right)\,,
\\
    \delta_L^\alpha &= \frac{1}{32 \pi^2 m_{\ell_\alpha}^2} \left( \sum_i \left[|\Gamma^{\alpha i}_R|^2 \left(-m_{\ell_\alpha}^2 \left(M_{\chi_{_i}}^2+m_{\ell_\alpha}^2-M_{\eta^\pm}^2\right) \operatorname{B^\prime_0}(m_{\ell_\alpha}^2,M_{\eta^\pm}^2,M_{\chi_{_i}}^2)\right. \right. \right. \nonumber\\
    &+ \left. \left. \left.\left(M_{\chi_{_i}}^2-M_{\eta^\pm}^2\right) \operatorname{B_0}(m_{\ell_\alpha}^2,M_{\eta^\pm}^2,M_{\chi_{_i}}^2)-\operatorname{A_0}(M_{\chi_{_i}}^2)+\operatorname{A_0}(M_{\eta^\pm}^2)\right)\right. \right. \nonumber\\
    &- \left. \left.|\Gamma^{\alpha i}_L|^2 \left(m_{\ell_\alpha}^2 \left(M_{\chi_{_i}}^2+m_{\ell_\alpha}^2-M_{\eta^\pm}^2\right) \operatorname{B^\prime_0}(m_{\ell_\alpha}^2,M_{\eta^\pm}^2,M_{\chi_{_i}}^2)+m_{\ell_\alpha}^2 \operatorname{B_0}(m_{\ell_\alpha}^2,M_{\eta^\pm}^2,M_{\chi_{_i}}^2)\right)\right. \right. \nonumber\\
    &- \left. \left. 4 m_{\ell_\alpha}^3 M_{\chi_{_i}} \text{Re}\left(\Gamma^{\alpha i}_L \left(\Gamma^{\alpha i}_R\right)^*\right) \operatorname{B^\prime_0}(m_{\ell_\alpha}^2,M_{\eta^\pm}^2,M_{\chi_{_i}}^2)+2 i M_{\chi_{_i}} m_{\ell_\alpha} \text{Im}\left(\Gamma^{\alpha i}_L \left(\Gamma^{\alpha i}_R\right)^*\right) \operatorname{B_0}(m_{\ell_\alpha}^2,M_{\eta^\pm}^2,M_{\chi_{_i}}^2)\right] \right. \nonumber\\
    & +  \left. \sum_k \left[|\Gamma^{\alpha k}_R|^2 \left(-m_{\ell_\alpha}^2 \left(M_\psi^2+m_{\ell_\alpha}^2-M_{\phi_{_ k}}^2\right) \operatorname{B^\prime_0}(m_{\ell_\alpha}^2,M_\psi^2,M_{\phi_{_ k}}^2)+\left(M_\psi^2-M_{\phi_{_ k}}^2\right) \operatorname{B_0}(m_{\ell_\alpha}^2,M_\psi^2,M_{\phi_{_ k}}^2)\right. \right. \right. \nonumber\\
    &- \left. \left. \left.\operatorname{A_0}(M_\psi^2)+\operatorname{A_0}(M_{\phi_{_ k}}^2)\right)-|\Gamma^{\alpha k}_L|^2 \left(m_{\ell_\alpha}^2 \left(M_\psi^2+m_{\ell_\alpha}^2-M_{\phi_{_ k}}^2\right) \operatorname{B^\prime_0}(m_{\ell_\alpha}^2,M_\psi^2,M_{\phi_{_ k}}^2)\right. \right. \right. \nonumber\\
    &+ \left. \left. \left.m_{\ell_\alpha}^2 \operatorname{B_0}(m_{\ell_\alpha}^2,M_\psi^2,M_{\phi_{_ k}}^2)\right)-4 m_{\ell_\alpha}^3 M_\psi \text{Re}\left(\Gamma^{\alpha k}_L \left(\Gamma^{\alpha k}_R\right)^*\right) \operatorname{B^\prime_0}(m_{\ell_\alpha}^2,M_\psi^2,M_{\phi_{_ k}}^2)\right. \right. \nonumber\\
    &+ \left. \left.2 i M_\psi m_{\ell_\alpha} \text{Im}\left(\Gamma^{\alpha k}_L \left(\Gamma^{\alpha k}_R\right)^*\right) \operatorname{B_0}(m_{\ell_\alpha}^2,M_\psi^2,M_{\phi_{_ k}}^2)\right] \right)\,,
\\
    \delta_R^\alpha &= \delta_L^\alpha (L \leftrightarrow R)\,,
\end{align}
in which $\operatorname{B^\prime_0}$ corresponds to the first derivative of the Passarino-Veltman function $\operatorname{B_0}$ with respect to its first argument. Let us notice that in the current study the $B_0$ function is always real\footnote{This is not always the case when the masses of the particles in the loop are comparable to the external momentum (squared); for completeness we notice that the $A_0$ function is always real.} as the NP particles have a mass significantly larger than the charged lepton
masses. In order to deal with complex couplings which can appear in the expression for the correction to the charged lepton propagator\footnote{This is due to the fact that both the $\mathcal{G}$ matrix and the $g_R$ couplings are complex.}, we follow \cite{Espriu:2002pz, Drauksas:2021xwz} (noticing that in order to do so one must adhere to their conventions and ensure that the mass matrix for the charged leptons is diagonal in the interaction basis). We also mention that for the diagonal terms~\cite{Espriu:2002pz} used a different convention leading to an overall negative sign for the definitions. Finally we note that the renormalisation quantities of the anti-fermionic fields are defined as $\overline{\delta^{\beta \alpha}_{L/R}} \equiv\left(\delta^{\alpha \beta}_{L/R} \right)^*$.

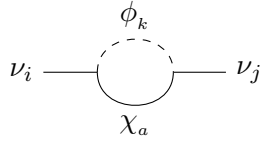
\begin{figure}[h!]
    \centering
    \begin{subfigure}[b]{0.96\textwidth}
        \centering
        \begin{tikzpicture}
        \begin{feynman}
        \vertex (a) at (0,0) {\(\nu_i\)};
        \vertex (b) at (1,0);
        \vertex (c) at (2,0);
        \vertex (d) at (3,0) {\(\nu_j\)};
        \diagram* {
        (a) -- [plain] (b),
        (b) -- [plain, half right, edge label'=\( \chi_{_a}\)] (c),
        (b) -- [scalar, half left, edge label=\( \phi_{_k}\)] (c),
        (c) -- [plain] (d)
        };
        \end{feynman}
        \end{tikzpicture}
    \end{subfigure}
    \caption{Contributions to the correction of  the neutrino propagator.}
    \label{fig:NSE}
\end{figure}

\paragraph{Correction to the neutrino propagator}
As depicted in Fig.~\ref{fig:NSE}, a single diagram is at the origin of the corrections to the neutrino propagator.

\noindent For the off-diagonal contributions one finds:
\begin{align}\label{eq:NoffdiagCT}
    \delta^{j i}_L & = -\frac{1}{16 \pi^2 m_{\nu_{_i}}(m_{\nu_{_i}}^2 - m_{\nu_{_j}}^2)} \sum_{ka}\left(\operatorname{B_0}(m_{\nu_{_i}}^2,M_{\phi_{_ k}}^2,M_{\chi_{_a}}^2) \left(\left(\Gamma^i_{a k}\right)^* \left(2 m_{\nu_{_i}}^2 M_{\chi_{_a}} \left(\Gamma^j_{a k}\right)^* \right. \right. \right. \nonumber \\
    & + \left. \left. \left. m_{\nu_{_j}} \Gamma^j_{a k} \left(M_{\chi_{_a}}^2+m_{\nu_{_i}}^2-M_{\phi_{_ k}}^2\right)\right)+m_{\nu_{_i}} \Gamma^i_{a k} \left(\left(\Gamma^j_{ak}\right)^* \left(M_{\chi_{_a}}^2+m_{\nu_{_i}}^2-M_{\phi_{_ k}}^2\right)+2 m_{\nu_{_j}} M_{\chi_{_a}} \Gamma^j_{a k}\right)\right)\right. \nonumber \\
    & + \left. \left(m_{\nu_{_i}} \Gamma^i_{a k} \left(\Gamma^j_{a k}\right)^*+m_{\nu_{_j}} \Gamma^j_{a k} \left(\Gamma^i_{a k}\right)^*\right) \left( \operatorname{A_0} \left(M_{\phi_{_ k}}^2\right)-\operatorname{A_0} \left(M_{\chi_{_a}}^2\right)\right)\right)\,,
\\
    \delta^{j i}_R & = \left( \delta^{j i}_L \right)^*  \,,
\end{align}
and for the diagonal contributions
\begin{align}\label{eq:NdiagCT}
    \delta^{i}_L & = -\frac{1}{32 \pi^2 m_{\nu_{_i}}^2} \sum_{ka}\left(\operatorname{B_0}(m_{\nu_{_i}}^2,M_{\phi_{_ k}}^2,M_{\chi_{_a}}^2) \left(\left(M_{\chi_{_a}} \left(\Gamma^i_{a k}\right)^*+m_{\nu_{_i}} \Gamma^i_{a k}\right) \left(m_{\nu_{_i}} \left(\Gamma^j_{a k}\right)^*-M_{\chi_{_a}} \Gamma^j_{a k}\right) \right. \right. \nonumber \\
    & + \left. \left. \Gamma^j_{a k} M_{\phi_{_ k}}^2 \left(\Gamma^i_{a k}\right)^*\right)+2 m_{\nu_{_i}}^2 \operatorname{B_0^\prime}(m_{\nu_{_i}}^2,M_{\phi_{_ k}}^2,M_{\chi_{_a}}^2) \left(Re\left(\Gamma^i_{a k} \left(\Gamma^j_{a k}\right)^*\right) \left(M_{\chi_{_a}}^2+m_{\nu_{_i}}^2-M_{\phi_{_ k}}^2\right) \right. \right. \nonumber \\
    & + \left. \left. 2 m_{\nu_{_i}} M_{\chi_{_a}} \text{Re}\left(\Gamma^i_{a k} \Gamma^j_{a k}\right)\right)+\Gamma^j_{a k} \left(\Gamma^i_{ak}\right)^* \left(\operatorname{A_0}(M_{\chi_{_a}}^2)-\operatorname{A_0}(M_{\phi_{_k}}^2)\right)\right)\,,
\\
    \delta^{i}_R & = \left( \delta^{i}_L \right)^* \,.
\end{align}

\subsection{$\pmb{Z}$-boson self-energy corrections}
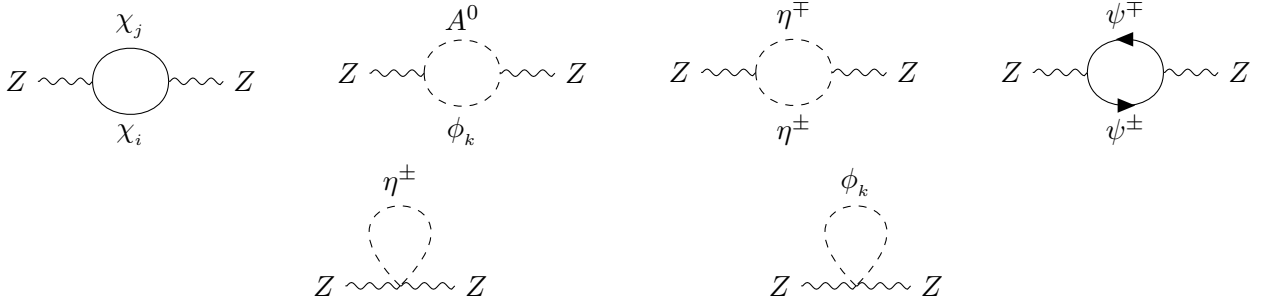
\begin{figure}[h!]
    \centering
    \begin{subfigure}[b]{0.23\textwidth}
        \centering
        \begin{tikzpicture}
        \begin{feynman}
        \vertex (a) at (0,0) {\(Z\)};
        \vertex (b) at (1,0);
        \vertex (c) at (2,0);
        \vertex (d) at (3,0) {\(Z\)};
        \diagram* {
        (a) -- [boson] (b),
        (b) -- [plain, half right, edge label'=\( \chi_{_i}\)] (c),
        (b) -- [plain, half left, edge label=\( \chi_{_j}\)] (c),
        (c) -- [boson] (d)
        };
        \end{feynman}
        \end{tikzpicture}
    \end{subfigure}
    \hfill
    \begin{subfigure}[b]{0.23\textwidth}
        \centering
        \begin{tikzpicture}
        \begin{feynman}
        \vertex (a) at (0,0) {\(Z\)};
        \vertex (b) at (1,0);
        \vertex (c) at (2,0);
        \vertex (d) at (3,0) {\(Z\)};
        \diagram* {
        (a) -- [boson] (b),
        (b) -- [scalar, half right, edge label'=\( \phi_{_k}\)] (c),
        (b) -- [scalar, half left, edge label=\( A^0\)] (c),
        (c) -- [boson] (d)
        };
        \end{feynman}
        \end{tikzpicture}
    \end{subfigure}
    \hfill
    \begin{subfigure}[b]{0.23\textwidth}
        \centering
        \begin{tikzpicture}
        \begin{feynman}
        \vertex (a) at (0,0) {\(Z\)};
        \vertex (b) at (1,0);
        \vertex (c) at (2,0);
        \vertex (d) at (3,0) {\(Z\)};
        \diagram* {
        (a) -- [boson] (b),
        (b) -- [scalar, half right, edge label'=\( \eta^\pm\)] (c),
        (b) -- [scalar, half left, edge label=\( \eta^\mp\)] (c),
        (c) -- [boson] (d)
        };
        \end{feynman}
        \end{tikzpicture}
    \end{subfigure}
    \hfill
    \begin{subfigure}[b]{0.23\textwidth}
        \centering
        \begin{tikzpicture}
        \begin{feynman}
        \vertex (a) at (0,0) {\(Z\)};
        \vertex (b) at (1,0);
        \vertex (c) at (2,0);
        \vertex (d) at (3,0) {\(Z\)};
        \diagram* {
        (a) -- [boson] (b),
        (b) -- [fermion, half right, edge label'=\( \psi^\pm\)] (c),
        (c) -- [fermion, half right, edge label'=\( \psi^\mp\)] (b),
        (c) -- [boson] (d)
        };
        \end{feynman}
        \end{tikzpicture}
    \end{subfigure}
    \\
    \begin{subfigure}[b]{0.3\textwidth}
        \centering
        \begin{tikzpicture}
        \begin{feynman}
        \vertex (a) at (0,0) {\(Z\)};
        \vertex (b) at (1,0);
        \vertex (c) at (2,0) {\(Z\)};
        \diagram* {
        (a) -- [boson] (b),
        b -- [scalar, loop, min distance=2cm, in=45, out=135, edge label=\( \eta^\pm\)] b,
        (b) -- [boson] (c)
        };
        \end{feynman}
        \end{tikzpicture}
    \end{subfigure}
\quad \quad
    \begin{subfigure}[b]{0.3\textwidth}
        \centering
        \begin{tikzpicture}
        \begin{feynman}
        \vertex (a) at (0,0) {\(Z\)};
        \vertex (b) at (1,0);
        \vertex (c) at (2,0) {\(Z\)};
        \diagram* {
        (a) -- [boson] (b),
        b -- [scalar, loop, min distance=2cm, in=45, out=135, edge label=\( \phi_{_k}\)] b,
        (b) -- [boson] (c)
        };
        \end{feynman}
        \end{tikzpicture}
    \end{subfigure}
    \caption{Feynman diagrams of the NP contributions for the $Z$-boson self-energy.}
    \label{fig:ZZSE}
\end{figure}

The diagrams responsible for the corrections to the 
$Z$-boson self-energy are given in Fig.~\ref{fig:ZZSE}.

\noindent One can thus derive the counter-terms, 
\begin{align}\label{eq:ZZCT}
    \delta_{M_Z^2} & = \frac{e^2}{576 M_Z^2 c_w^2 s_w^2 \pi^2} \left( \sum_{ij} \left[ |\Gamma^{ij}_{Z, L}|^2 \left(-3 M_Z^2 M_{\chi_{_i}}^2 \operatorname{B_0}(M_Z^2,M_{\chi_{_i}}^2,M_{\chi_{_j}}^2) \right. \right. \right. \nonumber \\
    &- \left. \left. \left. 3 M_{\chi_{_i}}^4 \operatorname{B_0}(M_Z^2,M_{\chi_{_i}}^2,M_{\chi_{_j}}^2) - 3 M_{\chi_{_j}}^4 \operatorname{B_0}(M_Z^2,M_{\chi_{_i}}^2,M_{\chi_{_j}}^2)+3 M_{\chi_{_j}}^2 \left(2 M_Z^2-\left(M_Z^2 \right. \right. \right. \right. \right. \nonumber \\
    &- \left. \left. \left. \left. \left. 2 M_{\chi_{_i}}^2\right) \operatorname{B_0}(M_Z^2,M_{\chi_{_i}}^2,M_{\chi_{_j}}^2)\right) + 6 M_Z^4 \operatorname{B_0}(M_Z^2,M_{\chi_{_i}}^2,M_{\chi_{_j}}^2) - 3 \left(-M_{\chi_{_i}}^2+M_{\chi_{_j}}^2+2 M_Z^2\right) \operatorname{A_0}(M_{\chi_{_i}}^2) \right. \right. \right. \nonumber \\
    &- \left. \left. \left. 3 \left(M_{\chi_{_i}}^2-M_{\chi_{_j}}^2+2 M_Z^2\right) \operatorname{A_0}(M_{\chi_{_j}}^2) + 6 M_Z^2 M_{\chi_{_i}}^2-2 M_Z^4\right) \right. \right. \nonumber \\
    &- \left. \left. 18 \text{Re}\left(\left(\Gamma^{ij}_{Z, L}\right)^2\right) M_Z^2 M_{\chi_{_i}} M_{\chi_{_j}} \operatorname{B_0}(M_Z^2,M_{\chi_{_i}}^2,M_{\chi_{_j}}^2) \right] \right. \nonumber \\
    &+ \left. \sum_k \left[ \left(U_\phi^{2k}\right)^2 \left(3 M_{\phi_{_k}}^4 \operatorname{B_0}(M_Z^2,M_{A^0}^2,M_{\phi_{_ k}}^2)+3 \left(M_{A^0}^2-M_Z^2\right)^2 \operatorname{B_0}(M_Z^2,M_{A^0}^2,M_{\phi_{_ k}}^2) \right. \right. \right. \nonumber \\
    &- \left. \left. \left. 6 M_{\phi_{_ k}}^2 \left(\left(M_{A^0}^2+M_Z^2\right) \operatorname{B_0}(M_Z^2,M_{A^0}^2,M_{\phi_{_ k}}^2)+M_Z^2\right)-3 \operatorname{A_0}(M_{A^0}^2) \left(-M_{\phi_{_ k}}^2+M_{A^0}^2+M_Z^2\right) \right. \right. \right. \nonumber \\
    &+ \left. \left. \left. 3 \left(-M_{\phi_{_ k}}^2+M_{A^0}^2-M_Z^2\right) \operatorname{A_0}(M_{\phi_{_ k}}^2)-6 M_{A^0}^2 M_Z^2+2 M_Z^4\right)+9 |U_\phi^{2k}+i U_\phi^{3k}|^2 M_Z^2 \operatorname{A_0}(M_{\phi_{_ k}}^2) \right] \right. \nonumber \\
    &+ \left. M_Z^2 \left(-3 \left(4 c_w^2 s_w^2-1\right) \left(M_Z^2-4 M_{\eta^\pm}^2\right) \operatorname{B_0}(M_Z^2,M_{\eta^\pm}^2,M_{\eta^\pm}^2) \right. \right. \nonumber \\
    &+ \left. \left. 12 \left(c_w^2-s_w^2\right)^2 \left(2 M_\psi^2+M_Z^2\right) \operatorname{B_0}(M_Z^2,M_\psi^2,M_\psi^2) - 12 \left(4 c_w^2 s_w^2-1\right) \operatorname{A_0}(M_{\eta^\pm}^2) \right. \right. \nonumber \\
    &- \left. \left. 24 \left(c_w^2-s_w^2\right)^2 \operatorname{A_0}(M_\psi^2) + 2 \left(2 c_w^4 \left(6 M_\psi^2-M_Z^2\right)+24 c_w^2 s_w^2 \left(M_{\eta^\pm}^2-M_\psi^2\right) - 6 M_{\eta^\pm}^2 \right. \right. \right. \nonumber \\
    &+ \left. \left. \left. 2 s_w^4 \left(6 M_\psi^2-M_Z^2\right)+M_Z^2\right)\right) \right) \,,
    \\
    \delta_{Z Z} & = \frac{e^2}{576 M_Z^4 c_w^2 s_w^2 \pi^2} \left( \sum_{ij} \left[ |\Gamma^{ij}_{Z, L}|^2 \left(-3 \left(\left(M_{\chi_{_i}}^2-M_{\chi_{_j}}^2\right)^2 \right. \right. \right. \right. \nonumber \\
    &+ \left. \left. \left. \left. 2 M_Z^4\right) \operatorname{B_0}(M_Z^2,M_{\chi_{_i}}^2,M_{\chi_{_j}}^2) + 3 \left(M_{\chi_{_j}}^2 \left(M_Z^2-2 M_{\chi_{_i}}^2\right)+M_Z^2 M_{\chi_{_i}}^2+M_{\chi_{_i}}^4+M_{\chi_{_j}}^4 \right. \right. \right. \right. \nonumber \\
    &- \left. \left. \left. \left. 2 M_Z^4\right) M_Z^2 \operatorname{B^\prime_0}(M_Z^2,M_{\chi_{_i}}^2,M_{\chi_{_j}}^2) + 3 \left(M_{\chi_{_i}}^2-M_{\chi_{_j}}^2\right) \left(\operatorname{A_0}(M_{\chi_{_i}}^2)-\operatorname{A_0}(M_{\chi_{_j}}^2)\right)+2 M_Z^4\right) \right. \right. \nonumber \\
    &+ \left. \left. 18 \text{Re}\left(\left( \Gamma^{ij}_{Z, L}\right)^2\right) M_Z^4 M_{\chi_{_i}} M_{\chi_{_j}} \operatorname{B^\prime_0}(M_Z^2,M_{\chi_{_i}}^2,M_{\chi_{_j}}^2) \right] \right. \nonumber \\
    &+ \left. \sum_k \left[ \left(U_\phi^{2k}\right)^2 \left(3 \left(M_{\phi_{_ k}}^4 \left(\operatorname{B_0}(M_Z^2,M_{A^0}^2,M_{\phi_{_ k}}^2)-M_Z^2 \operatorname{B^\prime_0}(M_Z^2,M_{A^0}^2,M_{\phi_{_ k}}^2)\right) \right. \right. \right. \right. \nonumber \\
    &+ \left. \left. \left. \left. 2 M_{\phi_{_ k}}^2 \left(\left(M_{A^0}^2+M_Z^2\right) M_Z^2 \operatorname{B^\prime_0}(M_Z^2,M_{A^0}^2,M_{\phi_{_ k}}^2)-M_{A^0}^2 \operatorname{B_0}(M_Z^2,M_{A^0}^2,M_{\phi_{_ k}}^2)\right) \right. \right. \right. \right. \nonumber \\
    &- \left. \left. \left. \left. \left(M_Z^2-M_{A^0}^2\right)^2 M_Z^2 \operatorname{B^\prime_0}(M_Z^2,M_{A^0}^2,M_{\phi_{_ k}}^2)+\left(M_{A^0}^4-M_Z^4\right) \operatorname{B_0}(M_Z^2,M_{A^0}^2,M_{\phi_{_ k}}^2) \right. \right. \right. \right. \nonumber \\
    &+ \left. \left. \left. \left. \left(M_{\phi_{_ k}}^2-M_{A^0}^2\right) \operatorname{A_0}(M_{A^0}^2)+\left(M_{A^0}^2-M_{\phi_{_ k}}^2\right) \operatorname{A_0}(M_{\phi_{_ k}}^2)\right)-2 M_Z^4\right) \right] \right. \nonumber \\
    &+ \left. M_Z^2 \left(3 M_Z^2 \left(4 c_w^2 s_w^2-1\right) \operatorname{B_0}(M_Z^2,M_{\eta^\pm}^2,M_{\eta^\pm}^2) \right. \right. \nonumber \\
    &- \left. \left. 3 \left(4 c_w^2 s_w^2-1\right) \left(4 M_{\eta^\pm}^2-M_Z^2\right) \operatorname{B^\prime_0}(M_Z^2,M_{\eta^\pm}^2,M_{\eta^\pm}^2) M_Z^2-12 M_Z^2 \left(c_w^2-s_w^2\right)^2 \operatorname{B_0}(M_Z^2,M_\psi^2,M_\psi^2) \right. \right. \nonumber \\
    &- \left. \left. 12 \left(c_w^2-s_w^2\right)^2 \left(2 M_\psi^2+M_Z^2\right) \operatorname{B^\prime_0}(M_Z^2,M_\psi^2,M_\psi^2) M_Z^2+2 M_Z^2 \left(2 c_w^4+2 s_w^4-1\right)\right) \right)\,.
\end{align}

\subsection{Higgs self-energy}
\begin{figure}[h!]
    \centering
    \begin{subfigure}[b]{0.31\textwidth}
        \centering
        \begin{tikzpicture}
        \begin{feynman}
        \vertex (a) at (0,0) {\(H\)};
        \vertex (b) at (1,0);
        \vertex (c) at (2,0);
        \vertex (d) at (3,0) {\(H\)};
        \diagram* {
        (a) -- [scalar] (b),
        (b) -- [plain, half right, edge label'=\( \chi_{_i}\)] (c),
        (b) -- [plain, half left, edge label=\( \chi_{_j}\)] (c),
        (c) -- [scalar] (d)
        };
        \end{feynman}
        \end{tikzpicture}
    \end{subfigure}
    \hfill
    \begin{subfigure}[b]{0.31\textwidth}
        \centering
        \begin{tikzpicture}
        \begin{feynman}
        \vertex (a) at (0,0) {\(H\)};
        \vertex (b) at (1,0);
        \vertex (c) at (2,0);
        \vertex (d) at (3,0) {\(H\)};
        \diagram* {
        (a) -- [scalar] (b),
        (b) -- [scalar, half right, edge label'=\( \phi_{_k}\)] (c),
        (b) -- [scalar, half left, edge label=\( \phi_{_l}\)] (c),
        (c) -- [scalar] (d)
        };
        \end{feynman}
        \end{tikzpicture}
    \end{subfigure}
    \hfill
    \begin{subfigure}[b]{0.31\textwidth}
        \centering
        \begin{tikzpicture}
        \begin{feynman}
        \vertex (a) at (0,0) {\(H\)};
        \vertex (b) at (1,0);
        \vertex (c) at (2,0);
        \vertex (d) at (3,0) {\(H\)};
        \diagram* {
        (a) -- [scalar] (b),
        (b) -- [scalar, half right, edge label'=\( \eta^\pm\)] (c),
        (b) -- [scalar, half left, edge label=\( \eta^\mp\)] (c),
        (c) -- [scalar] (d)
        };
        \end{feynman}
        \end{tikzpicture}
    \end{subfigure}
    \\
    \begin{subfigure}[b]{0.3\textwidth}
        \centering
        \begin{tikzpicture}
        \begin{feynman}
        \vertex (a) at (0,0) {\(H\)};
        \vertex (b) at (1,0);
        \vertex (c) at (2,0) {\(H\)};
        \diagram* {
        (a) -- [scalar] (b),
        b -- [scalar, loop, min distance=2cm, in=45, out=135, edge label=\( \eta^\pm\)] b,
        (b) -- [scalar] (c)
        };
        \end{feynman}
        \end{tikzpicture}
    \end{subfigure}
    \quad \quad
    \begin{subfigure}[b]{0.3\textwidth}
        \centering
        \begin{tikzpicture}
        \begin{feynman}
        \vertex (a) at (0,0) {\(H\)};
        \vertex (b) at (1,0);
        \vertex (c) at (2,0) {\(H\)};
        \diagram* {
        (a) -- [scalar] (b),
        b -- [scalar, loop, min distance=2cm, in=45, out=135, edge label=\( \phi_{_k}\)] b,
        (b) -- [scalar] (c)
        };
        \end{feynman}
        \end{tikzpicture}
    \end{subfigure}
    \caption{Contributions to the Higgs self-energy.}
    \label{fig:HSE}
\end{figure}
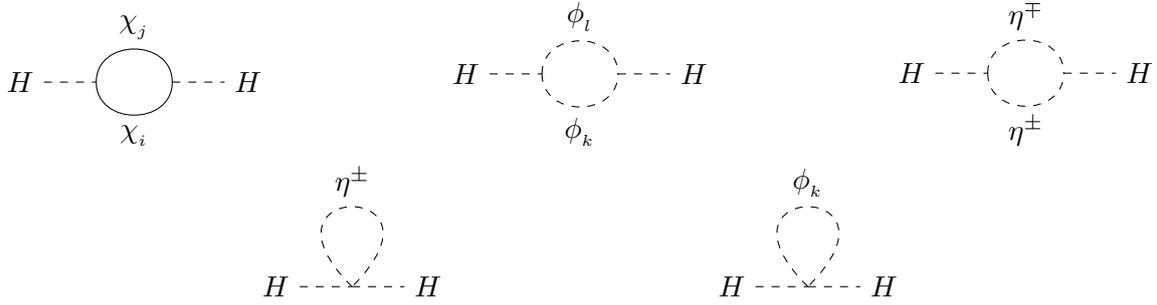

In Fig.~\ref{fig:HSE} we collect the diagrams for the Higgs self energy.

\noindent The counter-term is given by
\begin{align}\label{eq:HCT}
    \delta_H & = \frac{1}{16 \pi^2} \left( 2 \sum_{ij} \left[ \left(\operatorname{B^\prime_0}(M_H^2,M_{\chi_{_i}}^2,M_{\chi_{_j}}^2) \left(|\Gamma^{ij}_{H, L}|^2 \left(M_{\chi_{_i}}^2+M_{\chi_{_j}}^2-M_H^2\right) \right. \right. \right. \right. \nonumber\\
    &+ \left. \left. \left. \left. M_{\chi_{_i}} M_{\chi_{_j}} 
    \left(\left(\left(\Gamma^{ij}_{H, L}\right)^*\right)^2+\left(\Gamma^{ij}_{H, L}\right)^2\right)\right)-|\Gamma^{cij}_{H, L}|^2 \operatorname{B_0}(M_H^2,M_{\chi_{_i}}^2,M_{\chi_{_j}}^2)\right)  \right] \right. \nonumber\\
    &- \left. \frac{1}{8} \sum_{kl} \left[\left(\Gamma^{k l}_H\right)^2 \operatorname{B^\prime_0}(M_H^2,M_{\phi_{_ l}}^2,M_{\phi_{_ k}}^2) \right] \right. \nonumber\\
    &- \left. v^2 \lambda _{\eta }^2 \operatorname{B^\prime_0}(M_H^2,M_{\eta^\pm}^2,M_{\eta^\pm}^2) \right)\,.
\end{align}

\subsection{$\pmb{W}$-boson self-energy}
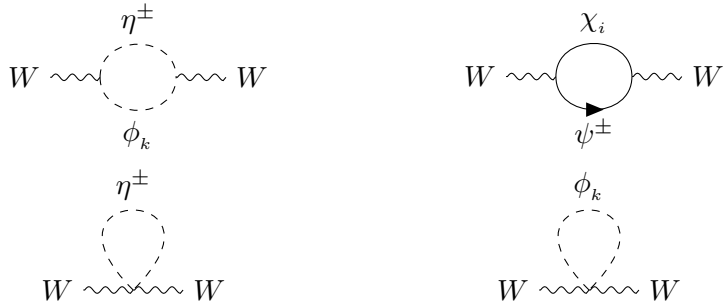
\begin{figure}[h!]
    \centering
    \begin{subfigure}[b]{0.3\textwidth}
        \centering
        \begin{tikzpicture}
        \begin{feynman}
        \vertex (a) at (0,0) {\(W\)};
        \vertex (b) at (1,0);
        \vertex (c) at (2,0);
        \vertex (d) at (3,0) {\(W\)};
        \diagram* {
        (a) -- [boson] (b),
        (b) -- [scalar, half right, edge label'=\( \phi_{_k}\)] (c),
        (b) -- [scalar, half left, edge label=\( \eta^\pm\)] (c),
        (c) -- [boson] (d)
        };
        \end{feynman}
        \end{tikzpicture}
    \end{subfigure}
    \quad \quad
    \begin{subfigure}[b]{0.3\textwidth}
        \centering
        \begin{tikzpicture}
        \begin{feynman}
        \vertex (a) at (0,0) {\(W\)};
        \vertex (b) at (1,0);
        \vertex (c) at (2,0);
        \vertex (d) at (3,0) {\(W\)};
        \diagram* {
        (a) -- [boson] (b),
        (b) -- [fermion, half right, edge label'=\( \psi^\pm\)] (c),
        (b) -- [plain, half left, edge label=\( \chi_{_i}\)] (c),
        (c) -- [boson] (d)
        };
        \end{feynman}
        \end{tikzpicture}
    \end{subfigure}
    \\
    \begin{subfigure}[b]{0.3\textwidth}
     \centering
        \begin{tikzpicture}
        \begin{feynman}
        \vertex (a) at (0,0) {\(W\)};
        \vertex (b) at (1,0);
        \vertex (c) at (2,0) {\(W\)};
        \diagram* {
        (a) -- [boson] (b),
        b -- [scalar, loop, min distance=2cm, in=45, out=135, edge label=\( \eta^\pm\)] b,
        (b) -- [boson] (c)
        };
        \end{feynman}
        \end{tikzpicture}
    \end{subfigure}
    \quad \quad
    \begin{subfigure}[b]{0.3\textwidth}
        \centering
        \begin{tikzpicture}
        \begin{feynman}
        \vertex (a) at (0,0) {\(W\)};
        \vertex (b) at (1,0);
        \vertex (c) at (2,0) {\(W\)};
        \diagram* {
        (a) -- [boson] (b),
        b -- [scalar, loop, min distance=2cm, in=45, out=135, edge label=\( \phi_{_k}\)] b,
        (b) -- [boson] (c)
        };
        \end{feynman}
        \end{tikzpicture}
    \end{subfigure}
    \hfill
    \caption{Feynman diagrams regarding the  contributions to the $W$-boson self-energy.}
    \label{fig:WSE}
\end{figure}

The diagrams for the $W$ self-energy are summarised in Fig.~\ref{fig:WSE}.

\noindent The expression for the counter-term is given by
\begin{align}\label{eq:WCT}
    \delta_{m_W^2}& = \frac{e^2}{576 c_w^2 s_w^2 M_Z^2 \pi^2} \left( \sum_i \left[ \left(|U_\chi^{3i}|^2+|U_\chi^{4 i}|^2\right) \left(2 \left(3 M_{\chi_{_i}}^2 \left(\operatorname{A_0}(M_{\chi_{_i}}^2)-\operatorname{A_0}(M_\psi^2)+2 c_w^2 M_Z^2\right) \right. \right. \right. \right. \nonumber\\
    & - \left. \left. \left. \left. 3 \left(2 c_w^2 M_Z^2+M_\psi^2\right) \operatorname{A_0}(M_{\chi_{_i}}^2)+3 \left(M_\psi^2-2 c_w^2 M_Z^2\right) \operatorname{A_0}(M_\psi^2)-2 c_w^4 M_Z^4+6 c_w^2 M_\psi^2 M_Z^2\right) \right. \right. \right. \nonumber\\
    & - \left. \left. \left. 6 \left(-2 c_w^4 M_Z^4+\left(c_w^2 M_Z^2-2 M_\psi^2\right) M_{\chi_{_i}}^2+c_w^2 M_\psi^2 M_Z^2+M_{\chi_{_i}}^4+M_\psi^4\right) \operatorname{B_0}(M_W^2,M_\psi^2,M_{\chi_{_i}}^2)\right) \right. \right. \nonumber\\
    & + \left. \left. 72 c_w^2 M_\psi M_{\chi_{_i}} M_Z^2 \text{Re}\left(U_\chi^{3 i} U_\chi^{4 i}\right) \operatorname{B_0}(M_W^2,M_\psi^2,M_{\chi_{_i}}^2) \right] \right. \nonumber \\
    &+ \left. \sum_k \left[ |U_\phi^{2k} + i U_\phi^{3k}|^2 \left(3 \left(-2 \left(c_w^2 M_Z^2+M_{\eta^\pm}^2\right) M_{\phi_{_ k}}^2+\left(M_{\eta^\pm}^2-c_w^2 M_Z^2\right)^2+M_{\phi_{_k}}^4\right) \operatorname{B_0}(M_W^2,M_{\eta^\pm}^2,M_{\phi_{_k}}^2) \right. \right. \right. \nonumber\\
    & - \left. \left. \left. 3 \operatorname{A_0}(M_{\eta^\pm}^2) \left(c_w^2 M_Z^2-M_{\phi_{_ k}}^2+M_{\eta^\pm}^2\right)+3 \left(2 c_w^2 M_Z^2-M_{\phi_{_ k}}^2+M_{\eta^\pm}^2\right) \operatorname{A_0}(M_{\phi_{_ k}}^2) \right. \right. \right. \nonumber\\
    & + \left. \left. \left. 2 c_w^2 M_Z^2 \left(c_w^2 M_Z^2-3 M_{\phi_{_ k}}^2-3 M_{\eta^\pm}^2\right)\right) \right] + 18 c_w^2 M_Z^2 \operatorname{A_0}(M_{\eta^\pm}^2) \right)\,,
\end{align}
where we have used ${M_W}/{M_Z} = c_w$ (at the tree level).

\subsection{Corrections to the photon propagator}
\begin{figure}[h!]
    \begin{subfigure}[b]{0.31\textwidth}
        \centering
        \begin{tikzpicture}
        \begin{feynman}
        \vertex (a) at (0,0) {\(\gamma\)};
        \vertex (b) at (1,0);
        \vertex (c) at (2,0);
        \vertex (d) at (3,0) {\(Z \, (\gamma)\)};
        \diagram* {
        (a) -- [boson] (b),
        (b) -- [scalar, half right, edge label'=\( \eta^\pm\)] (c),
        (b) -- [scalar, half left, edge label=\( \eta^\pm\)] (c),
        (c) -- [boson] (d)
        };
        \end{feynman}
        \end{tikzpicture}
    \end{subfigure}
    \hfill
    \begin{subfigure}[b]{0.31\textwidth}
        \centering
        \begin{tikzpicture}
        \begin{feynman}
        \vertex (a) at (0,0) {\(\gamma\)};
        \vertex (b) at (1,0);
        \vertex (c) at (2,0) {\(Z\, (\gamma)\)};
        \diagram* {
        (a) -- [boson] (b),
        b -- [scalar, loop, min distance=2cm, in=45, out=135, edge label=\( \eta^\pm\)] b,
        (b) -- [boson] (c)
        };
        \end{feynman}
        \end{tikzpicture}
    \end{subfigure}
    \hfill
    \begin{subfigure}[b]{0.31\textwidth}
        \centering
        \begin{tikzpicture}
        \begin{feynman}
        \vertex (a) at (0,0) {\(\gamma\)};
        \vertex (b) at (1,0);
        \vertex (c) at (2,0);
        \vertex (d) at (3,0) {\(Z\, (\gamma)\)};
        \diagram* {
        (a) -- [boson] (b),
        (b) -- [fermion, half right, edge label'=\( \psi^\pm\)] (c),
        (c) -- [fermion, half right, edge label'=\( \psi^\pm\)] (b),
        (c) -- [boson] (d)
        };
        \end{feynman}
        \end{tikzpicture}
    \end{subfigure}
    \caption{Contributions to the corrections to the photon propagator.}
    \label{fig:AZSE}
\end{figure}
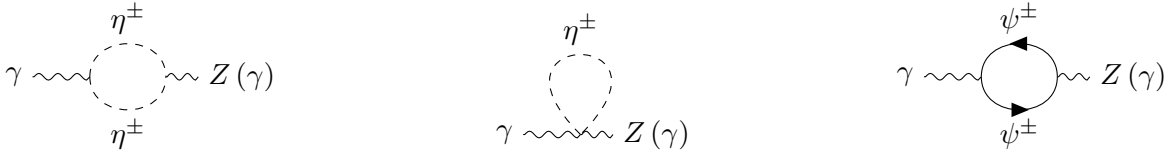

The diagrams leading to corrections to the photon propagator are collected in Fig.~\ref{fig:AZSE}.

\noindent The following quantities can be derived, for off-diagonal and self-energy,
\begin{align}\label{eq:AZCT}
    \delta_{\gamma Z} & = \frac{e^2 (c_w^2 - s_w^2)}{144 \pi^2 c_w s_w M_Z^2} \left(2 \left(-6 \left(2 M_\psi^2+M_Z^2\right) \operatorname{B_0}(M_Z^2,M_\psi^2,M_\psi^2)+6 M_{\eta^\pm}^2-12 M_\psi^2+M_Z^2\right) \right. \nonumber \\
    &+ \left. 3 \left(4 M_{\eta^\pm}^2-M_Z^2\right) \operatorname{B_0}(M_Z^2,M_{\eta^\pm}^2,M_{\eta^\pm}^2)-12 \operatorname{A_0}(M_{\eta^\pm}^2)+24 \operatorname{A_0}(M_\psi^2)\right) \,,
\\
    \delta_{Z \gamma} & = 0\,,
    \\
        \delta_{\gamma \gamma} & = \frac{e^2}{144 \pi^2} \left(12 M_{\eta^\pm}^2 \operatorname{B^\prime_0}(0,M_{\eta^\pm}^2,M_{\eta^\pm}^2)-3 \operatorname{B_0}(0,M_{\eta^\pm}^2,M_{\eta^\pm}^2) \right. \nonumber \\
    &- \left. 12 \operatorname{B_0}(0,M_\psi^2,M_\psi^2)-24 M_\psi^2 \operatorname{B^\prime_0}(0,M_\psi^2,M_\psi^2)+2\right)\,.
\end{align}

\section{Oblique parameters}\label{app:Oblique}
The non-vanishing parameters can be cast in an (approximate) analytical way as follows
\begin{align}\label{eq::OblParamDef}
        S &= \frac{4 s_W^2 c_W^2}{\bar \alpha} \left[ \frac{\partial \Sigma_T^{Z Z}(p^2)}{\partial p^2} \Biggr{\rvert}_{p^2=0} - \frac{c_W^2 - s_W^2}{s_W c_W} \frac{\partial \Sigma_T^{\gamma Z}(p^2)}{\partial p^2} \Biggr{\rvert}_{p^2=0} - \frac{\partial \Sigma_T^{\gamma \gamma}(p^2)}{\partial p^2} \Biggr{\rvert}_{p^2=0}\right]\,, \nonumber \\
        T &= \frac{1}{\bar \alpha} \left[ \frac{\Sigma_T^{W W}(0)}{M_W^2} - \frac{\Sigma_T^{Z Z}(0)}{M_Z^2} \right] \,, \nonumber\\
        U &= \frac{4 s_W^2}{\bar \alpha} \left[ \frac{\partial \Sigma_T^{W W}(p^2)}{\partial p^2} \Biggr{\rvert}_{p^2=0} - c_W^2  \frac{\partial \Sigma_T^{Z Z}(p^2)}{\partial p^2} \Biggr{\rvert}_{p^2=0} - 2 s_W c_W \frac{\partial \Sigma_T^{\gamma Z}(p^2)}{\partial p^2} \Biggr{\rvert}_{p^2=0} - s_W^2 \frac{\partial \Sigma_T^{\gamma \gamma}(p^2)}{\partial p^2} \Biggr{\rvert}_{p^2=0}\right]\,,
\end{align}
in which $\bar \alpha$ denotes the fine-structure constant $\bar \alpha = e^{2}/4\pi$ and $\Sigma^{XY}_T$ is the coefficient of the metric tensor $g^{\mu\nu}$ entering in the vacuum polarisation tensor~\cite{Grimus:2008nb},
\begin{equation}
    - i \Sigma^{X Y}_{\mu \nu}(p^2) = - i \left( g_{\mu \nu} - \frac{p_\mu p_\nu}{p^2}\right) \Sigma^{XY}_T (p^2) - i \frac{p_\mu p_\nu}{p^2} \Sigma^{XY}_L (p^2),
\end{equation}
with  $XY$ standing for $\gamma\gamma$, $Z \gamma$, $ZZ$, or $WW$. The full expressions of the $S$, $T$ and $U$ parameters can then be expressed as
\begin{align}\label{eq:Obliqueexpr}
    S & = \frac{1}{72 \pi} \left( \sum_{ij} \left[|\Gamma^{ij}_{Z, L}|^2 \left( \phantom{\rlap{$\frac{3 \left(M_{\chi_{_i}}^2-M_{\chi_{_j}}^2\right)^2}{M_Z^4}$}} 12 \operatorname{B_0}(0,M_{\chi_{_i}}^2,M_{\chi_{_j}}^2)-6 \left(M_{\chi_{_i}}^2+M_{\chi_{_j}}^2\right) \operatorname{B^\prime_0}(0,M_{\chi_{_i}}^2,M_{\chi_{_j}}^2) \right. \right. \right. \nonumber \\
    & - \left. \left. \left. \frac{3 \left(M_{\chi_{_i}}^2-M_{\chi_{_j}}^2\right)^2}{M_Z^4} f\left(\frac{M_{\chi_{_j}}^2}{M_{\chi_{_i}}^2},\frac{M_Z^2}{M_{\chi_{_i}}^2}\right)-4\right)-36 \text{Re}\left(\left(\Gamma^{ij}_{Z, L}\right)^2\right) M_{\chi_{_i}} M_{\chi_{_j}} \operatorname{B^\prime_0}(0,M_{\chi_{_i}}^2,M_{\chi_{_j}}^2) \right] \right. \nonumber \\
    & + \left. \sum_k \left[ \left(U_\phi^{2k}\right)^2 \left( \phantom{\rlap{$\frac{3 \left(M_{\chi_{_i}}^2-M_{\chi_{_j}}^2\right)^2}{M_Z^4}$}} 6 \operatorname{B_0}(0,M_{A^0}^2,M_{\phi_{_ k}}^2)-12 \left(M_{\phi_{_ k}}^2+M_{A^0}^2\right) \operatorname{B^\prime_0}(0,M_{A^0}^2,M_{\phi_{_ k}}^2) \right. \right. \right. \nonumber \\
    & + \left. \left. \left. \frac{3 \left(M_{A^0}^2-M_{\phi_{_ k}}^2\right)^2}{M_Z^4} f\left(\frac{M_{\phi_{_ k}}^2}{M_{A^0}^2},\frac{M_Z^2}{M_{A^0}^2}\right)+4\right) \right] \right. \nonumber \\
    & + \left. 24 M_{\eta^\pm}^2 \operatorname{B^\prime_0}(0,M_{\eta^\pm}^2,M_{\eta^\pm}^2)-6 \operatorname{B_0}(0,M_{\eta^\pm}^2,M_{\eta^\pm}^2)-24 \operatorname{B_0}(0,M_\psi^2,M_\psi^2) \right. \nonumber \\
    & - \left. 48 M_\psi^2 \operatorname{B^\prime_0}(0,M_\psi^2,M_\psi^2)+4 \phantom{\rlap{$\frac{3 \left(M_{\chi_{_i}}^2-M_{\chi_{_j}}^2\right)^2}{M_Z^4}$}}\right)\,,
    \\
    T & = \frac{1}{48 c_w^2 M_Z^2 \pi s_w^2} \left( \sum_i \left[ 2 \left(|U_\chi^{3 i}|^2+|U_\chi^{4 i}|^2\right) \left(-\left(M_{\chi_{_i}}^2+M_\psi^2\right) \operatorname{B_0}(0,M_\psi^2,M_{\chi_{_i}}^2) \right. \right. \right. \nonumber \\
    & - \left. \left. \left. \left(M_\psi^2-M_{\chi_{_i}}^2\right)^2 \operatorname{B^\prime_0}(0,M_\psi^2,M_{\chi_{_i}}^2)-2 \operatorname{A_0}(M_{\chi_{_i}}^2)-2 \operatorname{A_0}(M_\psi^2)+2 \left(M_{\chi_{_i}}^2+M_\psi^2\right)\right) \right. \right. \nonumber \\
    & + \left. \left. 24 M_\psi M_{\chi_{_i}} Re\left(U_\chi^{3 i} U_\chi^{4 i}\right) \operatorname{B_0}(0,M_\psi^2,M_{\chi_{_i}}^2) \right. \right. \nonumber \\
    & + \left. \left. \sum_j \left[ |\Gamma^{ij}_{Z, L}|^2 \left(M_{\chi_{_j}}^4 \operatorname{B^\prime_0}(0,M_{\chi_{_i}}^2,M_{\chi_{_j}}^2)+M_{\chi_{_j}}^2 \left(-2 M_{\chi_{_i}}^2 \operatorname{B^\prime_0}(0,M_{\chi_{_i}}^2,M_{\chi_{_j}}^2)+\operatorname{B_0}(0,M_{\chi_{_i}}^2,M_{\chi_{_j}}^2)-2\right) \right. \right. \right. \right. \nonumber \\
    & + \left. \left. \left. \left. M_{\chi_{_i}}^2 \left(M_{\chi_{_i}}^2 \operatorname{B^\prime_0}(0,M_{\chi_{_i}}^2,M_{\chi_{_j}}^2)+\operatorname{B_0}(0,M_{\chi_{_i}}^2,M_{\chi_{_j}}^2)-2\right)+2 \operatorname{A_0}(M_{\chi_{_i}}^2)+2 \operatorname{A_0}(M_{\chi_{_j}}^2)\right) \right. \right. \right. \nonumber \\
    & + \left. \left. \left. 6 \text{Re}\left(\left(\Gamma^{ij}_{Z, L}\right)^2\right) M_{\chi_{_i}} M_{\chi_{_j}} \operatorname{B_0}(0,M_{\chi_{_i}}^2,M_{\chi_{_j}}^2) \right] \right] \right. \nonumber \\
    & + \left. \sum_k \left[ \left(U_\phi^{2k}\right)^2 \left(2 \left(M_{\phi_{_ k}}^2+M_{A^0}^2\right) \operatorname{B_0}(0,M_{A^0}^2,M_{\phi_{_ k}}^2)-\left(M_{A^0}^2-M_{\phi_{_ k}}^2\right)^2 \operatorname{B^\prime_0}(0,M_{A^0}^2,M_{\phi_{_ k}}^2)+\operatorname{A_0}(M_{\phi_{_ k}}^2) \right. \right. \right. \nonumber \\
    & + \left. \left. \left. \operatorname{A_0}(M_{A^0}^2)+2 \left(M_{\phi_{_ k}}^2+M_{A^0}^2\right)\right)+|U_\phi^{2k}+i U_\phi^{3k}|^2 \left(-2 \left(M_{\phi_{_ k}}^2+M_{\eta^\pm}^2\right) \operatorname{B_0}(0,M_{\eta^\pm}^2,M_{\phi_{_ k}}^2) \right. \right. \right. \nonumber \\
    & + \left. \left. \left. \left(M_{\eta^\pm}^2-M_{\phi_{_ k}}^2\right)^2 \operatorname{B^\prime_0}(0,M_{\eta^\pm}^2,M_{\phi_{_ k}}^2)-\operatorname{A_0}(M_{\phi_{_ k}}^2)-\operatorname{A_0}(M_{\eta^\pm}^2)-2 \left(M_{\phi_{_ k}}^2+M_{\eta^\pm}^2\right)\right) \right] \right. \nonumber \\
    & + \left. 4 M_{\eta^\pm}^2 \left(1-4 c_w^2 s_w^2\right) \operatorname{B_0}(0,M_{\eta^\pm}^2,M_{\eta^\pm}^2)-8 M_\psi^2 \left(c_w^2-s_w^2\right)^2 \operatorname{B_0}(0,M_\psi^2,M_\psi^2)+2 \left(8 c_w^2 s_w^2+1\right) \operatorname{A_0}(M_{\eta^\pm}^2) \right. \nonumber \\
    & + \left. 8 \left(c_w^2-s_w^2\right)^2 \operatorname{A_0}(M_\psi^2)+4 M_{\eta^\pm}^2 \left(1-4 c_w^2 s_w^2\right)-8 M_\psi^2 \left(c_w^2-s_w^2\right)^2 \right)\,,
    \\
    U & = \frac{1}{72 \pi} \left( \sum_i \left[ 2 \left(|U_\chi^{3 i}|^2+|U_\chi^{4 i}|^2\right) \left( \phantom{\rlap{$\frac{3 \left(M_{\chi_{_i}}^2-M_{\chi_{_j}}^2\right)^2}{M_Z^4}$}} 12 \operatorname{B_0}(0,M_\psi^2,M_{\chi_{_i}}^2)-6 \left(M_{\chi_{_i}}^2+M_\psi^2\right) \operatorname{B^\prime_0}(0,M_\psi^2,M_{\chi_{_i}}^2) \right. \right. \right. \nonumber \\
    & - \left. \left. \left. \frac{3 \left(M_\psi^2-M_{\chi_{_i}}^2\right)^2}{M_Z^4} f\left(\frac{M_{\chi_{_i}}^2}{M_\psi^2},\frac{M_Z^2}{M_\psi^2}\right)-4\right)+144 M_\psi M_{\chi_{_i}} \text{Re}\left(U_\chi^{3 i} U_\chi^{4 i}\right) \operatorname{B^\prime_0}(0,M_\psi^2,M_{\chi_{_i}}^2) \right. \right. \nonumber \\
    & + \left. \left. \sum_j \left[ |\Gamma^{ij}_{Z, L}|^2 \left( \phantom{\rlap{$\frac{3 \left(M_{\chi_{_i}}^2-M_{\chi_{_j}}^2\right)^2}{M_Z^4}$}} -12 \operatorname{B_0}(0,M_{\chi_{_i}}^2,M_{\chi_{_j}}^2)+6 \left(M_{\chi_{_i}}^2+M_{\chi_{_j}}^2\right) \operatorname{B^\prime_0}(0,M_{\chi_{_i}}^2,M_{\chi_{_j}}^2) \right. \right. \right. \right. \nonumber \\
    & + \left. \left. \left. \left. \frac{3 \left(M_{\chi_{_i}}^2-M_{\chi_{_j}}^2\right)^2}{M_Z^4} f\left(\frac{M_{\chi_{_j}}^2}{M_{\chi_{_i}}^2},\frac{M_Z^2}{M_{\chi_{_i}}^2}\right)+4\right)+36 \text{Re}\left(\left(\Gamma^{ij}_{Z, L}\right)^2\right) M_{\chi_{_i}} M_{\chi_{_j}} \operatorname{B^\prime_0}(0,M_{\chi_{_i}}^2,M_{\chi_{_j}}^2) \right]\right] \right. \nonumber \\
    & - \left. \sum_k \left[ \left(\left(U_\phi^{2k}\right)^2 \left( \phantom{\rlap{$\frac{3 \left(M_{\chi_{_i}}^2-M_{\chi_{_j}}^2\right)^2}{M_Z^4}$}} 6 \operatorname{B_0}(0,M_{A^0}^2,M_{\phi_{_ k}}^2)-12 \left(M_{\phi_{_ k}}^2+M_{A^0}^2\right) \operatorname{B^\prime_0}(0,M_{A^0}^2,M_{\phi_{_ k}}^2) \right. \right. \right. \right. \nonumber \\
    & + \left. \left. \left. \left. \frac{3 \left(M_{A^0}^2-M_{\phi_{_ k}}^2\right)^2}{M_Z^4} f\left(\frac{M_{\phi_{_ k}}^2}{M_{A^0}^2},\frac{M_Z^2}{M_{A^0}^2}\right)+4\right)-|U_\phi^{2k}+i U_\phi^{3k}|^2 \left( \phantom{\rlap{$\frac{3 \left(M_{\chi_{_i}}^2-M_{\chi_{_j}}^2\right)^2}{M_Z^4}$}} 6 \operatorname{B_0}(0,M_{\eta^\pm}^2,M_{\phi_{_ k}}^2) \right. \right. \right. \right. \nonumber \\
    & - \left. \left. \left. \left. 12 \left(M_{\phi_{_ k}}^2+M_{\eta^\pm}^2\right) \operatorname{B^\prime_0}(0,M_{\eta^\pm}^2,M_{\phi_{_ k}}^2)+ \frac{3 \left(M_{\eta^\pm}^2-M_{\phi_{_ k}}^2\right)^2}{M_Z^4} f\left(\frac{M_{\phi_{_ k}}^2}{M_{\eta^\pm}^2},\frac{M_Z^2}{M_{\eta^\pm}^2}\right)+4\right)\right) \right] \right. \nonumber \\
    & + \left. 24 M_{\eta^\pm}^2 \operatorname{B^\prime_0}(0,M_{\eta^\pm}^2,M_{\eta^\pm}^2)-6 \operatorname{B_0}(0,M_{\eta^\pm}^2,M_{\eta^\pm}^2)-24 \operatorname{B_0}(0,M_\psi^2,M_\psi^2) \right. \nonumber\\
    &- \left. 48 M_\psi^2 \operatorname{B^\prime_0}(0,M_\psi^2,M_\psi^2)+4 \phantom{\rlap{$\frac{3 \left(M_{\chi_{_i}}^2-M_{\chi_{_j}}^2\right)^2}{M_Z^4}$}}  \right)\,.
\end{align}
In the above, $\operatorname{B^\prime_0}$ corresponds to the first derivative of the Passarino-Veltman function $\operatorname{B_0}$ with respect to its first argument and a new (loop) function must be further defined; it corresponds to the second derivative of $B_0$ with respect to the external momentum squared at $p^2=0$
\begin{align}\label{eq:loopCT2}
    f(x,y) & = \frac{y^2 ((x-1) [(x+10) x+1]-6 x (x+1) \log (x))}{3 (x-1)^5}\, , \quad 
        \forall x \neq 1\, ;
       \quad 
        f(1,y) = \frac{y^2}{30} \,.
\end{align}

{
\small
\bibliographystyle{JHEP}
\bibliography{DLOT2025}
}

\end{document}